\documentclass{article}

\usepackage{arxiv}

\usepackage[utf8]{inputenc} 
\usepackage[T1]{fontenc}    
\usepackage{placeins}
\usepackage{hyperref}       
\usepackage{url}            
\usepackage{booktabs}       
\usepackage{amsfonts}       
\usepackage{nicefrac}       
\usepackage{microtype}      
\usepackage{lipsum}		
\usepackage{graphicx}
\usepackage{natbib}
\usepackage{doi}

\usepackage{rotating}
\usepackage{graphicx}
\newcommand\sbullet[1][.5]{\mathbin{\vcenter{\hbox{\scalebox{#1}{$\bullet$}}}}}
\usepackage{soul}

\usepackage{afterpage}
\usepackage[inline]{enumitem}
\usepackage{float}
\usepackage{amsmath}
\usepackage{hyperref}
\hypersetup{
    colorlinks=true,
    linkcolor=red,
    filecolor=magenta,      
    urlcolor=cyan,
    pdftitle={Overleaf Example},
    pdfpagemode=FullScreen,
    }
\usepackage{cancel}
\usepackage[inline]{enumitem}
\usepackage{xcolor}
\usepackage{caption}
\usepackage{subcaption}
\usepackage{amssymb}
\usepackage{booktabs}
\usepackage{siunitx}
\usepackage{longtable}
\usepackage{algorithm}
\usepackage{algpseudocode}
\newcolumntype{L}{@{}l@{}} 

\usepackage{appendix}
\graphicspath{ {./imgs/}}

\DeclareCaptionLabelFormat{andtable}{#1~#2  \&  \tablename~\thetable}

\usepackage{authblk}

\usepackage{xspace}

\newcommand*{\ie}{i.e.\@\xspace}

\title{\textit{LDA2Net}: Digging under the surface of COVID-19 topics in scientific literature}

\author[1,2]{Giorgia Minello}
\author[3,4]{Carlo R.M.A. Santagiustina}
\author[1]{Massimo Warglien}

\affil[1]{\centering \vspace{0.1cm}Department of Management, Ca' Foscari University, Venice, Italy }

\affil[2]{ \centering \vspace{0.1cm}Strategic Management Department, IESE Business School, Barcelona, Spain}

\affil[3]{\centering \vspace{0.1cm}Department of Philosophy and Cultural Heritage, Ca' Foscari University, Venice, Italy}

\affil[4]{\centering \vspace{0.1cm}Venice International University, Venice, Italy  }

\affil[  ]{\vspace{0.2cm}{\ttfamily\{\,giorgia.minello, carlo.santagiustina, warglien\,\}\,@unive.it} \vspace{0.2cm}}

\renewcommand{\shorttitle}{\textit{LDA2Net}: Digging under the surface of COVID-19 topics in literature}

\hypersetup{
pdftitle={},
pdfsubject={q-bio.NC, q-bio.QM},
pdfauthor={Minello Santagiustina Warglien},
pdfkeywords={LDA, Network, Bigram, Covid },
}

\begin{document}

\maketitle

\begin{abstract}
During the COVID-19 pandemic, 
the scientific literature related to SARS-COV-2 has been growing dramatically, both in terms of the number of publications and of its impact on people's life. This literature encompasses a varied set of  sensible topics, ranging from vaccination, to protective equipment efficacy, to lockdown policy evaluation. Up to now, hundreds of thousands of papers have been uploaded on online repositories and published in scientific journals. As a result, the development of digital methods that allow an in-depth exploration of this growing literature has become a relevant issue, both to identify the topical trends of COVID-related research and to zoom-in its sub-themes. 
This work proposes a novel methodology, called \textit{LDA2Net}, which combines topic modelling and  network analysis to investigate topics under their surface. Specifically, \textit{LDA2Net} exploits the frequencies of pairs of consecutive words to reconstruct the network structure of topics discussed in the Cord-19 corpus. The results suggest that the effectiveness of topic models can be magnified by enriching them with word network representations, and by using the latter to display, analyse, and explore COVID-related topics at different levels of granularity.
\end{abstract}

\keywords{SARS-CoV-2 \and   COVID-19 \and  Cord-19 \and  Natural Language Processing \and  Text Mining  \and  Topic Modeling \and   Latent Dirichlet Allocation ·\and   Bigrams \and  Networks \and  Graphs \and  Centrality Measures \and  Community Detection \and  Automatic Labeling}

\section{Introduction}

 The massive response of the scientific community to the COVID-19 pandemic has produced an unprecedented amount of research and related publications. For example, the Cord-19 corpus \citep{wang2020cord} currently includes more than five hundred thousand published peer reviewed and pre-peer reviewed articles. This huge volume of COVID-related works and the fast emergence of new research branches far exceeds the human capability to meaningfully organize and explore such material. This makes it impossible, even to a specialist, to map, explore and summarise such a massive corpus of documents without the help of automated tools that can extract and classify useful semantic information from unstructured texts.  A key issue related to the analysis of large scientific corpora consists in identifying topics that may be spread across many disciplines or different research branches.  
 
 
Topic models are one of the most intuitive and transparent families of statistical methods developed to categorise documents and extract semantic information from textual corpora, in order to capture which subject matters are contained in each document.
Given the simplicity of topic modelling implementations, many papers, like \cite{colavizza2021scientometric}, have categorised and summarised COVID-related literature contained in the Cord-19 corpus. 
However, despite offering a categorization of documents, these analyses often remain at a surface-level, in particular for what concerns the exploration and understanding of topic contents and of their internal structure. 

Latent Dirichlet Allocation (LDA) \citep{blei2003latent} is indisputably the most frequently used topic modelling approach and, despite its simplicity, it has established itself as the state of the art Probabilistic Graphical Model in numerous applied research fields. 
Most criticism to LDA has been addressing some of its statistical limitations, such as the the lack of a unambiguous criteria for choosing the number of topics, the inability to capture correlations between topics, or its static nature. Many solutions to these shortcomings have been proposed in further developments of the basic probabilistic approach to topic modelling.  
Recent research about topic models has focused on extending or modifying LDA to account for syntax  \citep{boyd2010syntactic}, correlations between topics \citep{blei2006correlated}, semantic data \citep{guo2011semantic}, and metadata \citep{roberts2013structural}, mainly to overcome problems related to the simplifying assumptions of the LDA model. 
However, less attention has been devoted to address the limitations of LDA and related models deriving from their ``bag-of-words" approach that neglects word order. LDA returns a (weighted) list of words for each topic, disregarding the short-distance semantic information contained in the sequences in which words are arranged.
Even when extended to consider \textit{n-grams}, that is sequences of \textit{n} consecutive words, rather than words as basic units of analysis, it returns a weighted list of such entities, without considering their interrelations.
Attempts to include and model syntagmatic information (i.e. information concerning  sequential relations between words) in topic models have already been investigated in \cite{mackay1995hierarchical, wallach2006topic, wang2007topical, yan2013biterm, nallapati2007sparse}. 


In this paper, we address the ``bag-of-words" limit of  LDA by complementing it with the relational information provided by bigrams (pairs  of two consecutive words) associated to each topic. The idea of associating topic models and network analysis is not new.  \cite{gerlach2018network} have suggested to encode the relationships between words and documents in hypergraphs and extract topics as communities of the hypergraph. Our approach is different, as we integrate LDA with the bigrams graph of each topic with the aim of increasing the interpretability of the topics and allowing to discover their internal semantic structure at a finer level.

One of the key differences with previously mentioned studies is that, in this work the  enrichment of the topic model occurs only after topics ae inferred using unigrams - therefore topic prevalence in documents is not affected by bigrams. However bigrams are used to explore and summarise at a deeper level the syntagmatic structure of topics inferred with LDA. 
  
The proposed method, that we label  \textit{LDA2Net},  is deterministic and non parametric, and allows to incorporate in LDA bigram information without any statistical assumption on the data generation process for word sequences, being based on observed document-level bigram counts. 
Even though the method is here applied to a LDA model, it can be implemented with other topic models that can be estimated on words (\ie unigrams), such as CTM \citep{blei2006correlated} or STM \citep{roberts2013structural}.


The contribution of this paper is  fourfold:  (i)  we provide an intuitive method that enhances the interpretation of topics by exploiting document-level bigram information to reconstruct topic networks; (ii) we explore a set of metrics to evaluate the topic-specific relevance of words and of their associations; (iii) we map and cluster topics using aggregate statistical measures, which allow to differentiate cross-cutting topics from specialised ones; finally (iv) we automatically extract and label subtopics from highly modular topics.
The COVID-19 literature is vast and heterogeneous, and even specialists cannot easily navigate it – nor can they ignore connections to streams of literature which are unfamiliar to them.  \textit{LDA2Net} allows to summarise and make sense of topics more rapidly and in a human-friendly way, by allowing a transparent and intuitive interpretation of single topics through visual inspection of topic-specific word networks. Moreover, the bigrams in the word networks constrain human interpretation with respect to simple weighted lists of words, making it less arbitrary.  The proposed approach also allows to work on multiple levels of granularity, enabling the exploration of the subtopics of interest. Automatic topic label suggestions enormously facilitate navigating the large number of topics to find those of relevance for specific interest: it took hours to an expert to identify a label for all topics extracted from the Cord-19 corpus (see Table \ref{tab:topic_labels} in the Appendices), not to mention subtopics. Finally, \textit{LDA2Net} provides ground for developing more sophisticated search strategies, based on relevant sequences of words.

The remainder of this paper is organised as follows. 
In Section~\ref{sec:results},  \textit{LDA2Net} is applied to the Cord-19 corpus. The main results obtained are summarised and compared to the classical LDA  outputs. In Section~\ref{sec:discussion}, the core findings of the current work are discussed and related to classic LDA. Also, the limits and possible extensions of the proposed method are highlighted. Finally, in Section~\ref{sec:methods},  the methodological and computational details of each step of  \textit{LDA2Net} are reported.
Additional information is provided in the article's Appendices.

\begin{figure}[H]
    \centering
    \includegraphics[width=0.99\textwidth]{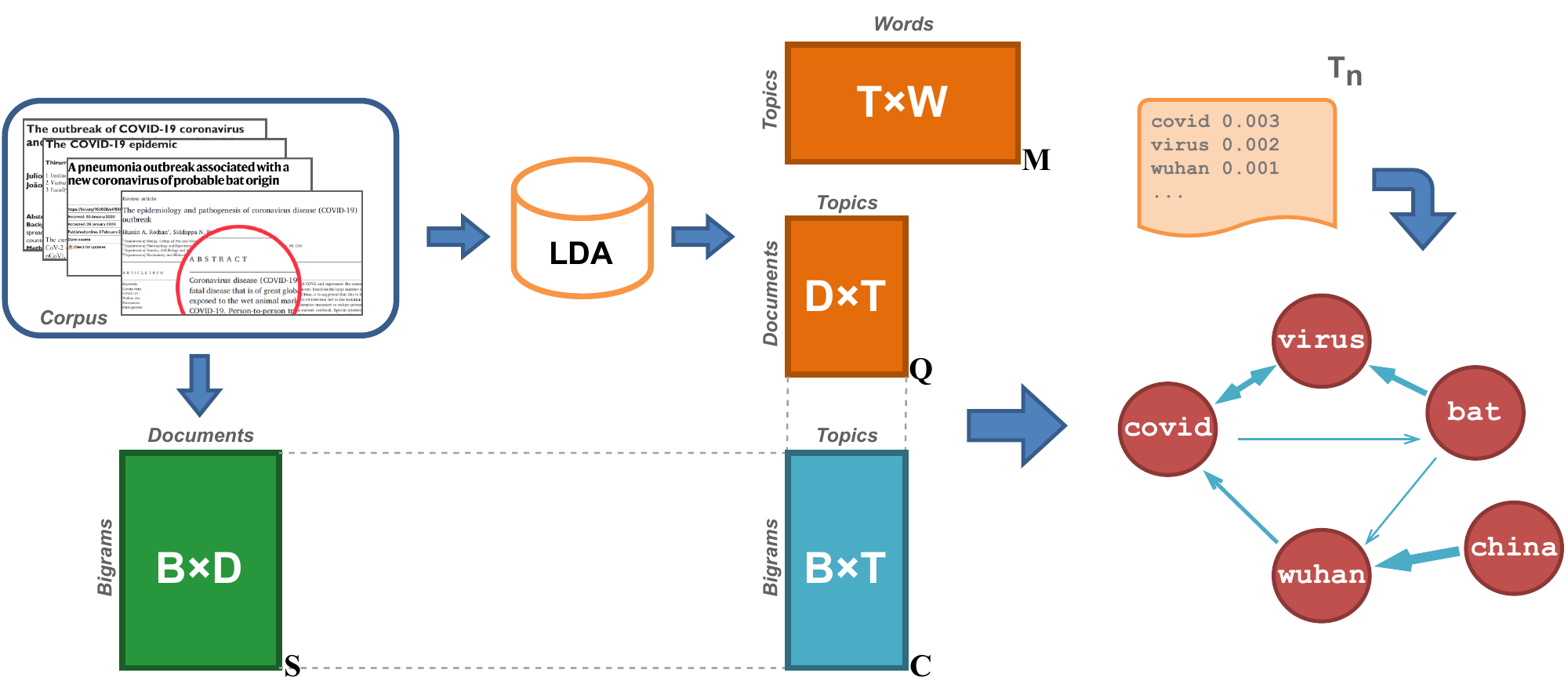}
    \caption{From LDA output to networks, a summary diagram. }
    \label{fig:ldatograph}
\end{figure}

\section{Results}\label{sec:results}

A LDA model with 120 topics is estimated using a subset of the abstracts of the Cord-19 corpus. The filtered corpus includes ($398 818$) documents published after December 31, 2019 (date in which the WHO was first-informed of COVID-related cases of pneumonia in Wuhan).
In the sections that follow, we present a very-compact summary of relevant aspects of the proposed methodology together with associated results and key findings. For corpus, data pre-processing, computational and methodological details please see Section~\ref{sec:methods} and  Appendices.



For a visual summary of each topic we  refer the reader   to  supplements  available at the following link:
\begin{center}
    \url{https://github.com/carlosantagiustina/underthesurfaceofCOVID19topics}
\end{center}

\paragraph{Topics as networks}

The first contribution of this work concerns the readability of topics.
LDA can be used to discover topics in a large collection of documents and provides a (weighted) list of words for each topic. 
However, the interpretation of these word lists is often difficult and arbitrary as LDA does not provide any information on word associations and sequences. Our approach addresses this very problem by transforming topics, \ie weighted list of words obtained via LDA, into (weighted and directed) networks that are capable to capture (short-distance) word relations.  Basically, \textit{LDA2Net} converts each LDA topic into a network where nodes represent words and edges represent relations between them.  Such a conversion is carried out by applying weights to edges between words, representing the strength and direction of their sequential arrangement in the corpus. These weights are based on the combination, through matrix multiplication, of observed frequencies of bigrams in documents and LDA's output matrices (see Figure \ref{fig:ldatograph}). A more technical  and exhaustive description of the proposed approach can be found in Section \ref{subsec:networkcreation}. In essence,  \textit{LDA2Net} makes topics more transparent and readable by binding their interpretation through observed word associations.
In  Figure  \ref{fig:ex1} and Figure \ref{fig:ex2}, we show both the list of the top 25 words by posterior topic-word probability distributions obtained through LDA, and the graph of the top 25 bigrams (\ie ordered word pairs) by \textit{LDA2Net} weights for two topics that have been selected for illustrative purposes, that is topic $\#50$ (Figure \ref{fig:ex1}) and $\#88$ (Figure \ref{fig:ex2}). 
Edge widths in the two figures are a function of topic-specific bigram weights obtained through \textit{LDA2Net}. The two Figures highlight the interpretive advantage offered by the proposed method. By observing the network made by the top 25 edges of topic $ \#50 $  (Figure~\ref{fig:ex1}) we can immediately notice the marginal role played by the word \texttt{CI}, which by contrast is the heaviest one in the topic-word distribution obtained through LDA. Interestingly, the word \texttt{risk}, which is ninth in terms of probability, appears to play a relevant role in the top bigrams network.
A reader using only probabilities would probably focus  his/her attention on the confidence interval acronym (\texttt{CI}), possibly missing that the core issue of the topic is the measurement of COVID-related risk factors.

\begin{minipage}{\textwidth}
\vspace{0.3cm}
  \resizebox{3cm}{!}{
  \begin{minipage}[b]{0.19\textheight}
    \centering
    \begin{tabular}{rlr}
      \hline
     rank & unigram & prob. \\ 
      \hline
    1 & CI & 0.2283 \\ 
      2 & prevalence & 0.0446 \\ 
      3 & ratio & 0.0380 \\ 
      4 & among & 0.0378 \\ 
      5 & interval & 0.0267 \\ 
      6 & adjusted & 0.0250 \\ 
      7 & study & 0.0227 \\ 
      8 & association & 0.0209 \\ 
      9 & risk & 0.0198 \\ 
      10 & compared & 0.0180 \\ 
      11 & RR & 0.0172 \\ 
      12 & higher & 0.0163 \\ 
      13 & respectively & 0.0158 \\ 
      14 & included & 0.0142 \\ 
      15 & analysis & 0.0113 \\ 
      16 & aor & 0.0110 \\ 
      17 & analyses & 0.0109 \\ 
      18 & multivariable & 0.0105 \\ 
      19 & ratios & 0.0105 \\ 
      20 & conducted & 0.0093 \\ 
      21 & estimated & 0.0087 \\ 
      22 & age & 0.0086 \\ 
      23 & total & 0.0083 \\ 
      24 & overall & 0.0082 \\ 
      25 & cohort & 0.0077 \\ 
      \hline
    \end{tabular}
       \captionof*{table}{LDA top 25 words by prob.}
    \end{minipage}}
 \begin{minipage}[b]{0.89\textwidth}
    \centering
    \includegraphics[width=0.65\textwidth]{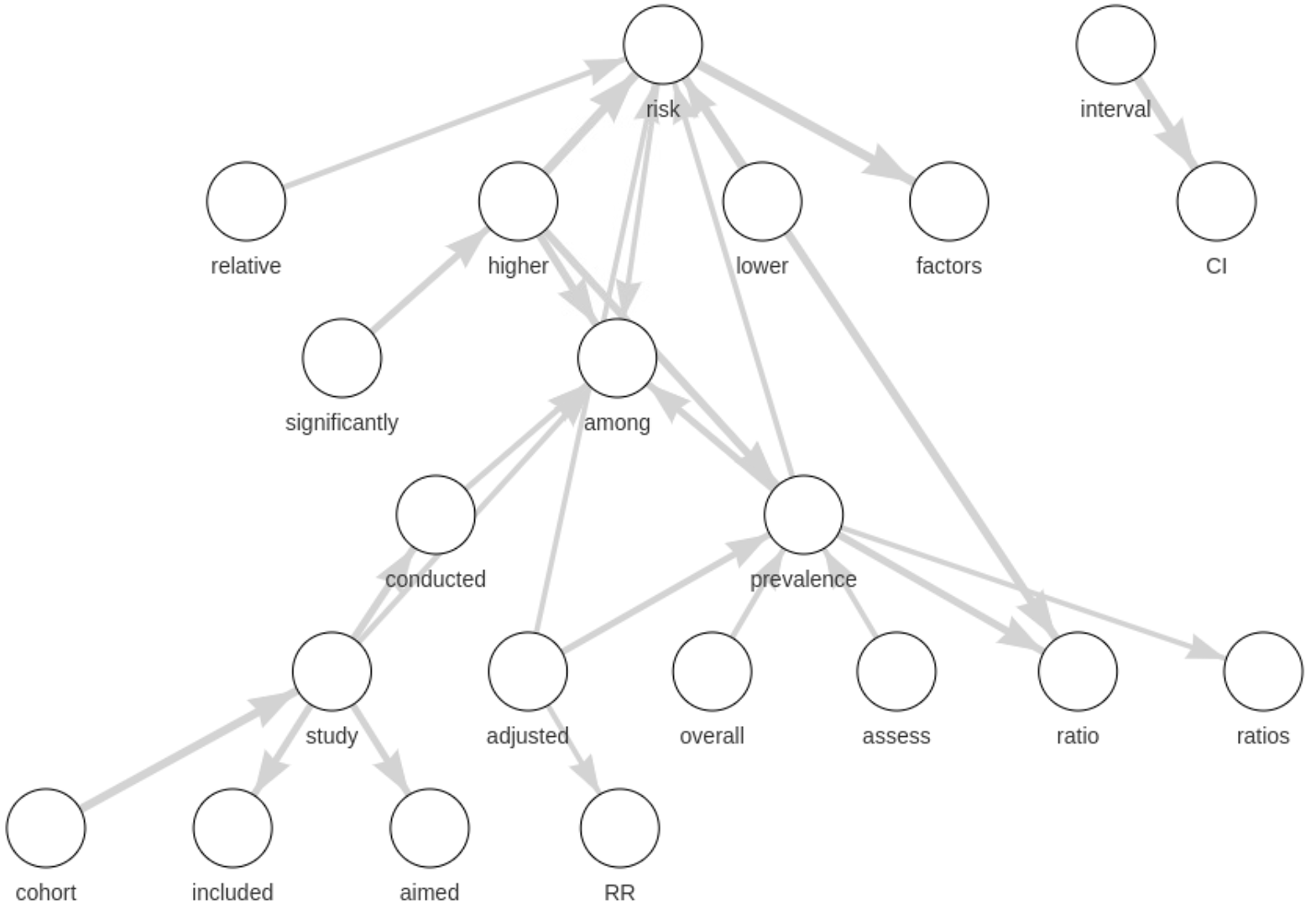}
  
    \captionof{figure}{Topic \#50. Network of top 25 bigrams by \textit{LDA2Net}-weight.}\label{fig:ex1}
  \end{minipage}
\end{minipage}
\vspace{0.3em}

 The  network made by the top 25 bigrams of topic $\#88$ (Figure~\ref{fig:ex2}) contains many disconnected components, which suggests that this topic is highly modular: we observe strong relationships between words within word communities and weak relationships between words in different communities.  Disconnected components also suggest the presence of many subtopics, each representing separable but related aspects of topic $\#88$.
For example, while \texttt{binding$\rightarrow$affinity} and \texttt{viral$\rightarrow$replication} are distinct issues they are often mentioned together in literature, in particular in paper abstracts. 
As both figures prove, top 25 bigrams networks can already facilitate the topic
interpretation process of LDA outputs, and allow to unveil the topic-specific word association structures which cannot be identified through top-word lists.
  
\vspace{0.3em}
\begin{minipage}{\textwidth}

  \resizebox{3cm}{!}{
  \begin{minipage}[b]{0.19\textheight}
    \centering

        \begin{tabular}{rlr}
          \hline
         rank & unigram & prob. \\ 
        \hline
        1 & compounds & 0.0289 \\ 
          2 & binding & 0.0226 \\ 
          3 & drug & 0.0220 \\ 
          4 & activity & 0.0173 \\ 
          5 & drugs & 0.0163 \\ 
          6 & inhibitors & 0.0162 \\ 
          7 & antiviral & 0.0152 \\ 
          8 & protease & 0.0151 \\ 
          9 & target & 0.0145 \\ 
          10 & main & 0.0111 \\ 
          11 & molecules & 0.0108 \\ 
          12 & viral & 0.0101 \\ 
          13 & targets & 0.0090 \\ 
          14 & replication & 0.0089 \\ 
          15 & mpro & 0.0083 \\ 
          16 & site & 0.0076 \\ 
          17 & inhibitor & 0.0076 \\ 
          18 & RNA & 0.0070 \\ 
          19 & inhibition & 0.0066 \\ 
          20 & affinity & 0.0065 \\ 
          21 & promising & 0.0064 \\ 
          22 & compound & 0.0063 \\ 
          23 & potent & 0.0063 \\ 
          24 & novel & 0.0062 \\ 
          25 & interactions & 0.0060 \\ 
          \hline
        \end{tabular}
       \captionof*{table}{ LDA top 25 words by prob.}
    \end{minipage}}
 \begin{minipage}[b]{0.89\textwidth}
    \centering
    \includegraphics[width=0.99\textwidth]{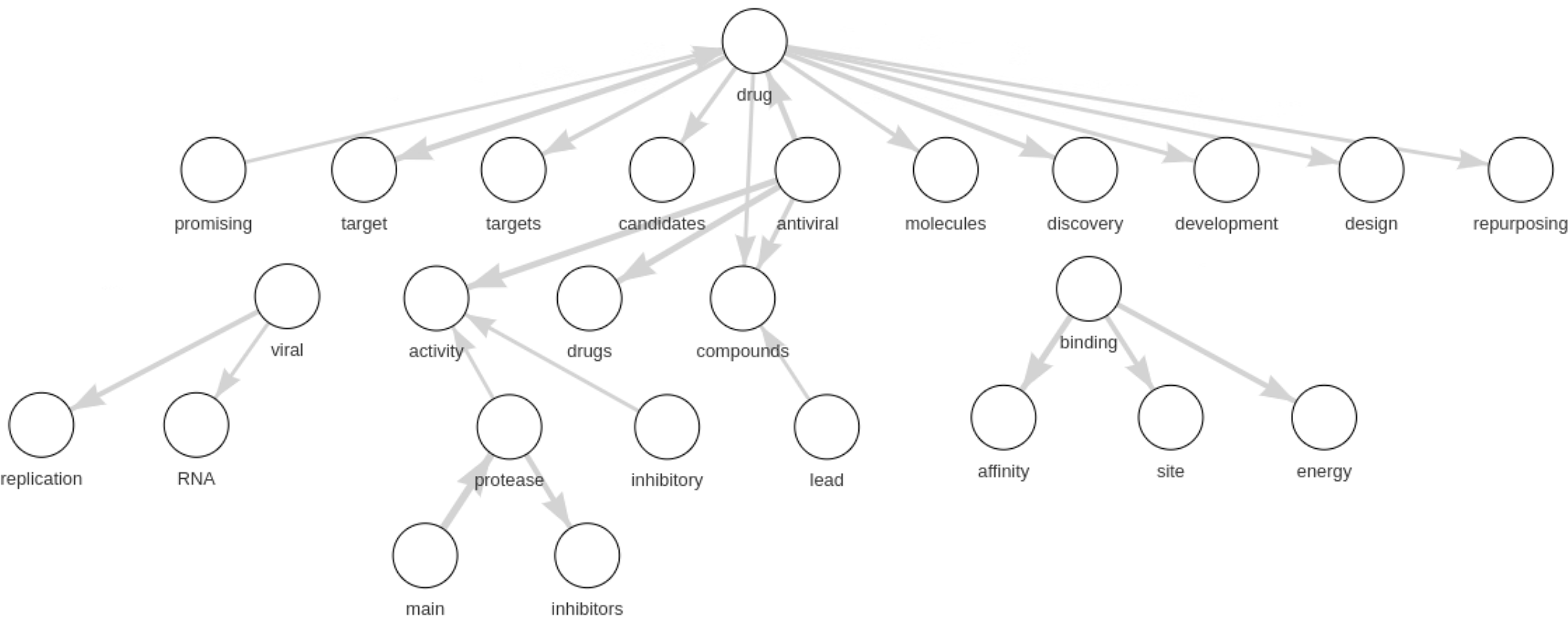}
  
    \captionof{figure}{Topic $\#88$. Network of top 25 bigrams by \textit{LDA2Net}-weight.}\label{fig:ex2}
  \end{minipage}
  \vspace{0.3cm}
\end{minipage}

We emphasize that the \textit{LDA2Net} method,  by enriching (a-posteriori) LDA topics with bigram information, does not affect the categorisation of documents in topics: document-topic distributions are the ones obtained from LDA. However, \textit{LDA2Net} adds to LDA topic-specific word relation and ordering information that can facilitate the exploration, analysis, interpretation and comparison of topics through their networks. 

\paragraph{Word metrics} 
Different aspects, related to the relevance of words in topics, can be captured and analysed through metrics based on topic network structures, which can be used in combination with the LDA word probabilities.
For this purpose, we here investigate the importance of words based on different metrics.  Recall that in \textit{LDA2Net} a network is composed by nodes, representing words, and edges, representing directed relations between pairs of words (namely bigrams).  For each word, node influence metrics, also called centralities, are derived from topic-specific networks and employed to assign word scores; details can be found in Appendix   \ref{sec:appendix_nettheory}. In Figure \ref{fig:wordcloud} we depict, through word clouds, words' importance based on six different metrics, five of which are weighted node centrality  measures. The larger is the font size the greater is the influence of that word within that topic, based on a given metric. With the exception of the first column, which refers to LDA word probabilities, each column represents a centrality measure, while  each row corresponds to a topic.
We observe, for instance, that for topic in the middle row, corresponding to topic $\#50$, all network-based metrics de-emphasize the role of the term \texttt{CI} providing complementary perspectives centered on associations to the word \texttt{risk}. The betweenness centrality helps making sense of the topic's relational context, while in- and out-degree allow to single-out directional relations between words. Finally Page-rank combines, by construction, the information provided by degree-measures and the local influence of words. 

\begin{figure}[ht!]
\centering

\begin{subfigure}[t]{0.15\textwidth}
\centering
\includegraphics[width=\textwidth]{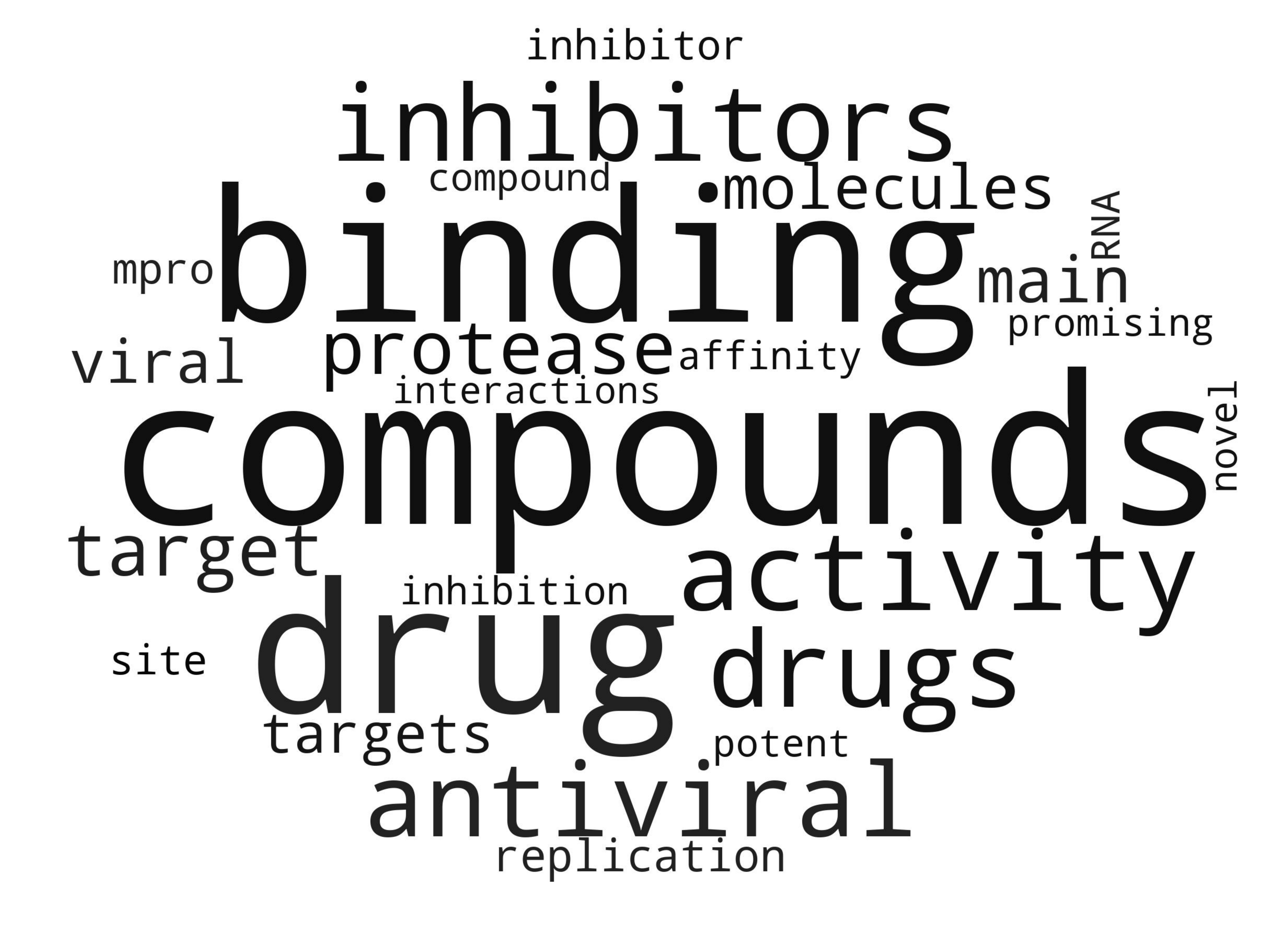}

\end{subfigure}%
~ 
\begin{subfigure}[t]{0.15\textwidth}
\centering
\includegraphics[width=\textwidth]{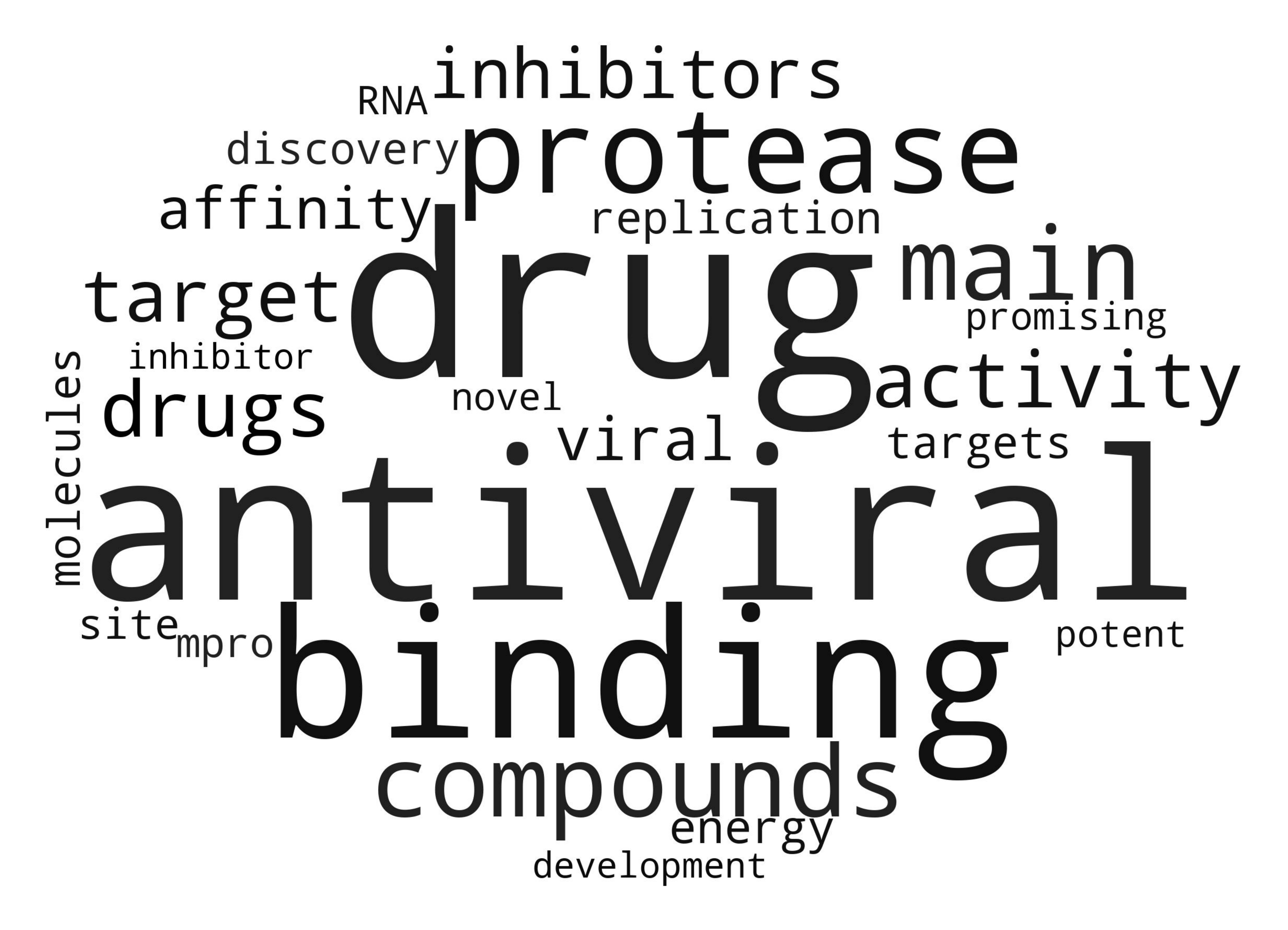}

\end{subfigure}
~ 
\begin{subfigure}[t]{0.15\textwidth}
\centering
\includegraphics[width=\textwidth]{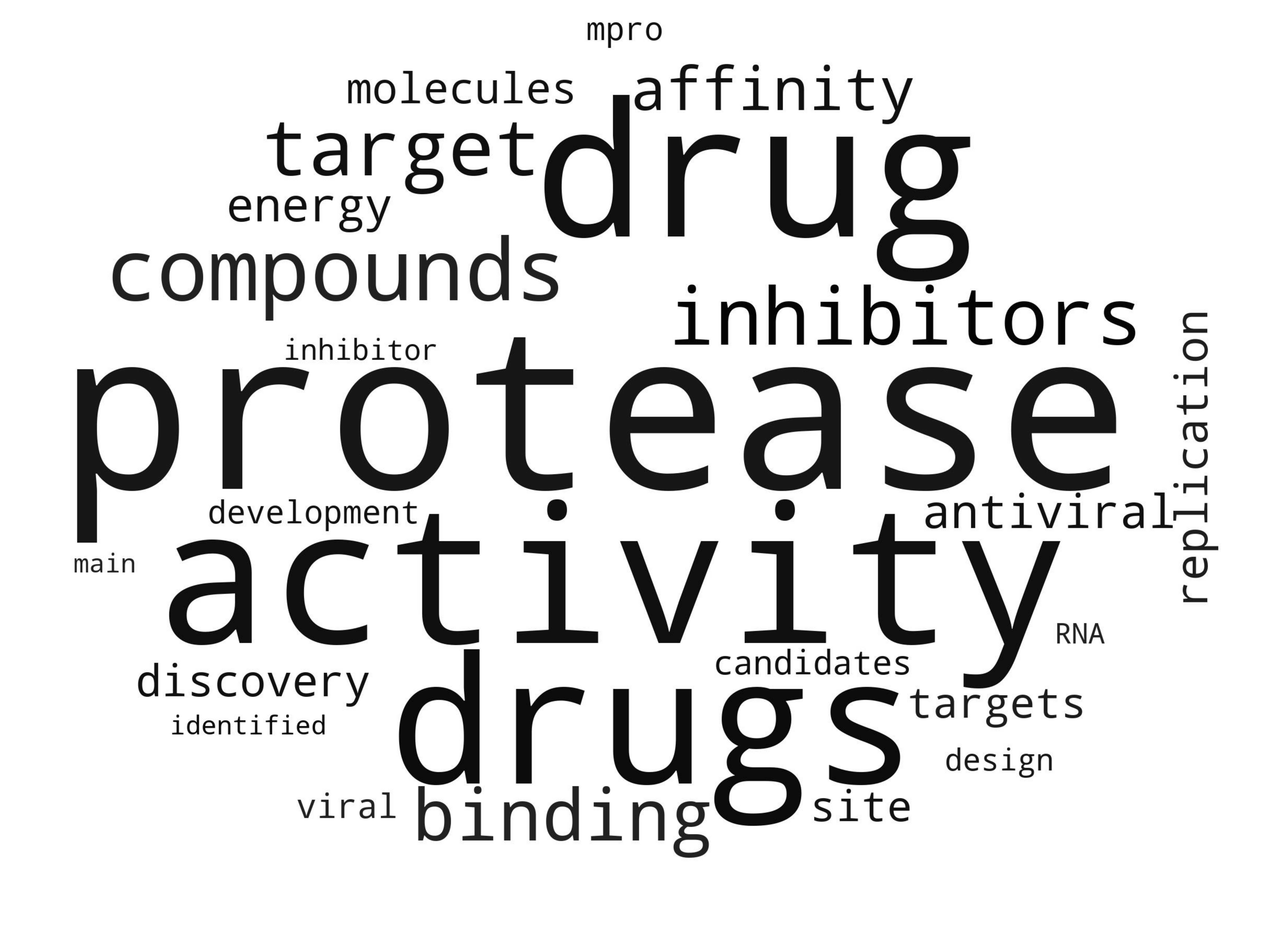}

\end{subfigure}
~ 
\begin{subfigure}[t]{0.15\textwidth}
\centering
\includegraphics[width=\textwidth]{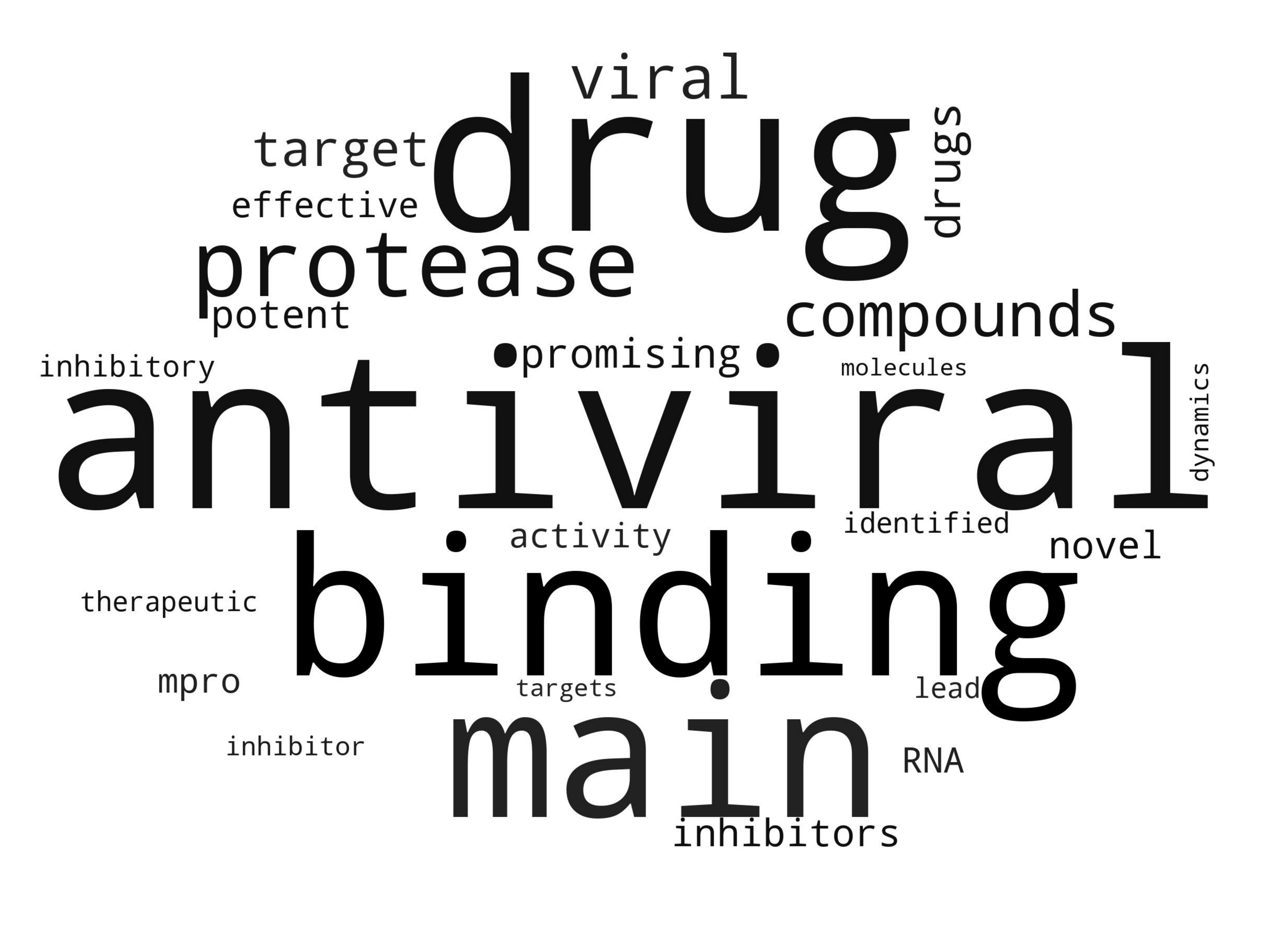}

\end{subfigure}
~ 
\begin{subfigure}[t]{0.15\textwidth}
\centering
\includegraphics[width=\textwidth]{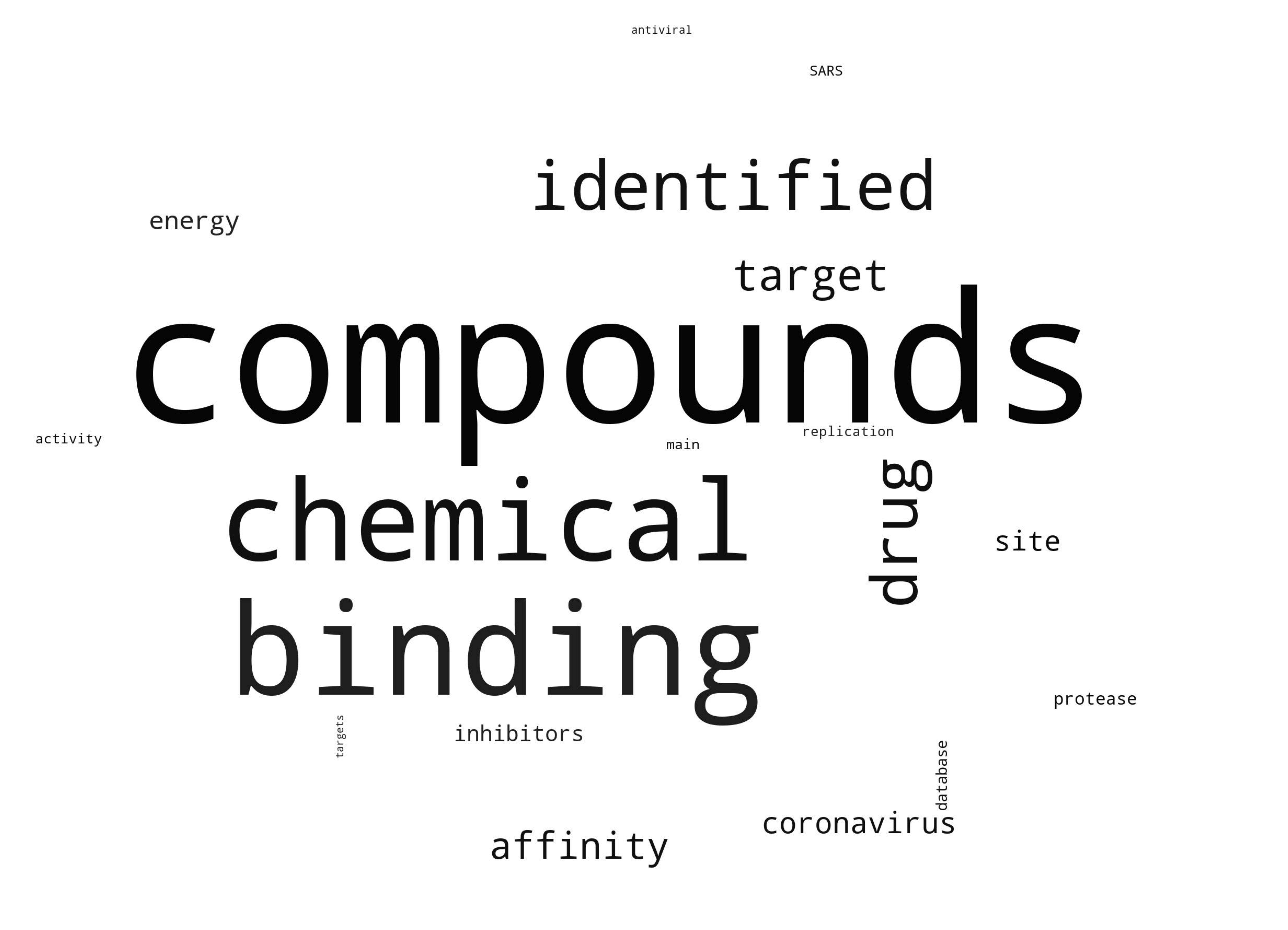}

\end{subfigure}
~ 
\begin{subfigure}[t]{0.15\textwidth}
\centering
\includegraphics[width=\textwidth]{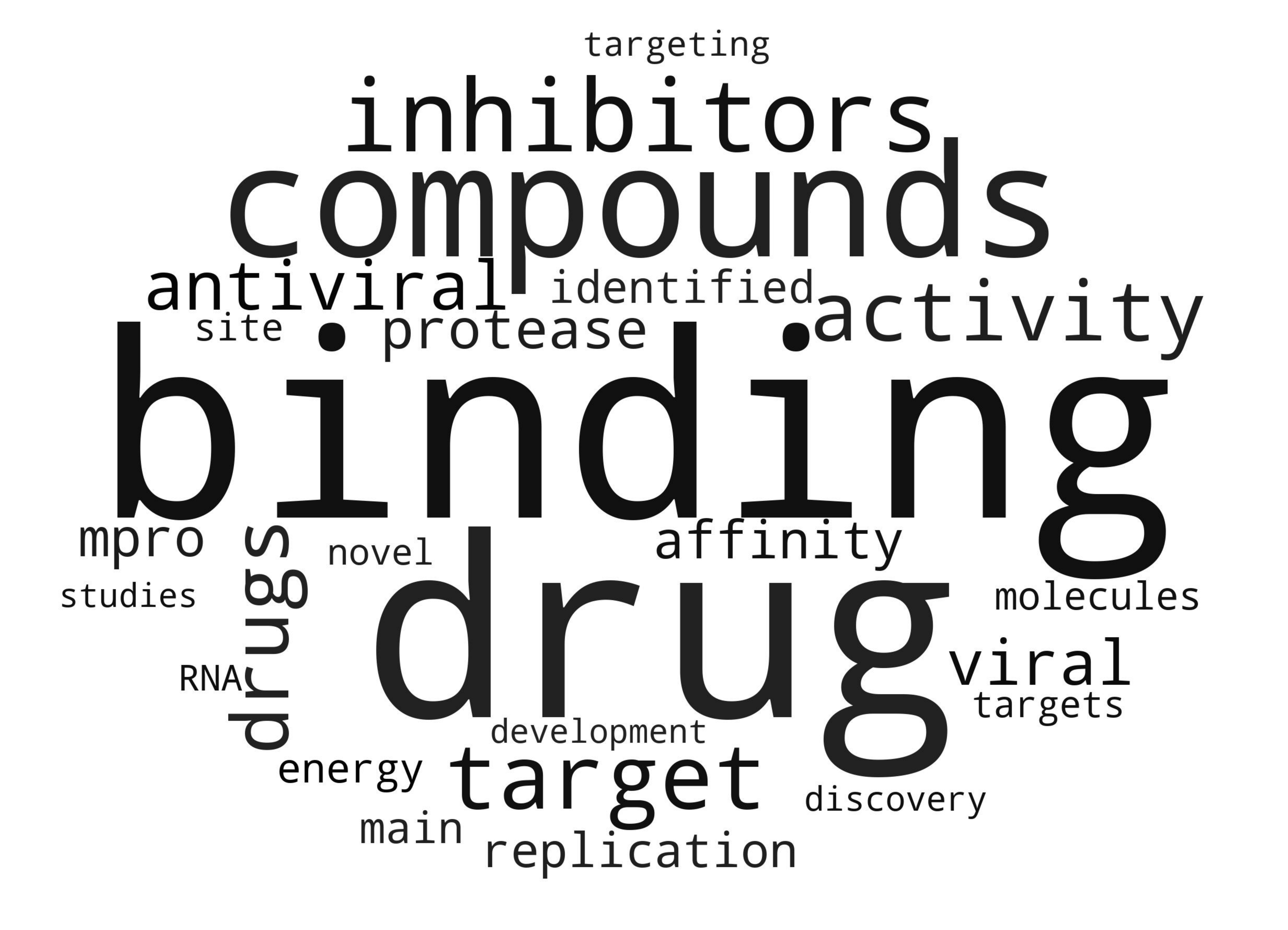}

\end{subfigure}

\begin{subfigure}[t]{0.15\textwidth}
\centering
\includegraphics[width=\textwidth]{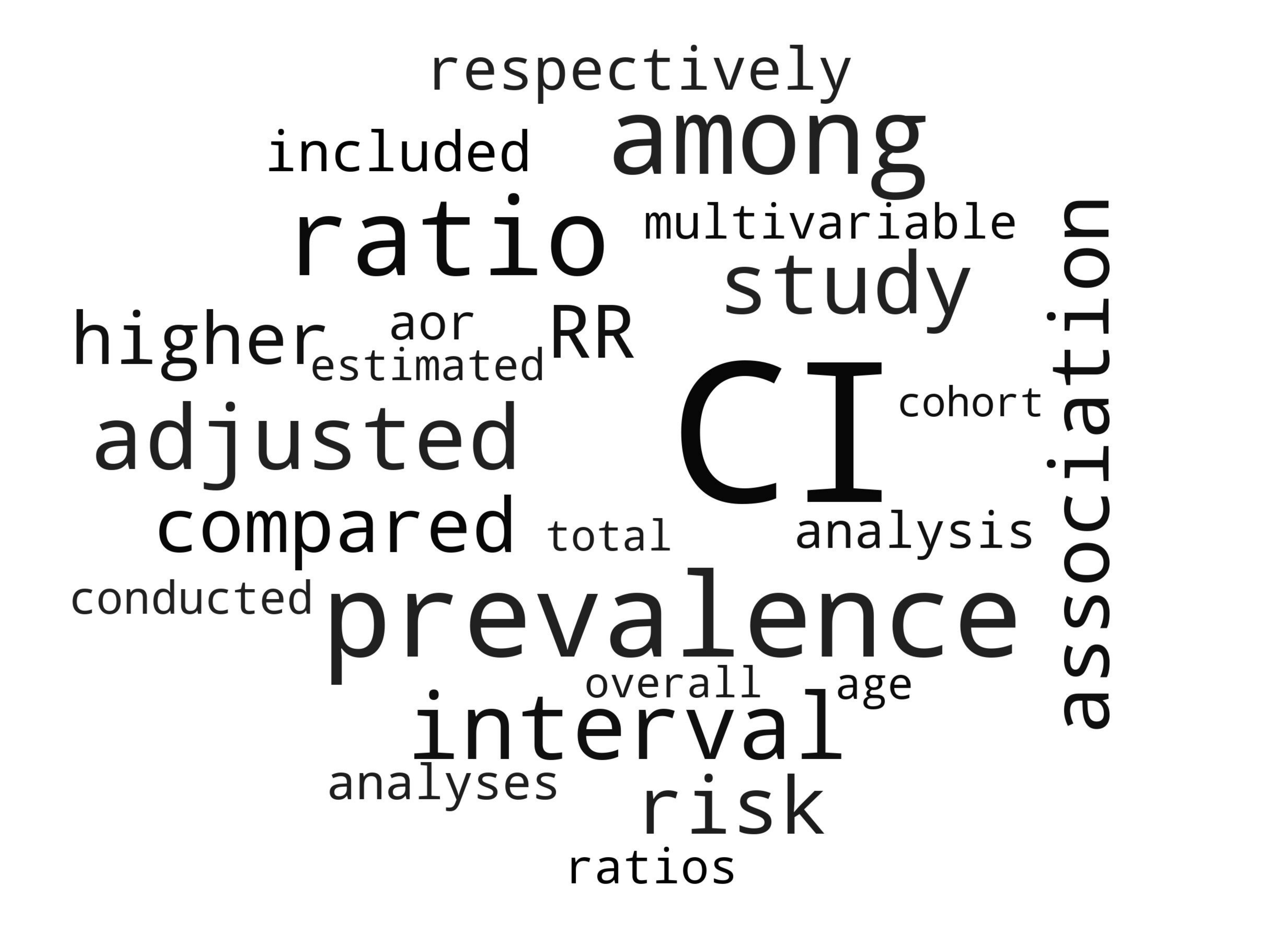}
\end{subfigure}%
~ 
\begin{subfigure}[t]{0.15\textwidth}
\centering
\includegraphics[width=\textwidth]{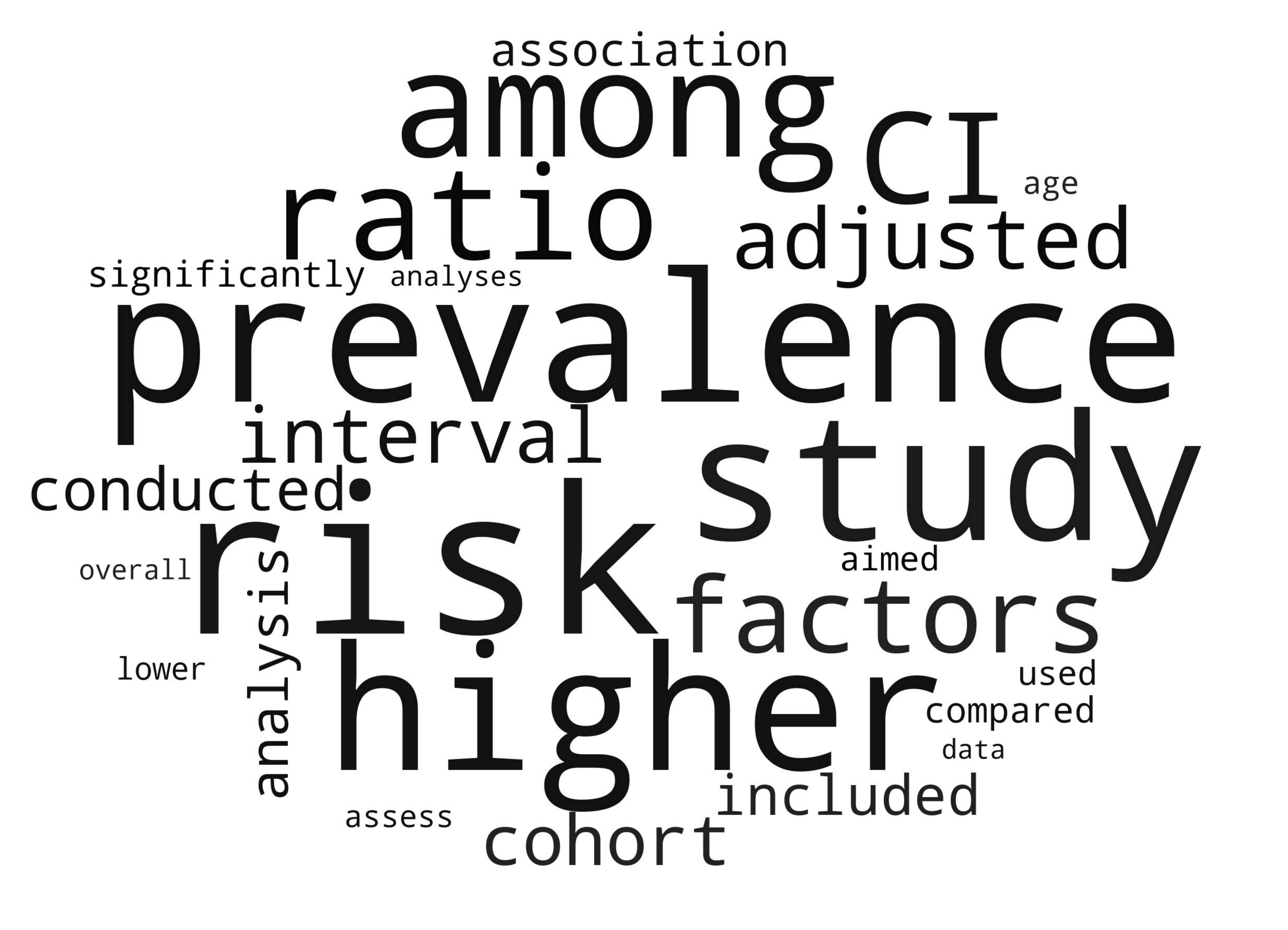}
\end{subfigure}
~ 
\begin{subfigure}[t]{0.15\textwidth}
\centering
\includegraphics[width=\textwidth]{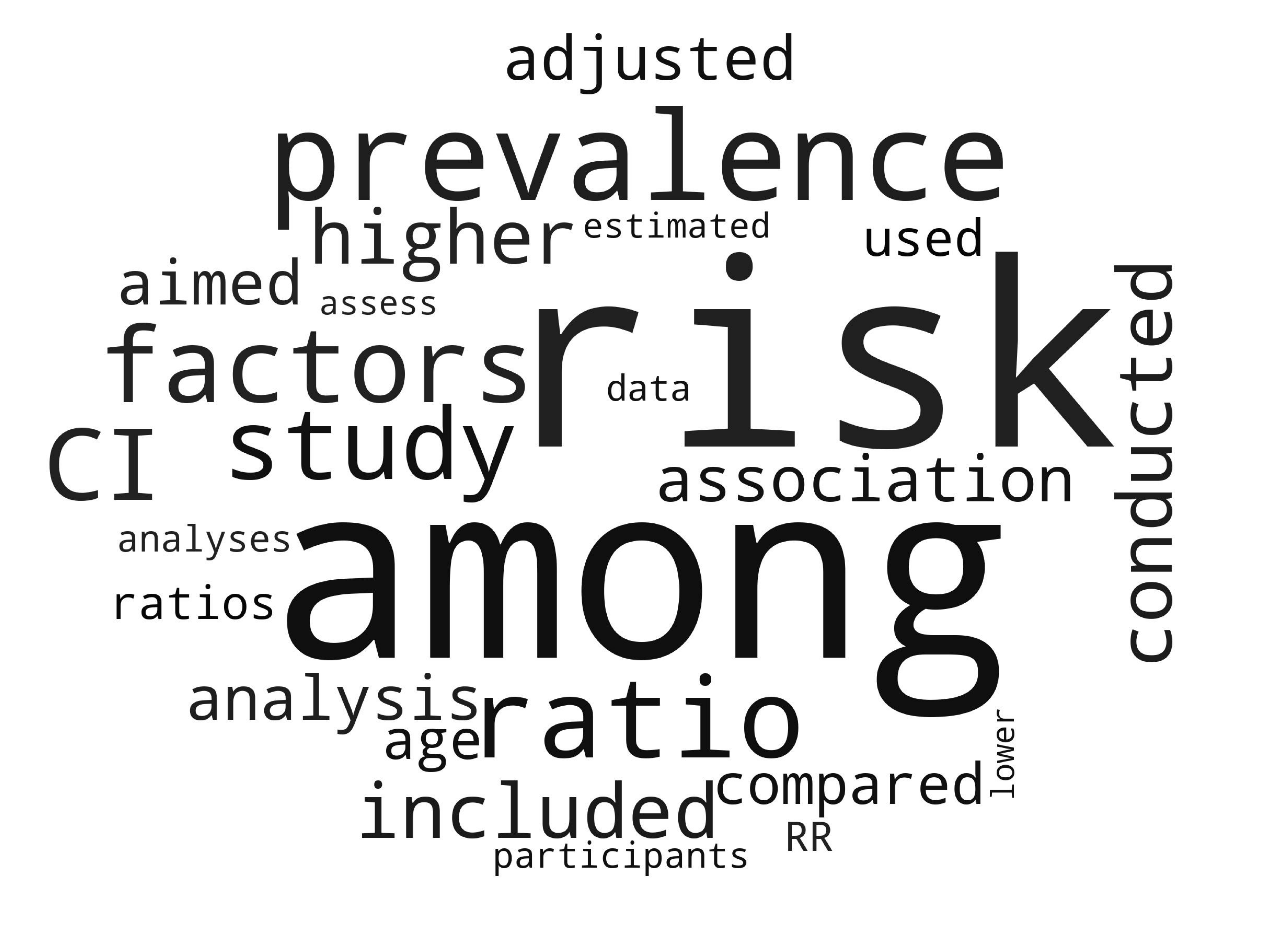}
\end{subfigure}
~ 
\begin{subfigure}[t]{0.15\textwidth}
\centering
\includegraphics[width=\textwidth]{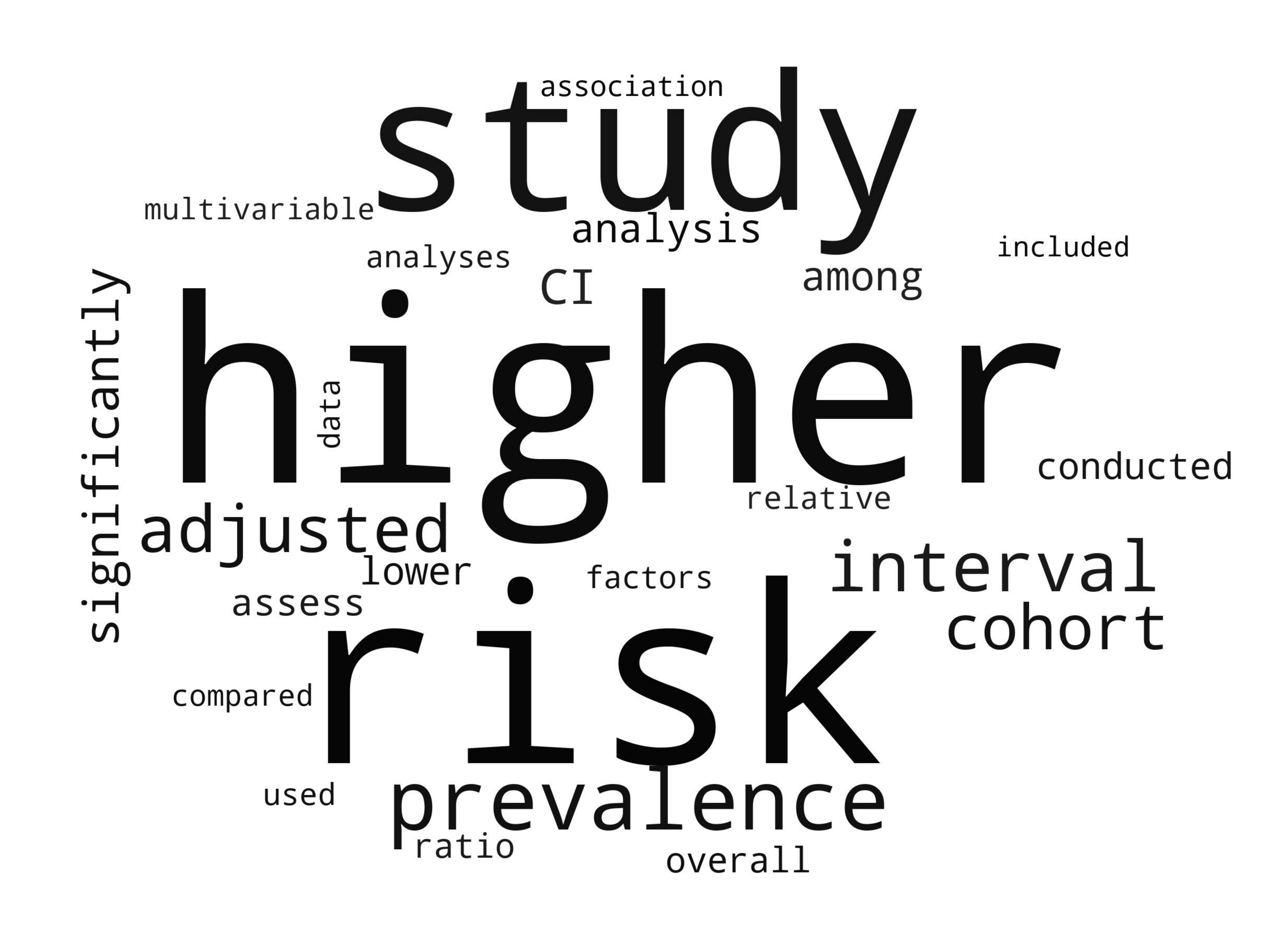}
\end{subfigure}
~ 
\begin{subfigure}[t]{0.15\textwidth}
\centering
\includegraphics[width=\textwidth]{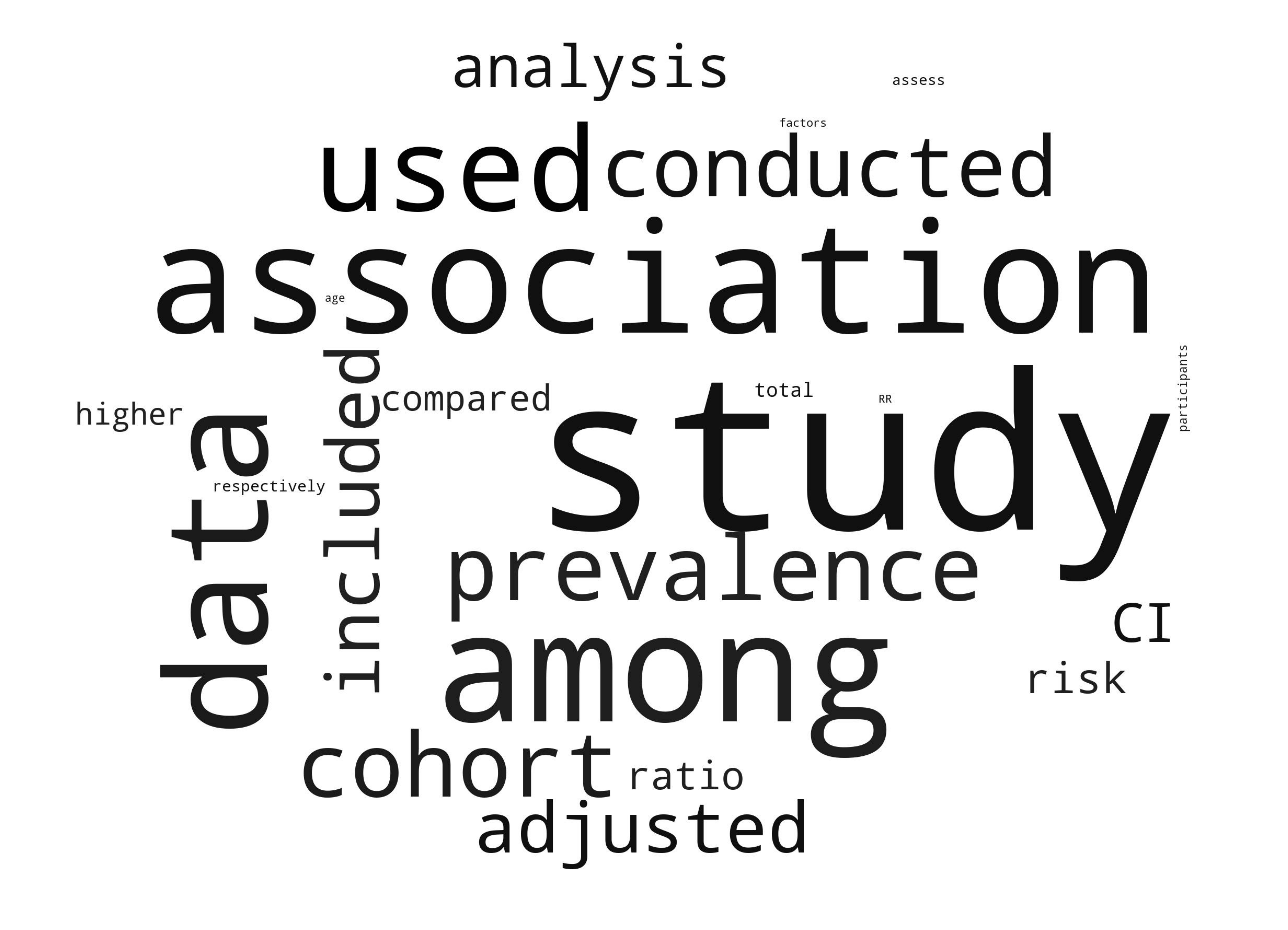}
\end{subfigure}
~ 
\begin{subfigure}[t]{0.15\textwidth}
\centering
\includegraphics[width=\textwidth]{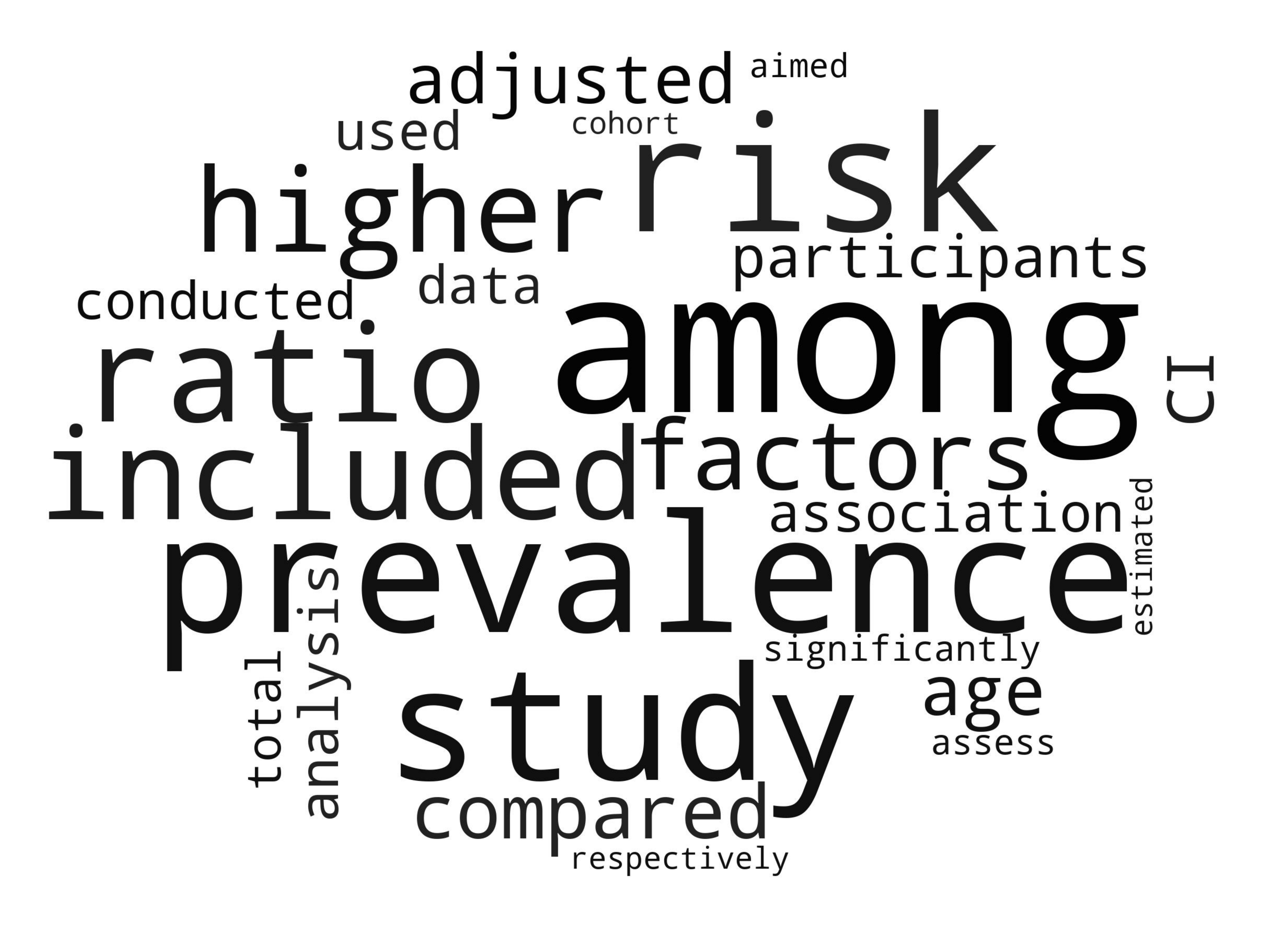}
\end{subfigure}

\begin{subfigure}[t]{0.15\textwidth}
\centering
\includegraphics[width=\textwidth]{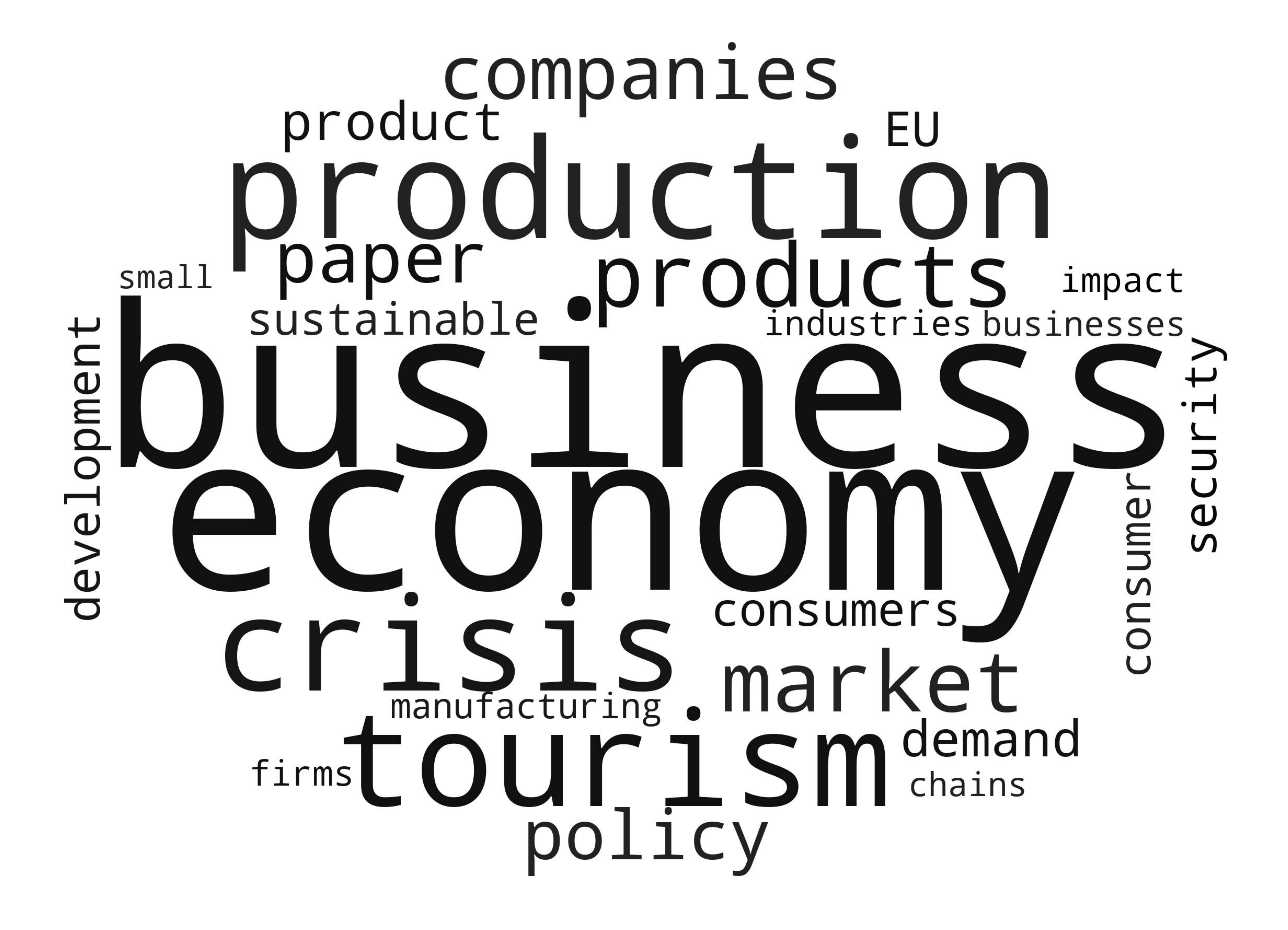}
\caption{probabilities}
\end{subfigure}%
~ 
\begin{subfigure}[t]{0.15\textwidth}
\centering
\includegraphics[width=\textwidth]{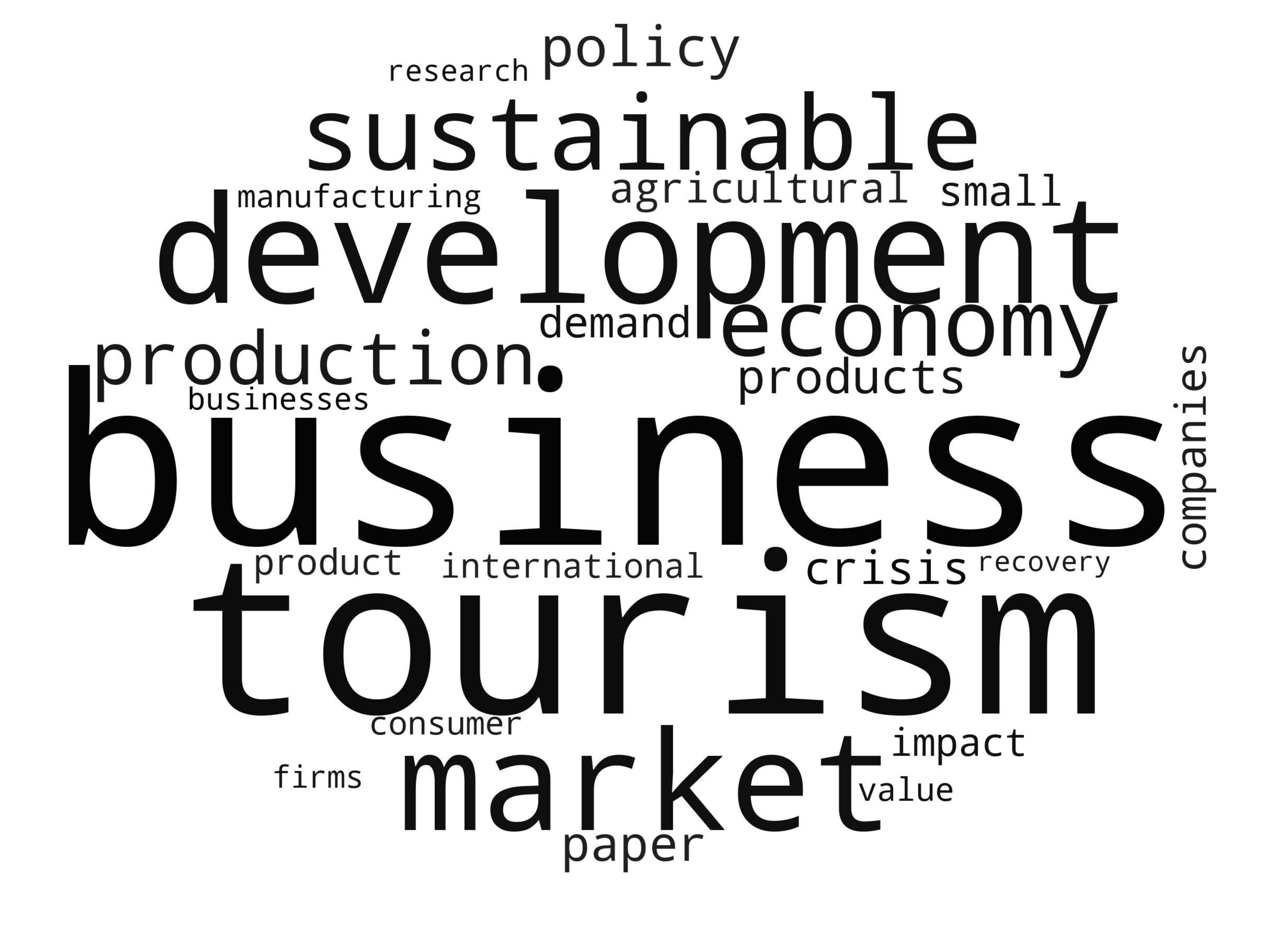}
\caption{degree}
\end{subfigure}
~ 
\begin{subfigure}[t]{0.15\textwidth}
\centering
\includegraphics[width=\textwidth]{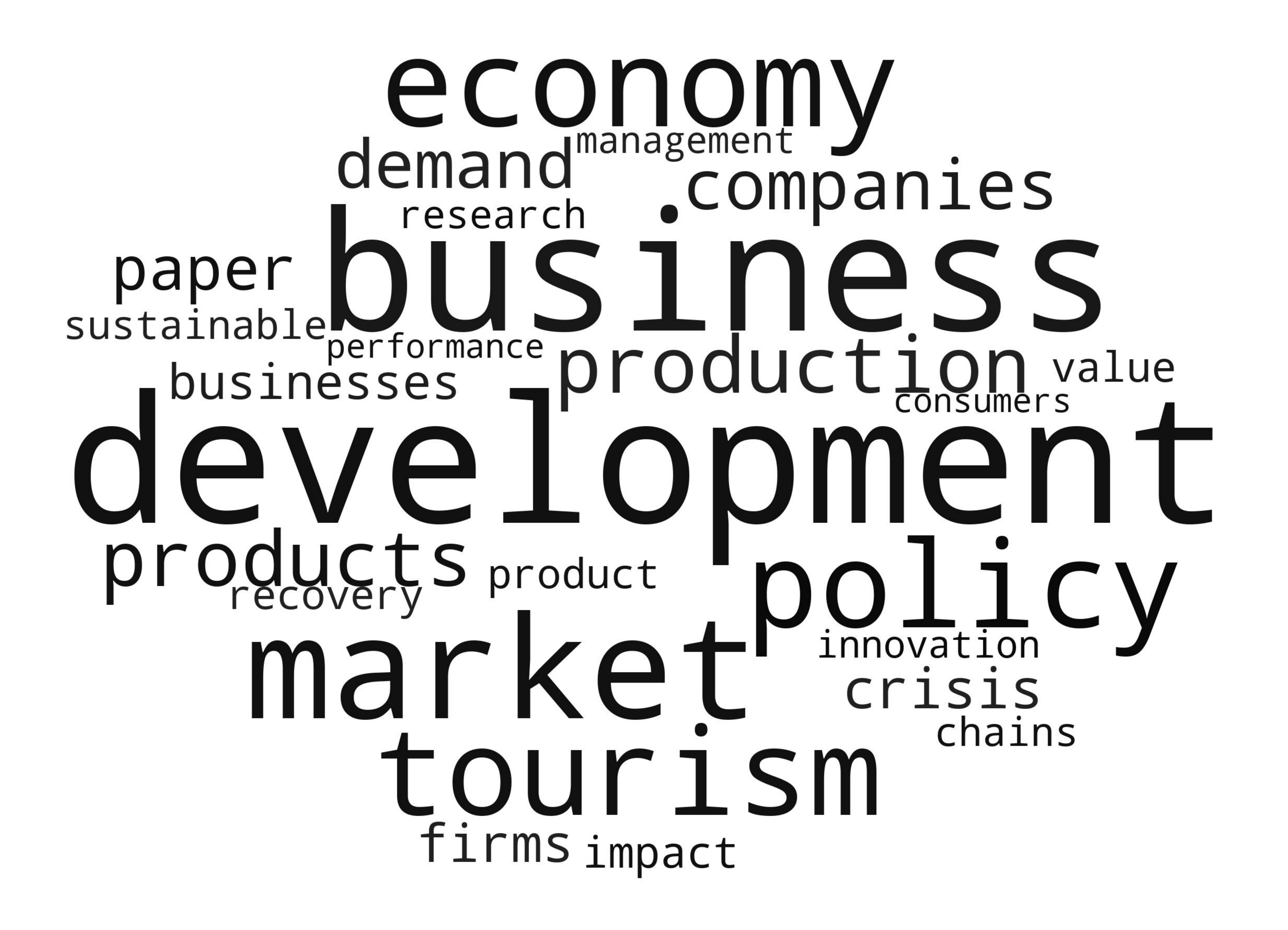}
\caption{in-degree}
\end{subfigure}
~ 
\begin{subfigure}[t]{0.15\textwidth}
\centering
\includegraphics[width=\textwidth]{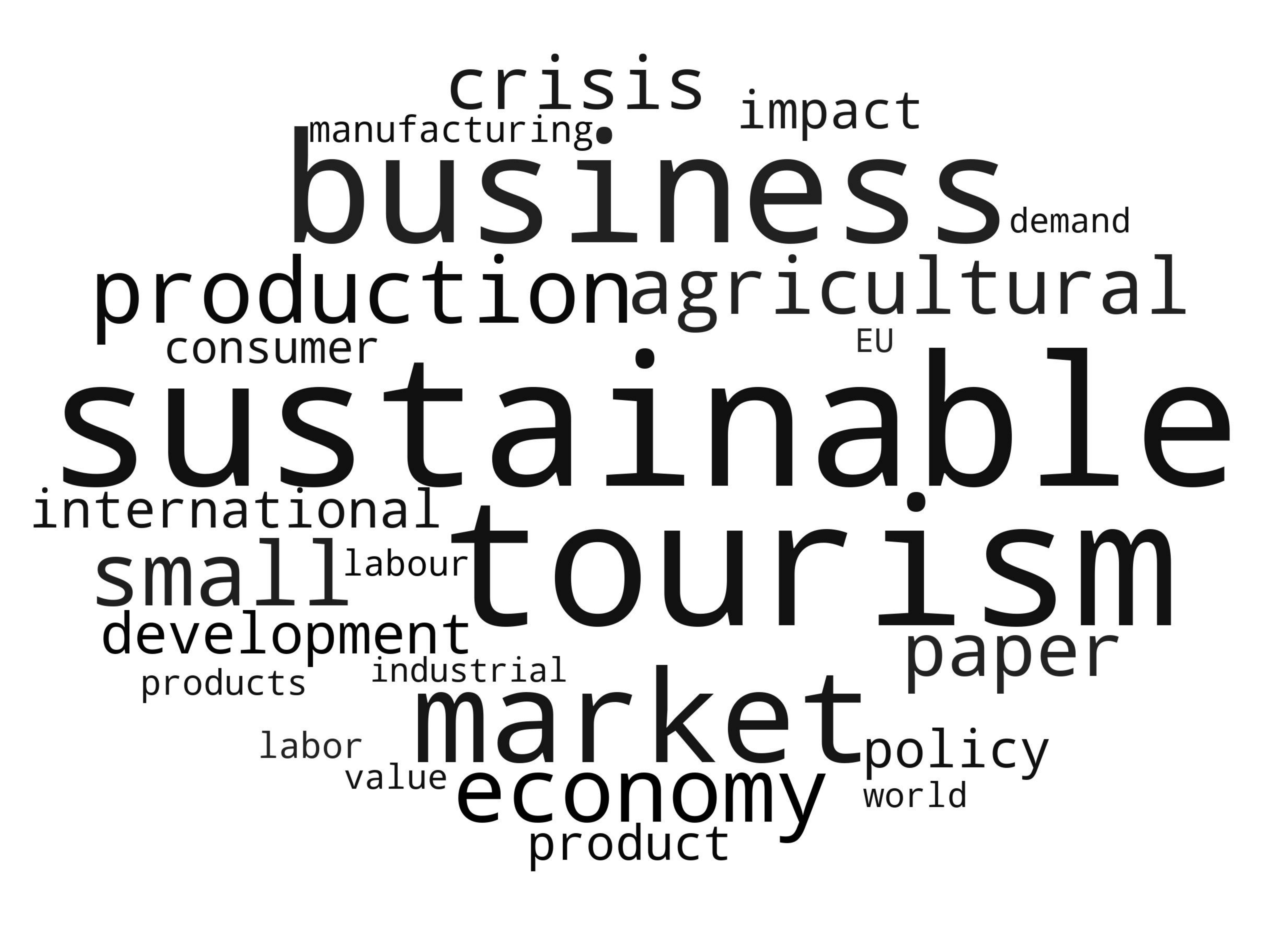}
\caption{out-degree}
\end{subfigure}
~ 
\begin{subfigure}[t]{0.15\textwidth}
\centering
\includegraphics[width=\textwidth]{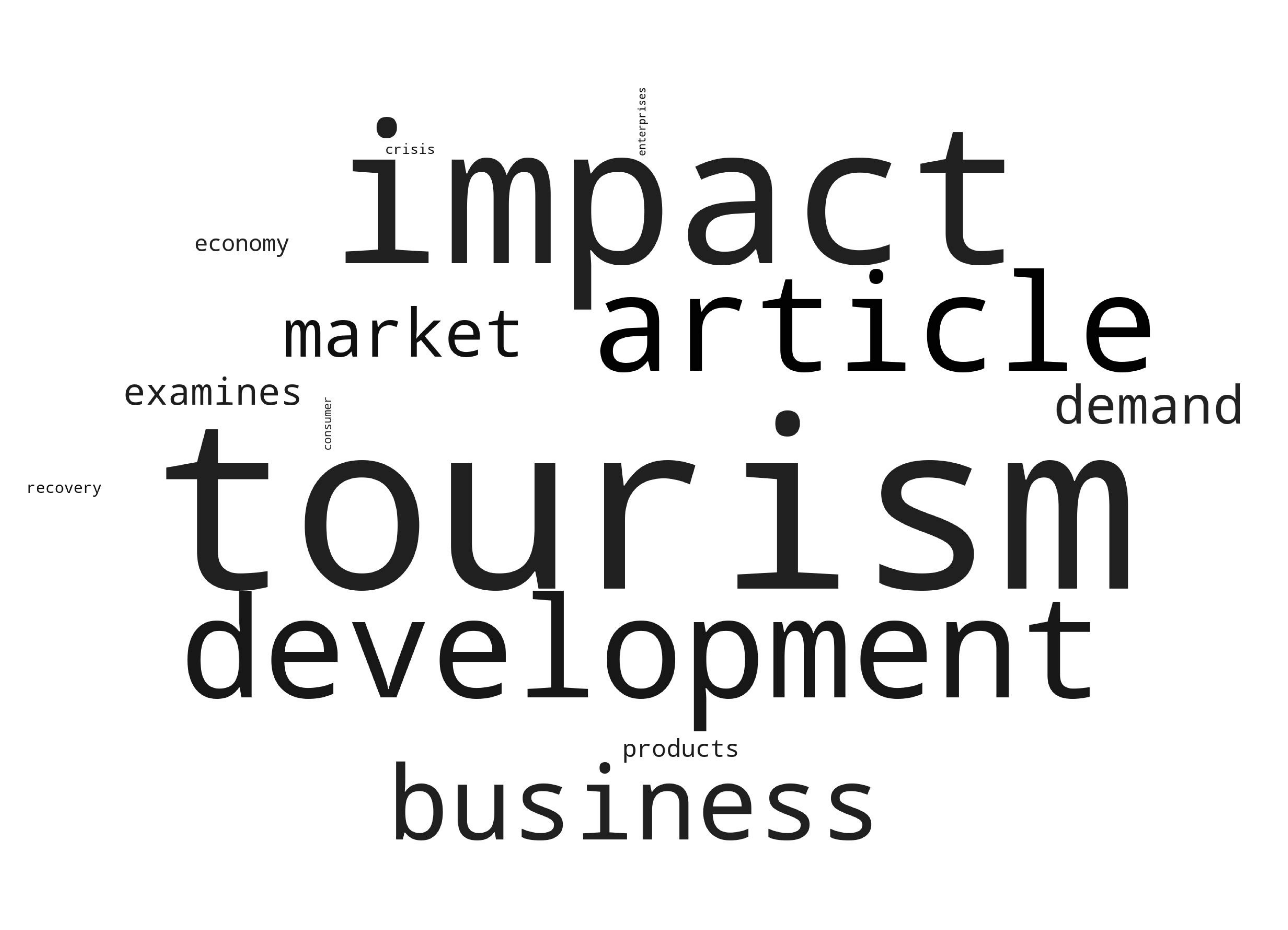}
\caption{betweenness}
\end{subfigure}
~ 
\begin{subfigure}[t]{0.15\textwidth}
\centering
\includegraphics[width=\textwidth]{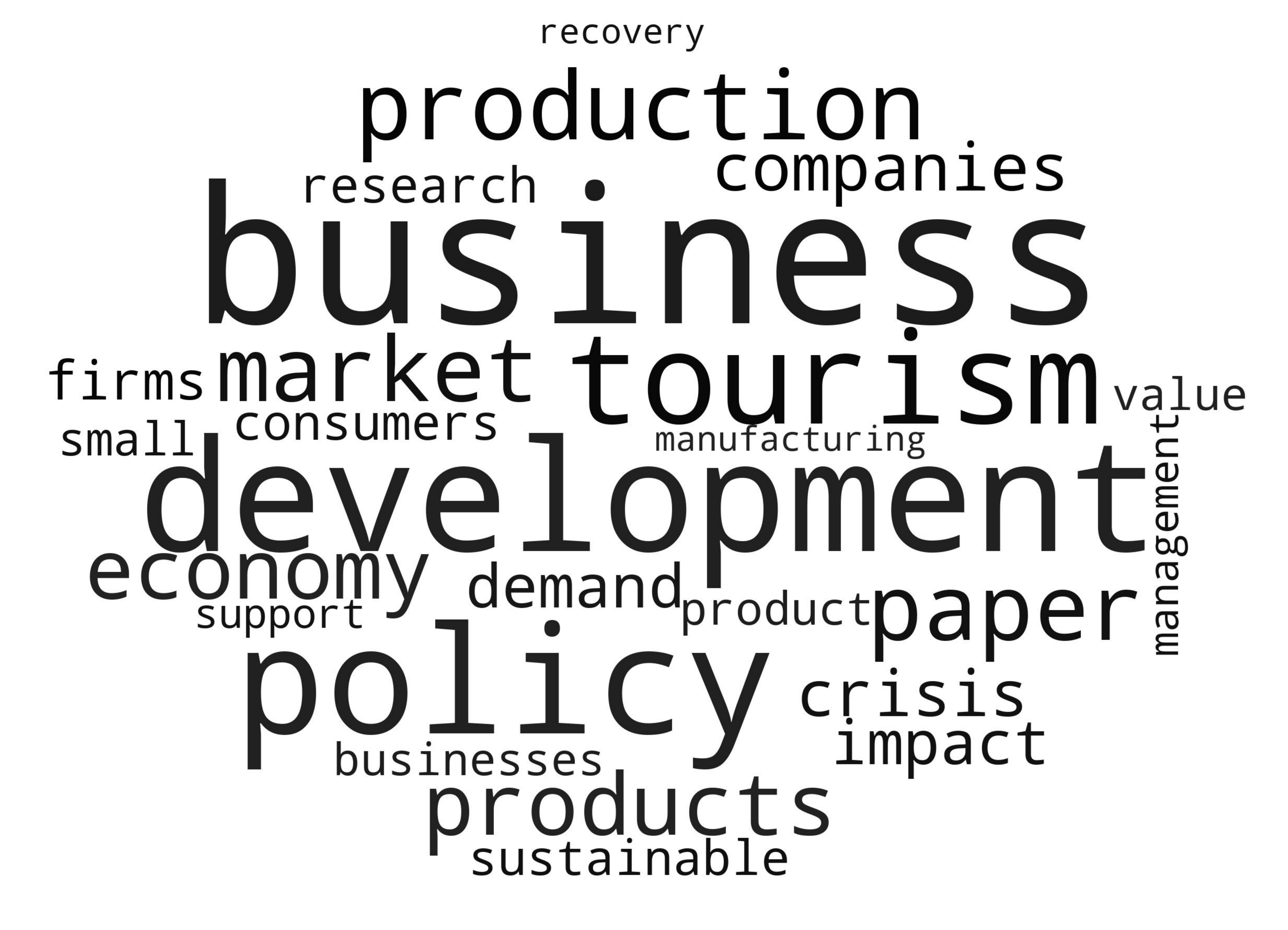}
\caption{pagerank}
\end{subfigure}
\caption{Top 25 words (i.e., nodes) by measure, for topics $\#88$ (upper row), $\#50$  (middle row) and $\#36$ (lower row)}\label{fig:wordcloud}
\end{figure}

\begin{figure}[h!]
\centering



\begin{subfigure}[t]{0.22\textwidth}
\centering
\includegraphics[width=\textwidth]{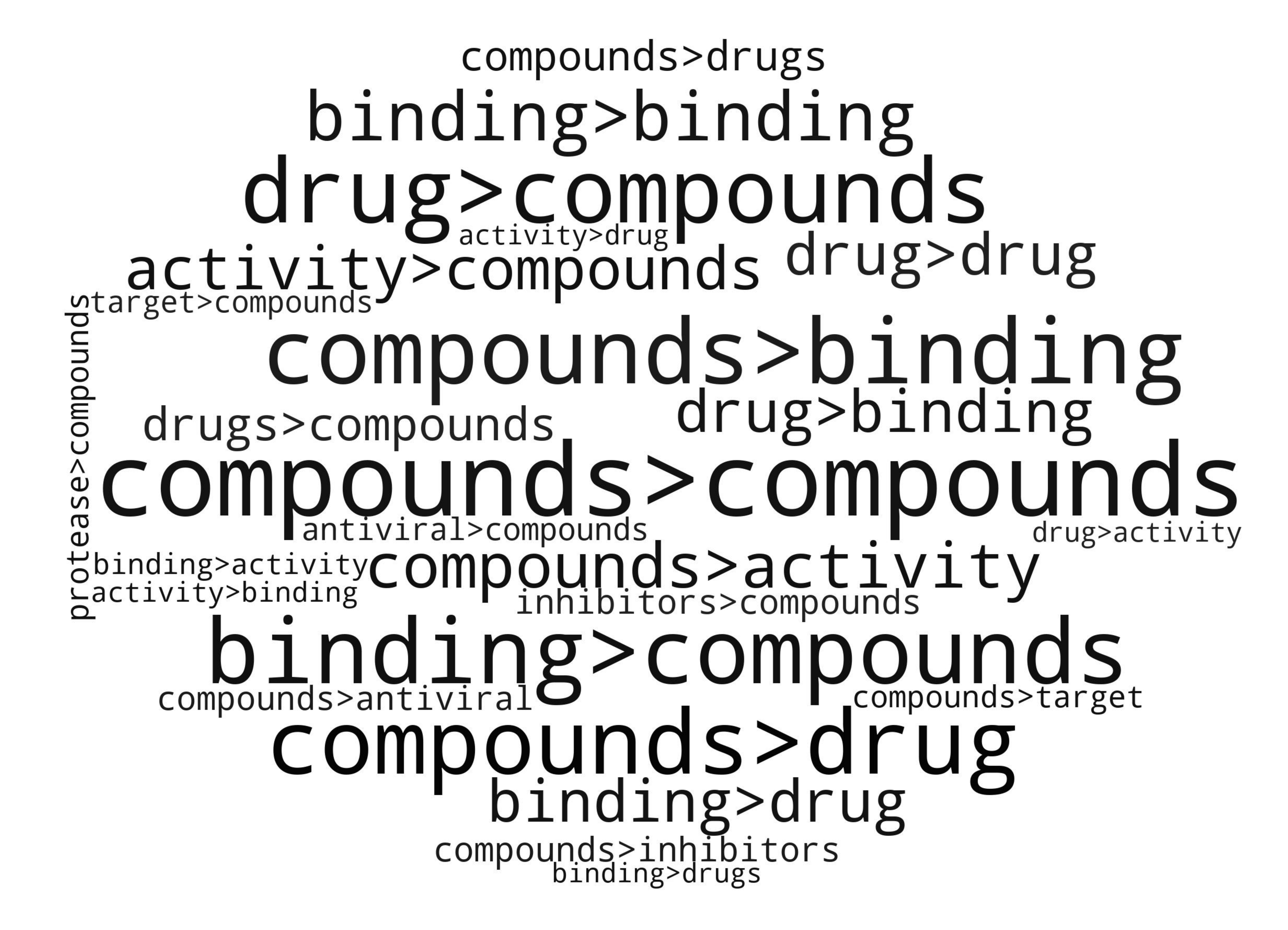}
\end{subfigure}%
~ 
\begin{subfigure}[t]{0.22\textwidth}
\centering
\includegraphics[width=\textwidth]{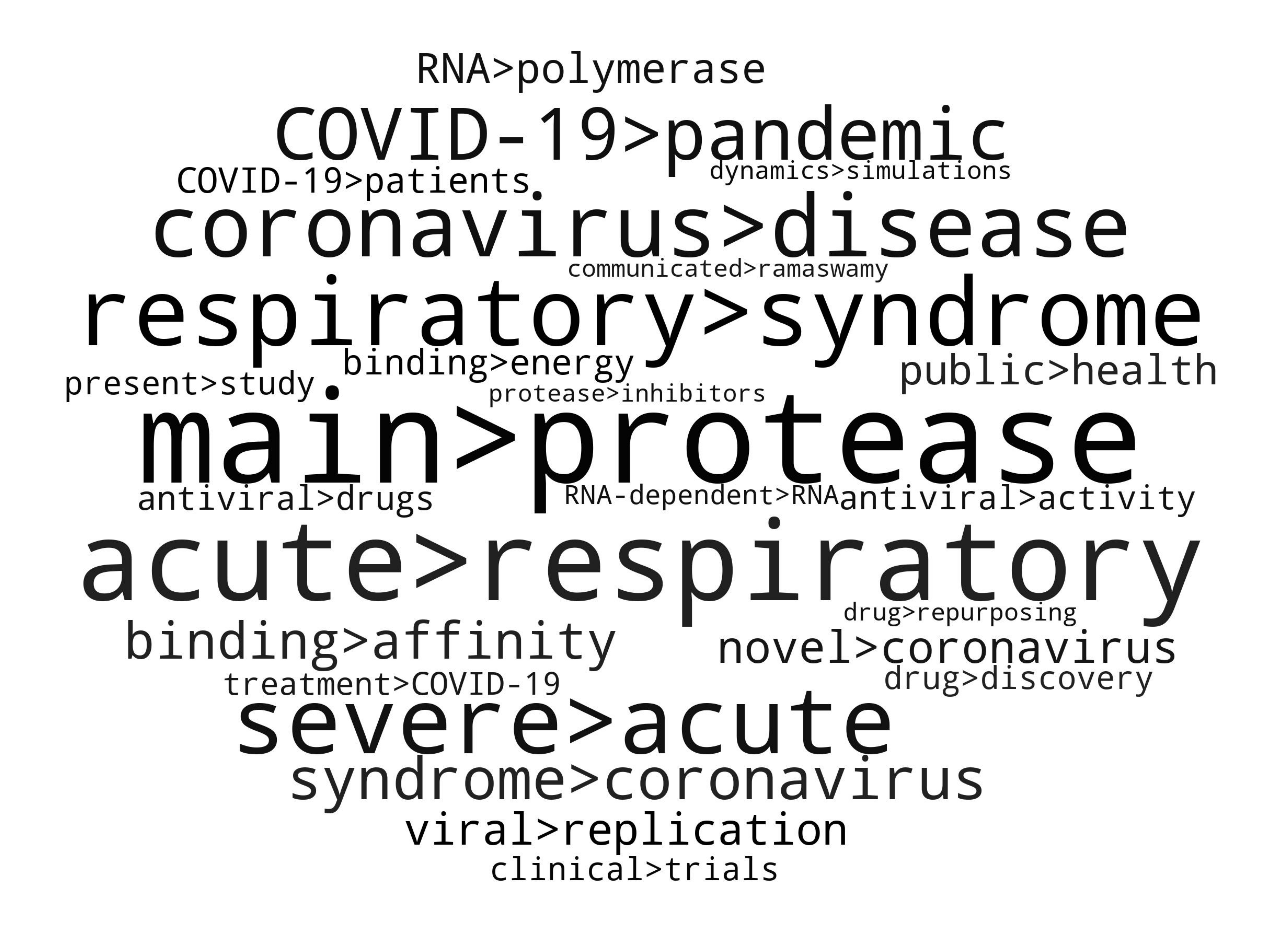}
\end{subfigure}
~ 
\begin{subfigure}[t]{0.22\textwidth}
\centering
\includegraphics[width=\textwidth]{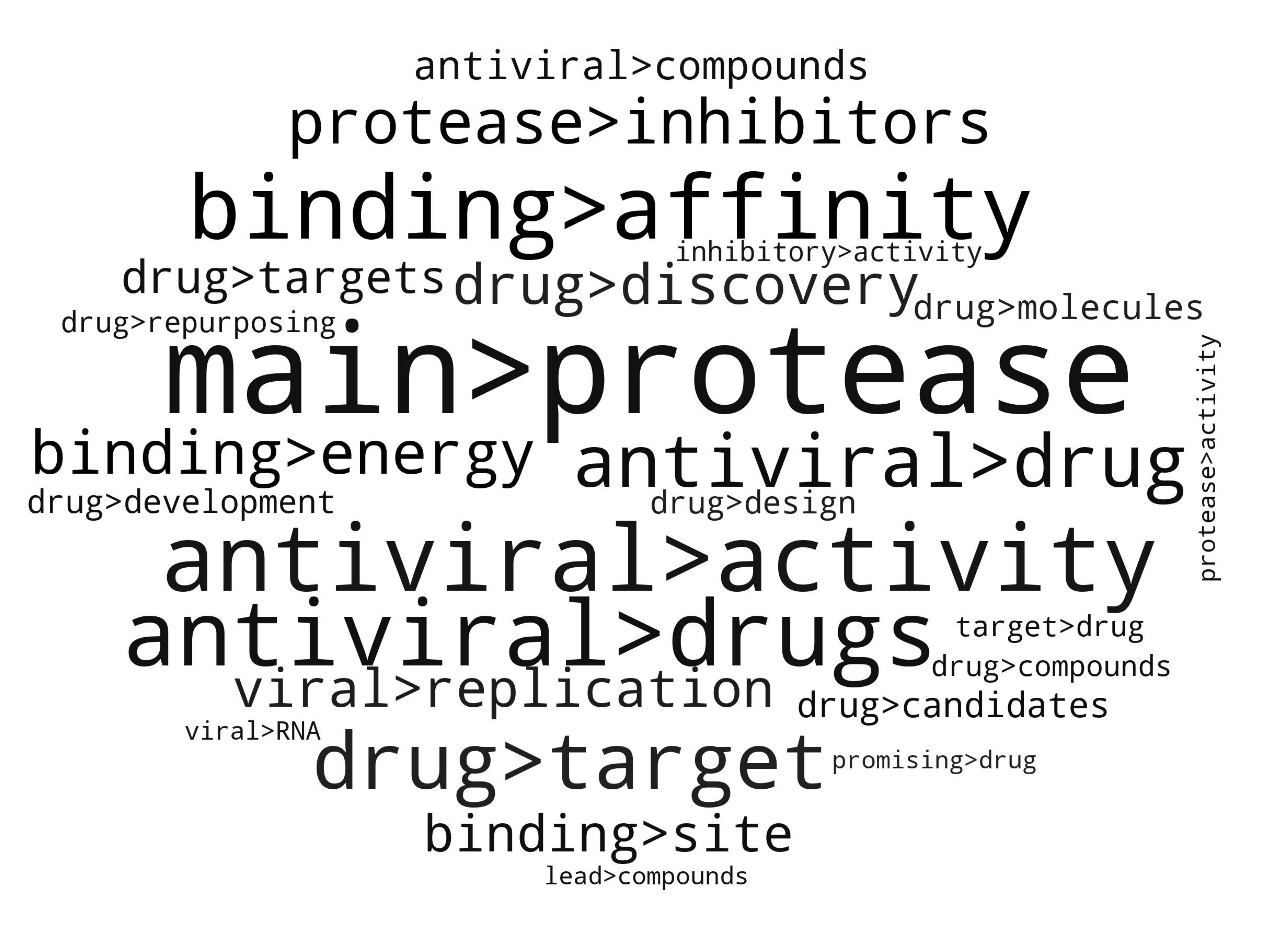}
\end{subfigure}

\begin{subfigure}[t]{0.22\textwidth}
\centering
\includegraphics[width=\textwidth]{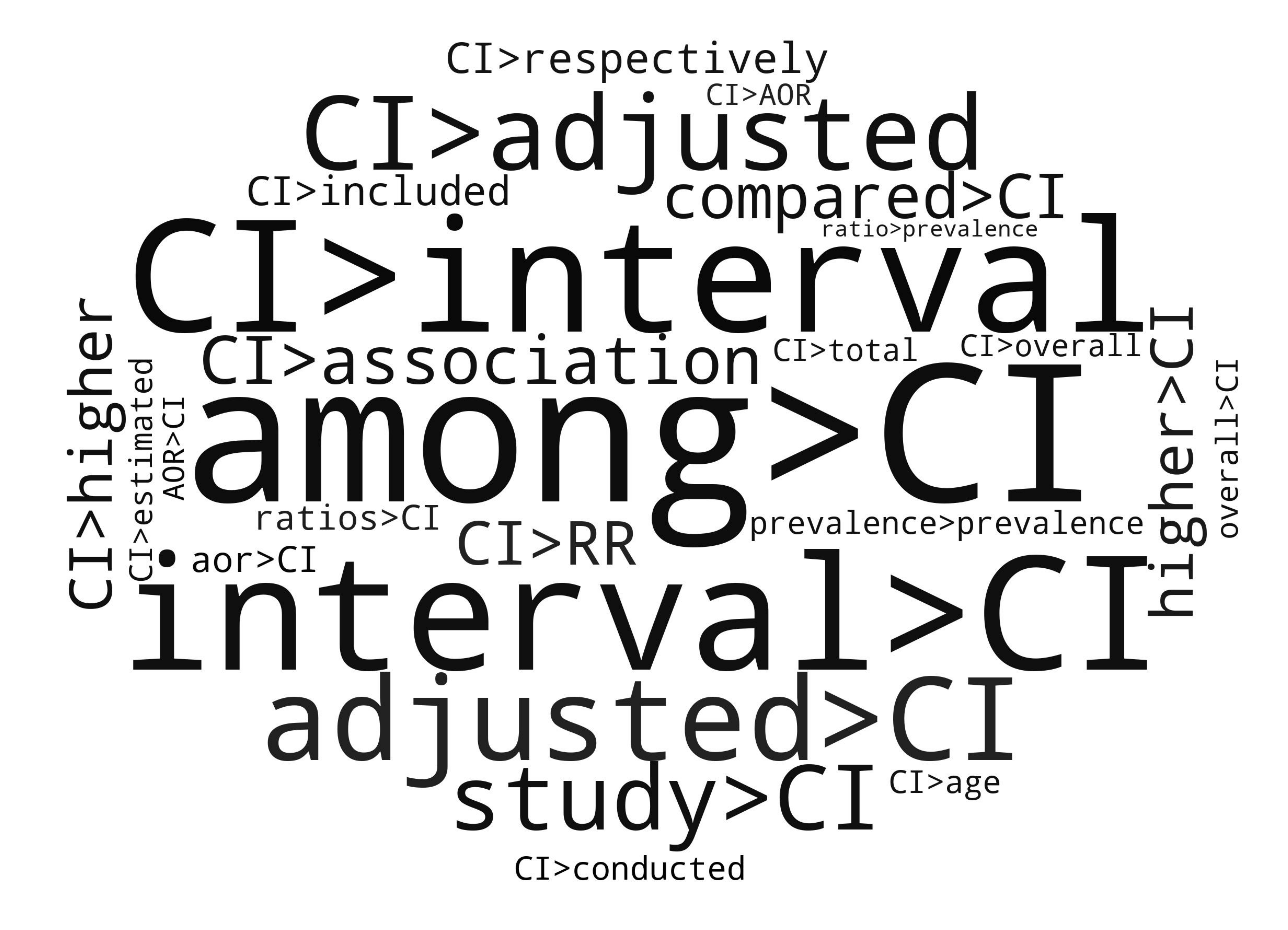}
\end{subfigure}%
~ 
\begin{subfigure}[t]{0.22\textwidth}
\centering
\includegraphics[width=\textwidth]{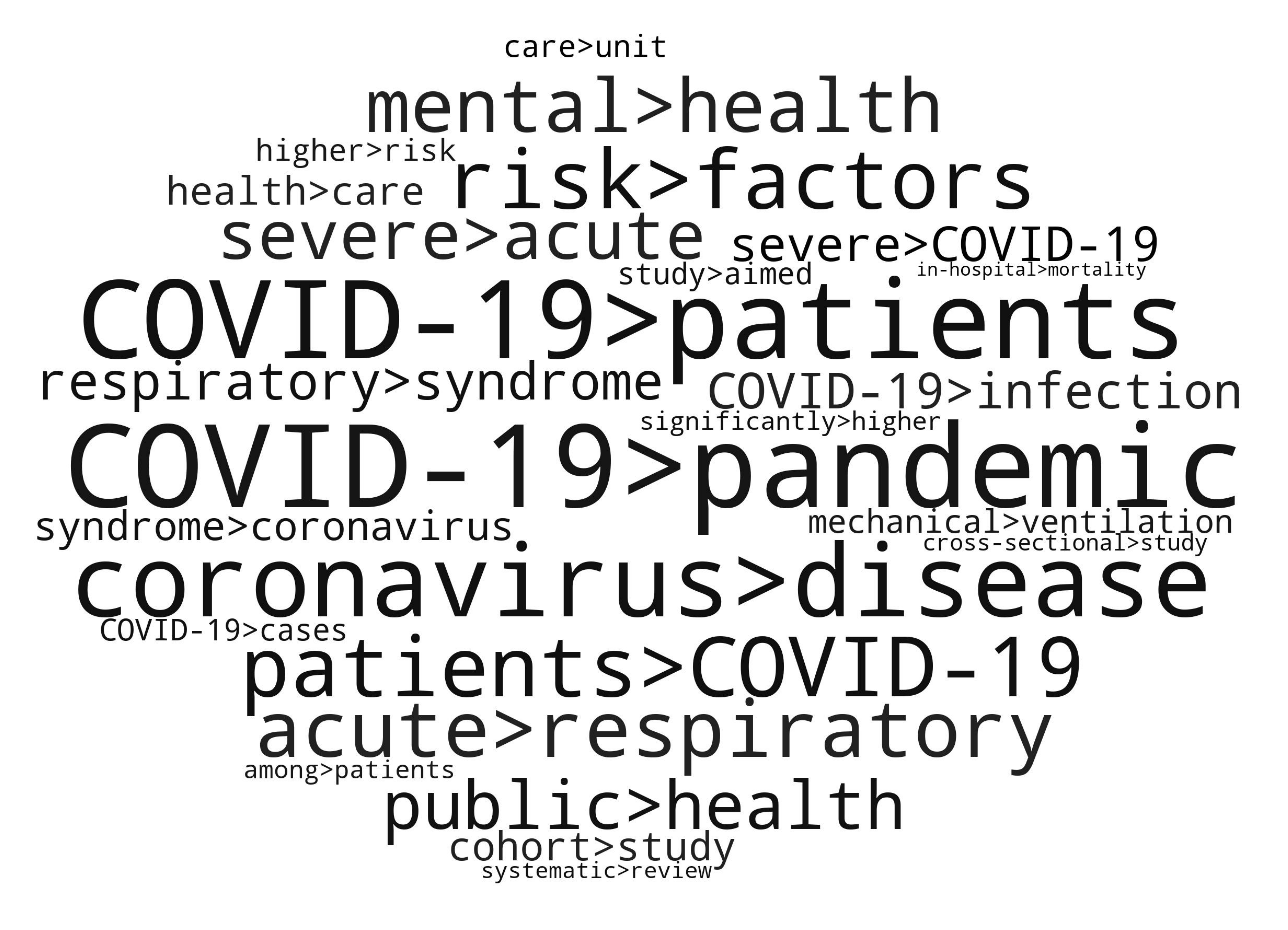}
\end{subfigure}
~ 
\begin{subfigure}[t]{0.22\textwidth}
\centering
\includegraphics[width=\textwidth]{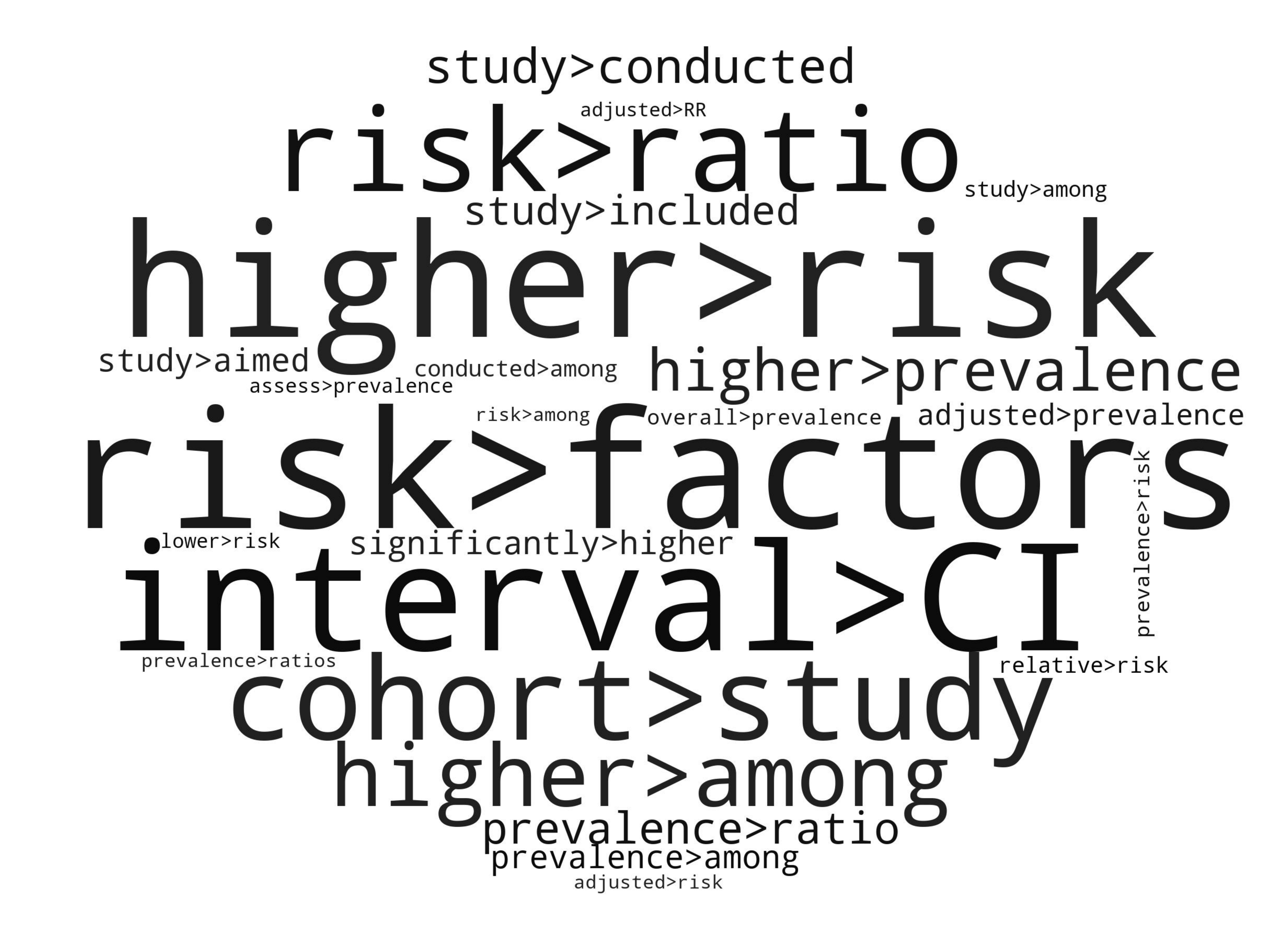}
\end{subfigure}

\begin{subfigure}[t]{0.22\textwidth}
\centering
\includegraphics[width=\textwidth]{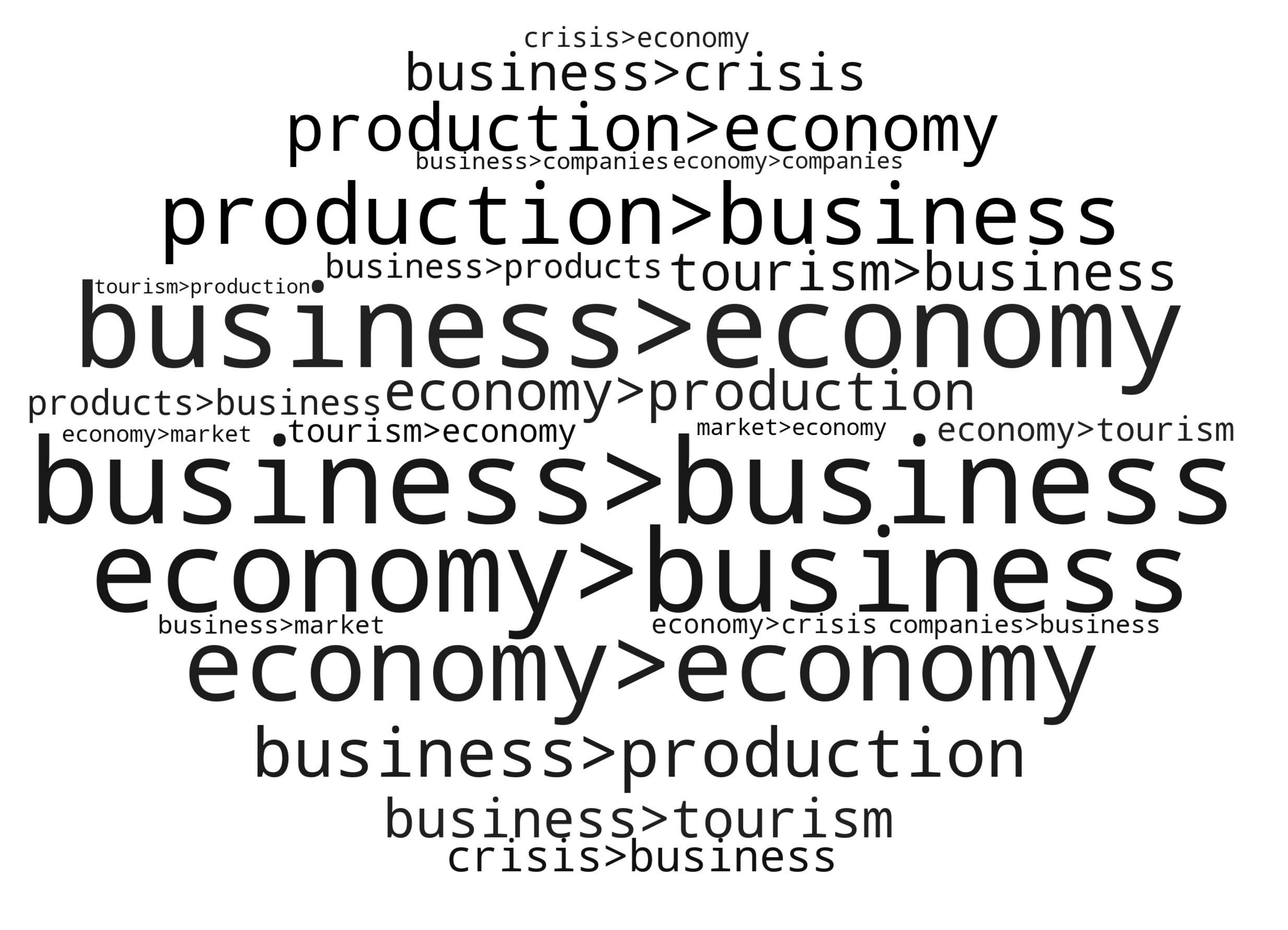}
\caption{\textit{probs}-weights}
\end{subfigure}%
~ 
\begin{subfigure}[t]{0.22\textwidth}
\centering
\includegraphics[width=\textwidth]{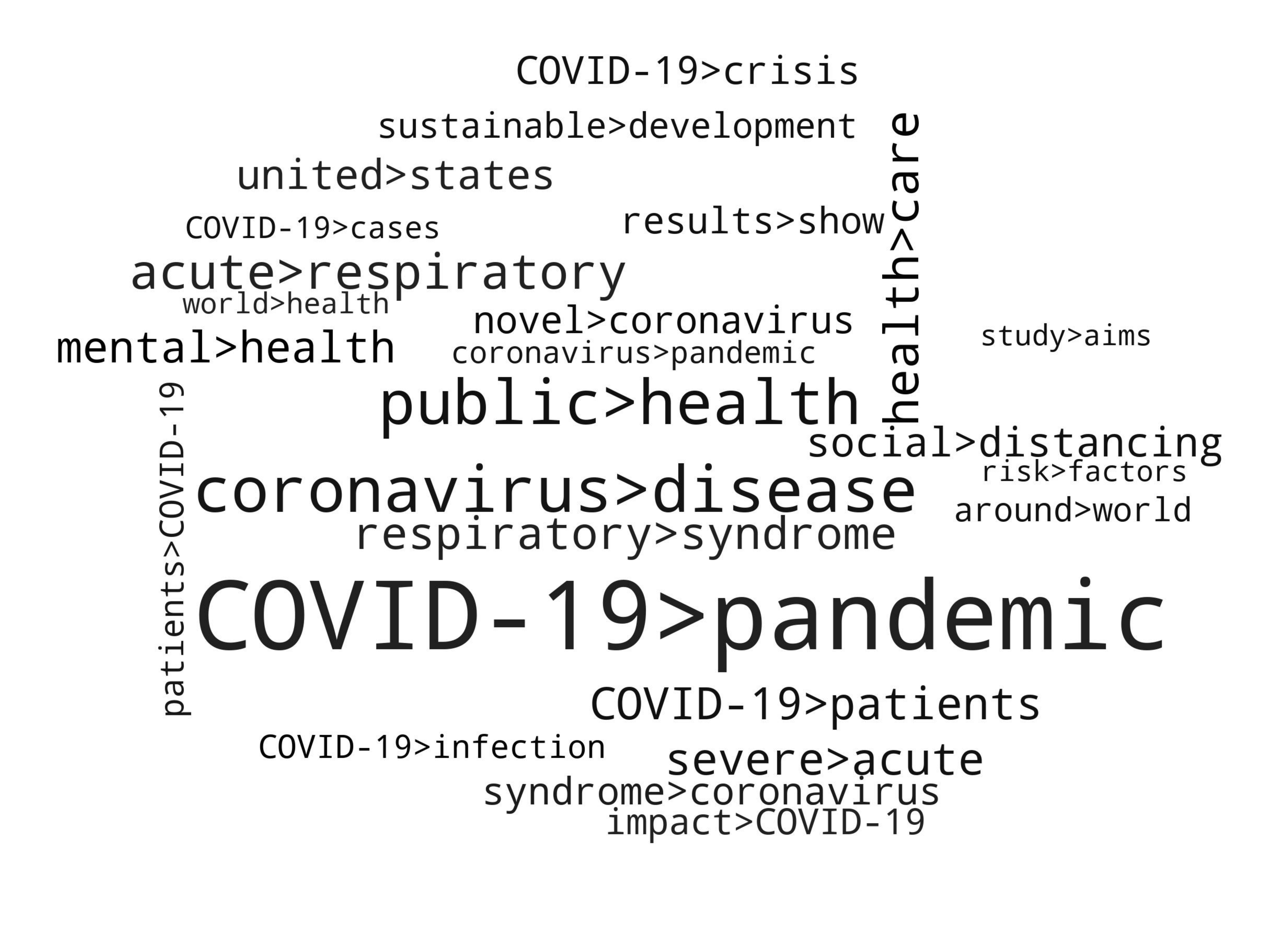}
\caption{\textit{counts}-weights}
\end{subfigure}
~ 
\begin{subfigure}[t]{0.22\textwidth}
\centering
\includegraphics[width=\textwidth]{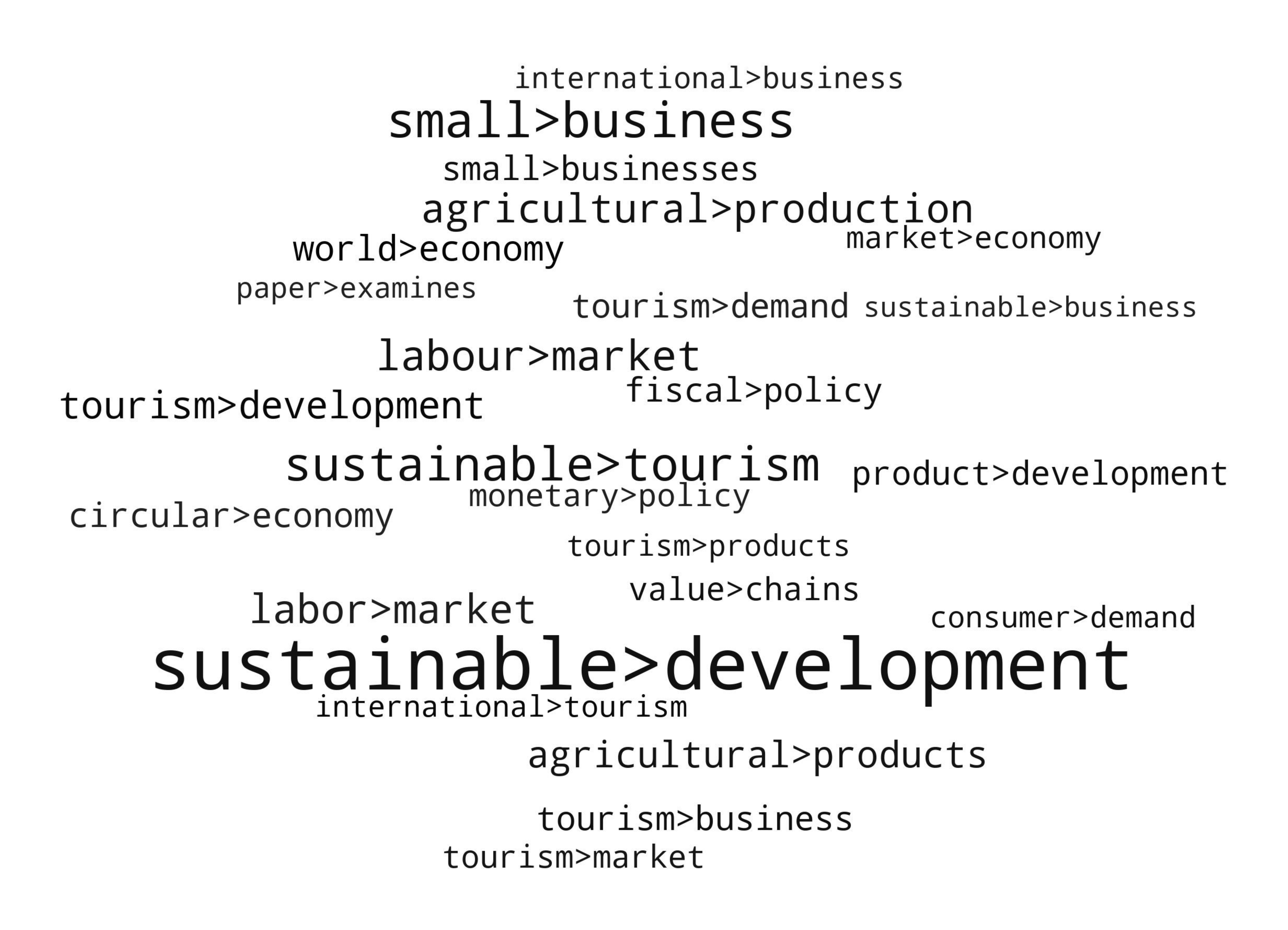}
\caption{ \textit{LDA2Net}-weights}
\end{subfigure}

\caption{Top 25 bigrams (i.e., edges) by measure, for topics $\#88$ (first row), $\#50$ (second row) and $\#36$ (third row). For details about edge weights see Section \ref{subsec:networkcreation}} \label{fig:wordcloud_edges}
\end{figure}

 \FloatBarrier

\paragraph{Bigram metrics}  
One can gain additional insights through \textit{LDA2Net} by moving from the analysis of words to the one of word associations, here captured through bigrams. Figure \ref{fig:wordcloud_edges}  compares two different bigram weighting strategies, called  \textit{probs}-weight and  \textit{counts}-weight and their combination in our approach, called \textit{LDA2Networks}-weight, for three topics. 
We refer the reader to Section \ref{subsec:networkcreation} for details.
As the figure illustrates, both \textit{probs}-weight and \textit{counts}-weight  have apparent weaknesses.
While topic-specific bigram weights based on the pairwise product of LDA word probabilities (\ie \textit{probs}-weight)  suffer from over-representing the combination of few words with high probabilities, the weights based on document-bigram counts assigned to a given topic (\ie \textit{counts}-weight) on the basis of LDA's document-topic distributions, appear to over-represent associations between generic words that appear in most documents of the corpus, like \texttt{COVID-19$\rightarrow$pandemic} and \texttt{coronavirus$\rightarrow$pandemic}. The combination of both components into the proposed \textit{LDA2Net}-weights summarises well the distinctive word associations characterising each topic. 

\begin{figure}[!ht]
\centering
\includegraphics[width=0.45\textwidth]{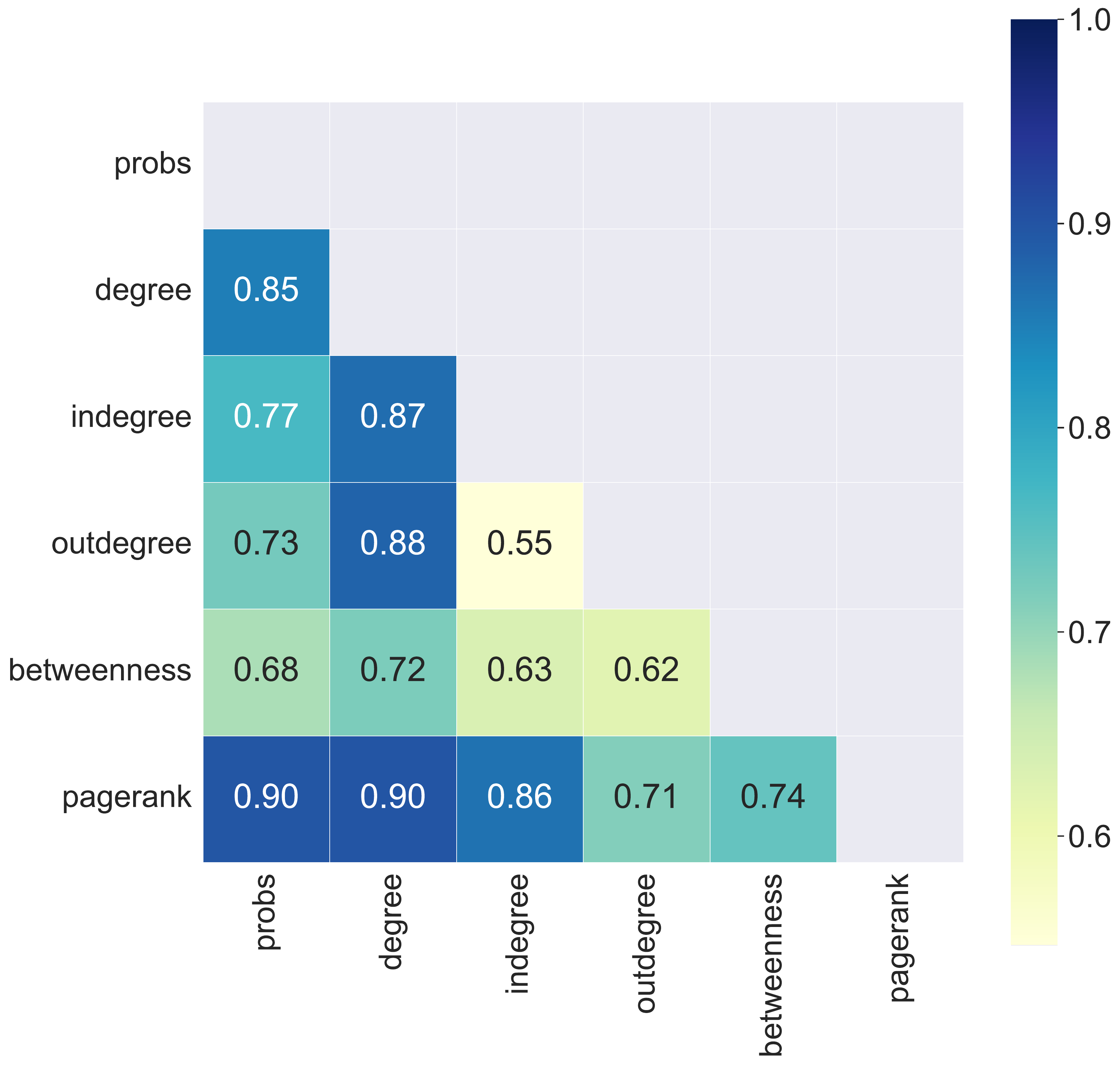}
\includegraphics[width=0.45\textwidth]{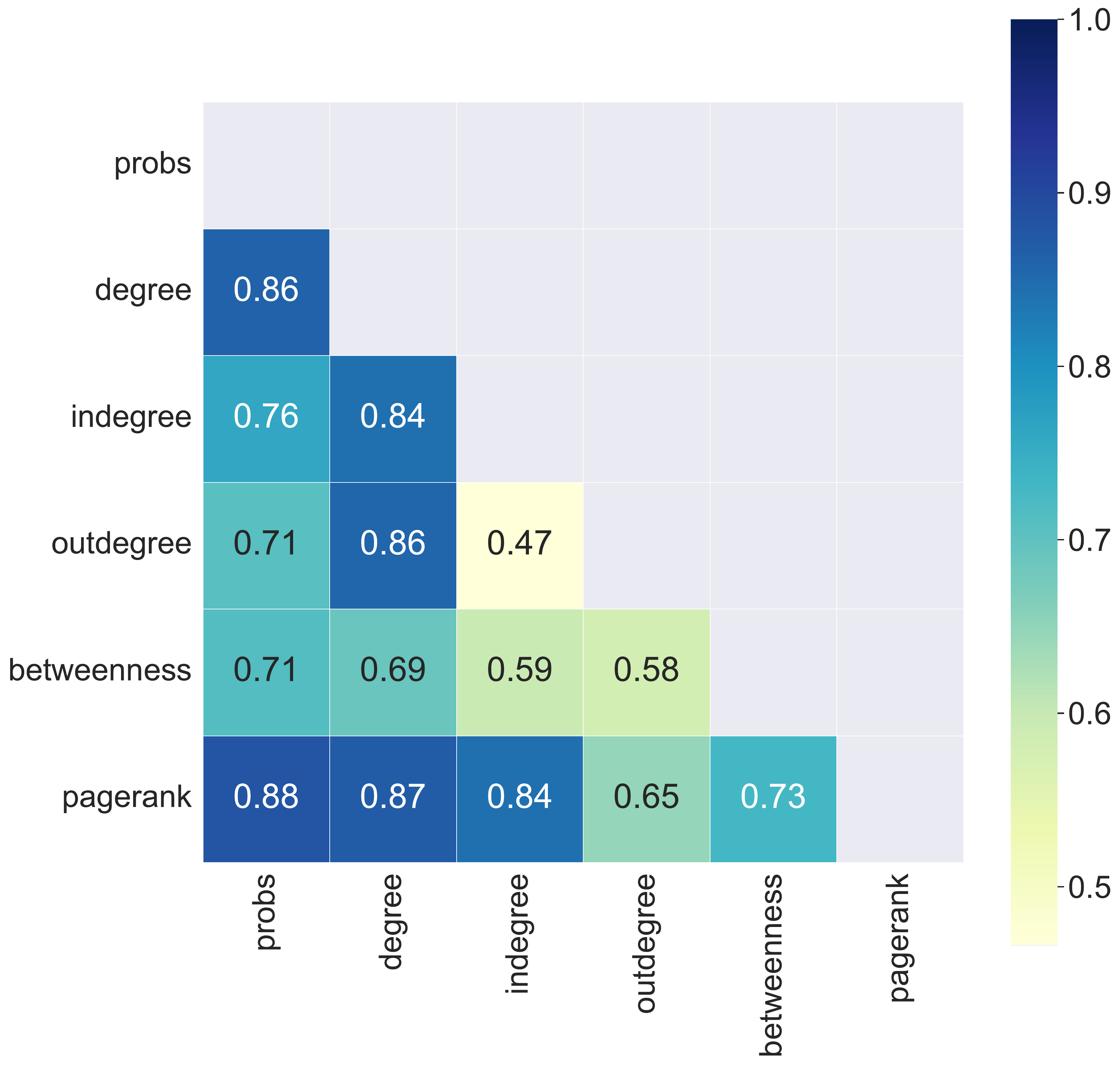}
\caption{Pearson correlation between node centrality measures computed over networks  and probabilities from LDA output. On the left correlation computed over all words; on the right correlation considering only the top 30 words by LDA probability (probs). 
}
\label{fig:correlation}
\end{figure}

\begin{figure}[!ht]
\centering
\includegraphics[width=0.45\textwidth]{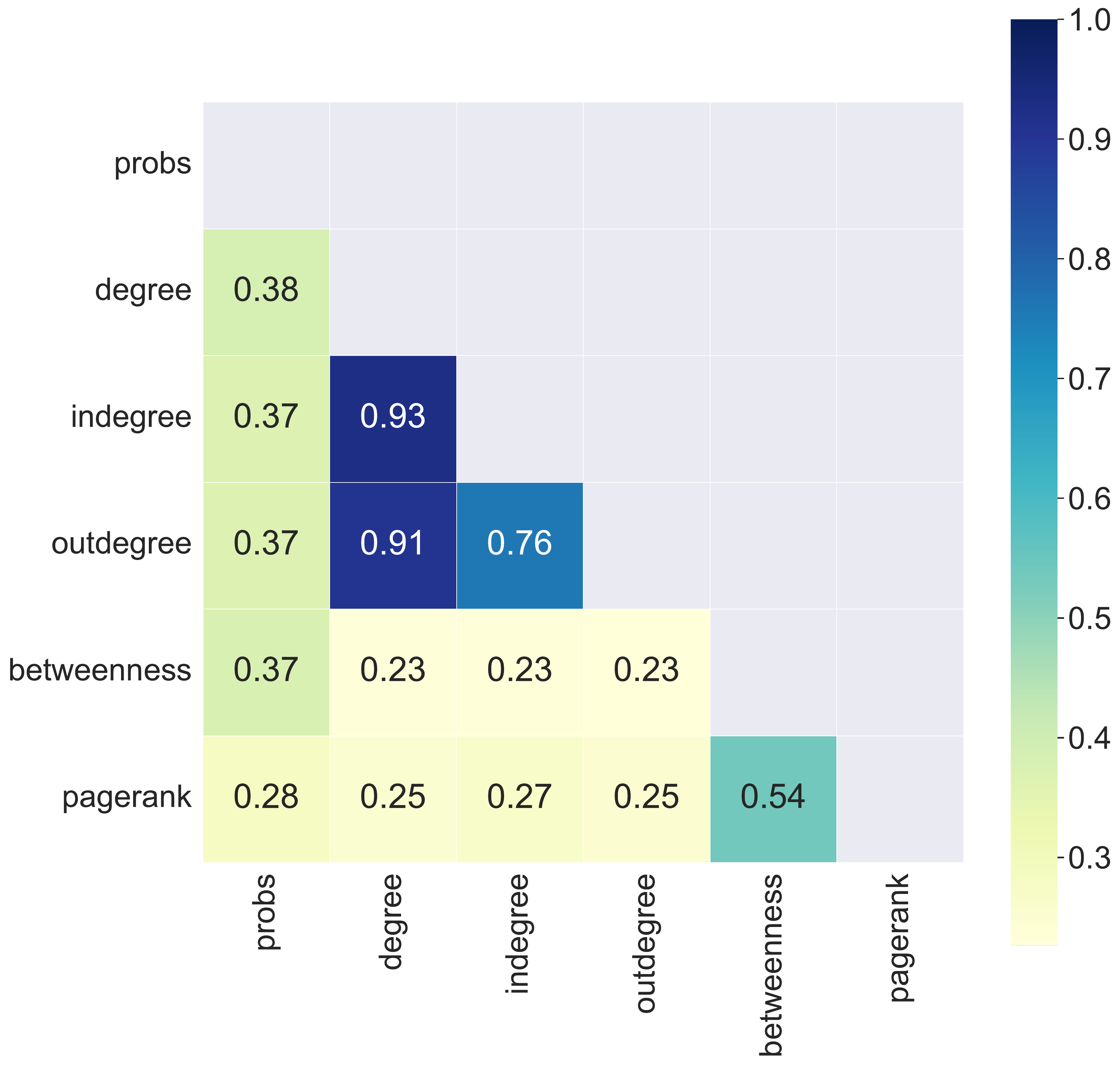}
\includegraphics[width=0.45\textwidth]{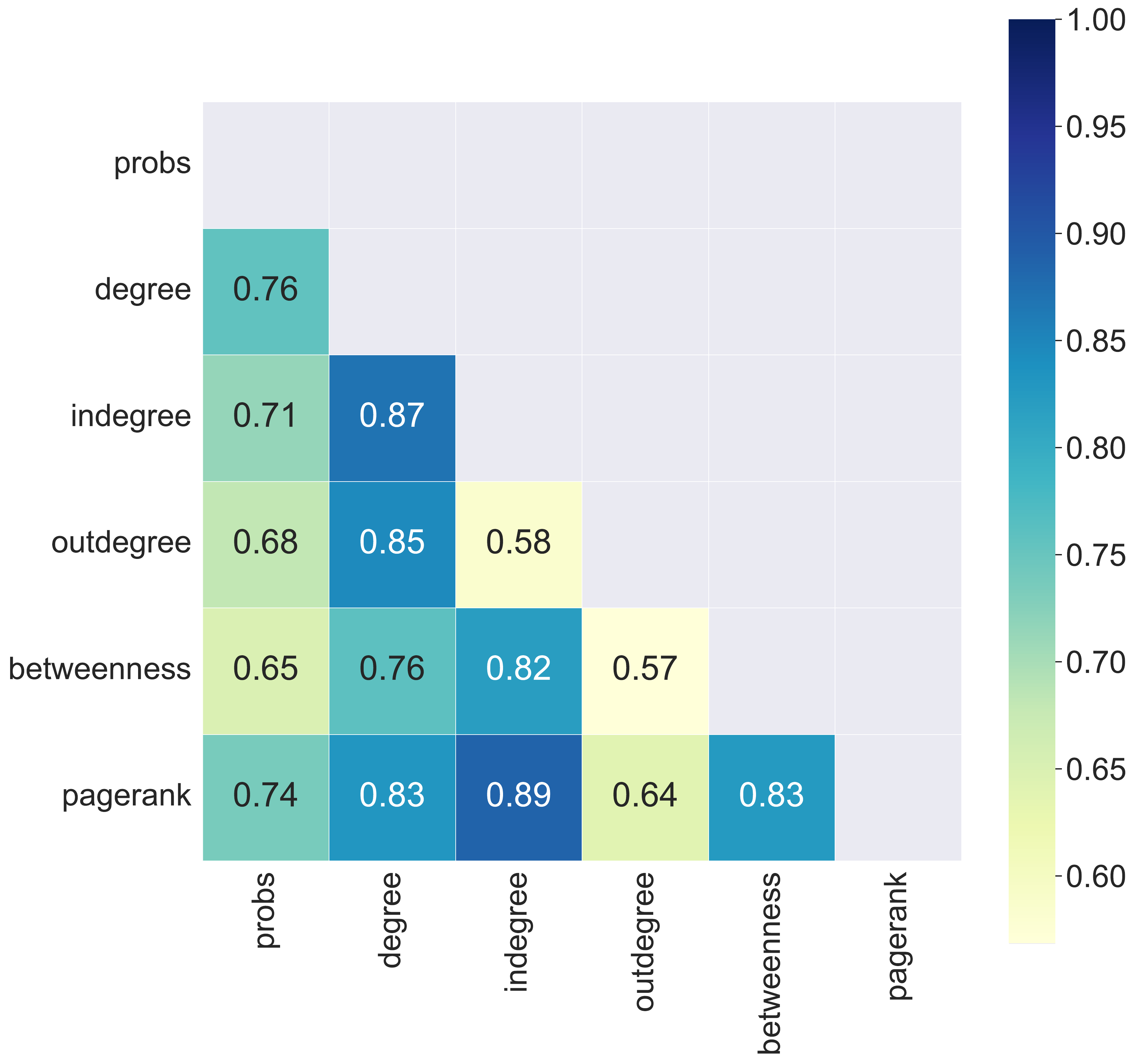}
\caption{Spearman (rank) correlation between node centrality measures computed over networks  and probabilities from LDA output. On the left correlation computed over all words; on the right correlation considering only the top 30 words by LDA probability (probs). 
}
\label{fig:rankcorrelation}
\end{figure}

\paragraph{Relations between metrics} As shown in Figure \ref{fig:correlation} and \ref{fig:rankcorrelation}, we can assess the enrichment effect of the proposed method  via Pearson (Figure  \ref{fig:correlation}) and Spearman (Figure \ref{fig:rankcorrelation}) correlation analysis among alternative word-relevance metrics. The definitions  of the different centralities we display can be found in Appendix \ref{sec:centralities}.
The matrices of Pearson correlations show that degree and other centrality metrics preserve information contained in LDA probabilities. However, there are relational properties such as betweenness  that have lower correlation with probabilities – which suggests that network measures carry relevant additional information.
A comparison of Pearson and Spearman (rank) correlation further shows that the aforementioned relations among measures are weaker below top words – and further information can be lost by using only LDA probabilities when dealing with lower rank words.

We can also measure the information added by the \textit{LDA2Net} method, by computing the Jensen–Shannon divergence (JSD) between the two components used to compute the \textit{LDA2Net}-weights of bigrams, that is  \textit{counts}-weights obtained using bigram frequencies (second column in figure \ref{fig:wordcloud_edges}) and the \textit{probs}-weights obtained by using the pairwise product of LDA's word probabilities (first column in figure \ref{fig:wordcloud_edges}). For a formal definition of JSD we refer the reader to Section \ref{sec:methods}.
This divergence captures the information added by word ordering constraints introduced by \textit{LDA2Net}, which combines LDA with observed sequences of words. The information brought by the proposed method is highlighted by the distribution of topic JSDs, which is heavily skewed towards the right of the range of possible JSD values, that is $[0,1]$.
\FloatBarrier

\begin{figure}[!h]
\centering
\includegraphics[width=0.65\textwidth]{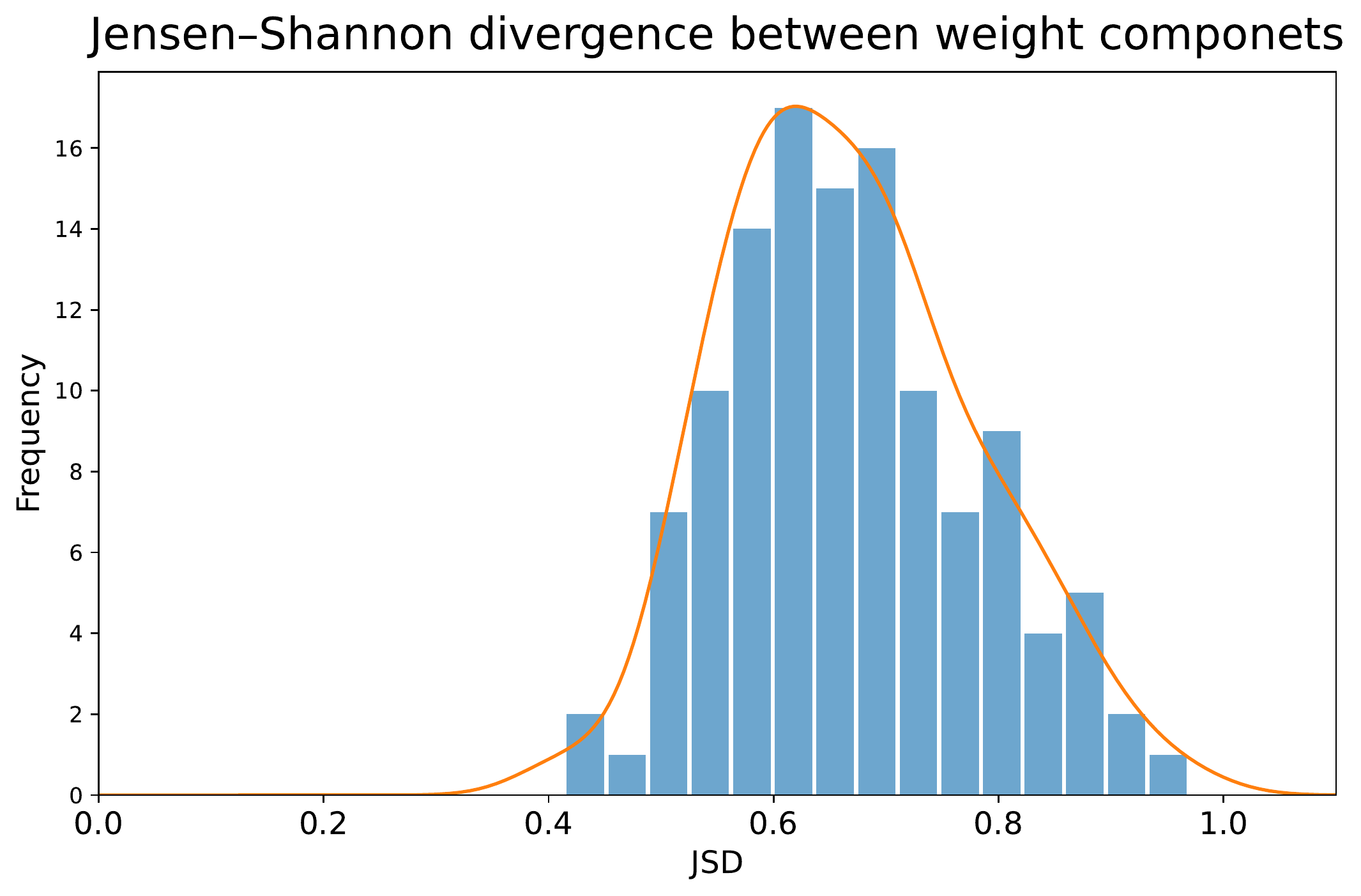}
 
\caption{Histogram of JSDs between the two normalized components used to compute bigram (i.e., edge) weights in topic networks.}
 
\end{figure}
\FloatBarrier

\begin{figure*}[h!]
    \centering
\includegraphics[width=0.85\textwidth]{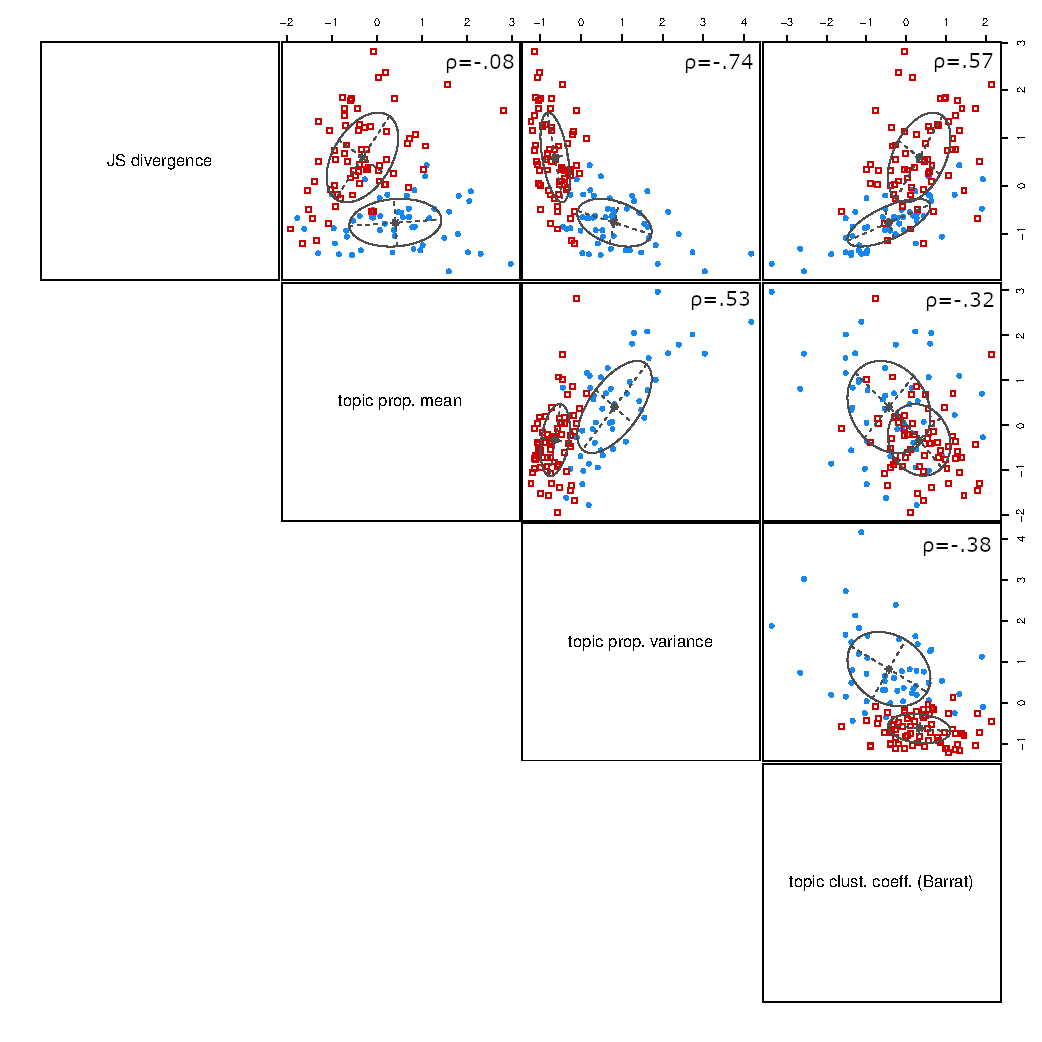}
    \caption{Topics clustering with Gaussian Finite Mixture Model. Automatic model (VEV: ellipsoidal \& equal volume) and number of clusters selection (2) based on BIC criterion (see \textsc{mclust} \texttt{R} library by \citet{scrucca2016mclust}). Two outlier topics ($\#54$:french and $\#106$:spanish) were removed. All measures were standardised (to have zero means and unitary SEs) before clusterization. Spearman correlations are displayed in the top right corner of each quadrant. 
    }
    \label{fig:clustering}
\end{figure*}

\paragraph{Mapping and clustering topics through aggregate topic metrics} In \textit{LDA2Net}, topics can be summarized at a macro level by means of a set of topic-specific measures that allow to distinguish different classes of topics based on some key properties.
To characterize topics four measures are here employed, refer to Section \ref{sec:methods} for details. Two measures are employed to summarise network-level properties of topics, that is the topic-specific JSD and average Barrat Clustering Coefficient (BCC) \citep{barrat2004architecture}, while the other two measures characterise topics by their distribution in the corpus, inferred through LDA. 
A gaussian finate mixture model \citep{scrucca2016mclust} is employed for clustering topics in an unsupervised way using as features the topic metrics.  This method allows to clusters topics into groups without having to make assumptions about the number of groups (mixing components) and the covariance parameterisation, which are selected automatically using the Bayesian Information Criterion (BIC). 
Two clusters are identified\footnote{Before clusterization we remove two linguistic outliers, that is, topic $\#54$ (French) and $\#106$ (Spanish).}. As shown in figure \ref{fig:clustering}, for the two clusters, the distribution of topics in the four selected dimensions suggests a clear interpretation of their structural differences: The first cluster (in red) contains cross-cutting topics, which are those that have low topic propensity variance with respect to the mean. Cross-cutting topics are also characterised by relatively higher JSD and BCC levels with respect to the other cluster (in blue), which includes specialised topics, that is, those that exhibit high topic propensity variance with respect to their mean. 
Inside the cross-cutting topics cluster we find, for example, topic $\#36$, which is about the economic effects of the COVID-19 pandemic.
On the other side, among the topics belonging to the specialised topics cluster we find topic $\#88$ and $\#50$, which focus respectively on COVID-related risk factors and on the inhibitors of SARS-CoV-2 main protease, among others.

These results suggest that the documents of the Cord-19 corpus are based on two structurally different clusters of topics, the first one includes topics that are more technical and focused on field-specific terms. For these topics, the information provided by the reconstructed topic-networks is relatively more in line with that provided by LDA's topic-word distributions. The second cluster is made by cross-cutting topics that employ a more varied and generic (\ie non domain specific) set of terms. For the latter topics, reconstructed topic networks appear to provide further information that cannot be captured by classical (bag-of-words) topic modelling approaches. 

In addition to the aforementioned findings, the proposed metrics can serve the readers to explore the set of topics in a more informed  and effective way \textit{LDA2Net} . One can,  even before looking at the topics' contents, differentiate specialised and cross-cutting topics. This information allows \textit{LDA2Net} users to understand which topics should be first analysed more closely, that is those that have a more complex network structure. Moreover, JSD can be used as a criteria to single out for which topics the word sequence information provided by \textit{LDA2Net} is more relevant (\ie those with the highest JSD values).
Hence, thanks to \textit{LDA2Net}'s aggregate topic metrics, topic models with hundreds of topics are easier to explore and understand.
\FloatBarrier
\paragraph{Detecting,  Exploring and Labelling Subtopics}\label{subsec:subtopics}

An additional advantage provided by \textit{LDA2Net} topic networks consists in the possibility of sampling random walks, and by so doing generate sequences of words that represent topic-specific phrase fragments. Random walks in topic-specific word networks can also be used for identifying topic communities (i.e., subtopics), for example through the walktrap community algorithm \citep{pons2005computing}.

\begin{figure}[ht!]
    \centering
    \includegraphics[width=0.99\textwidth]{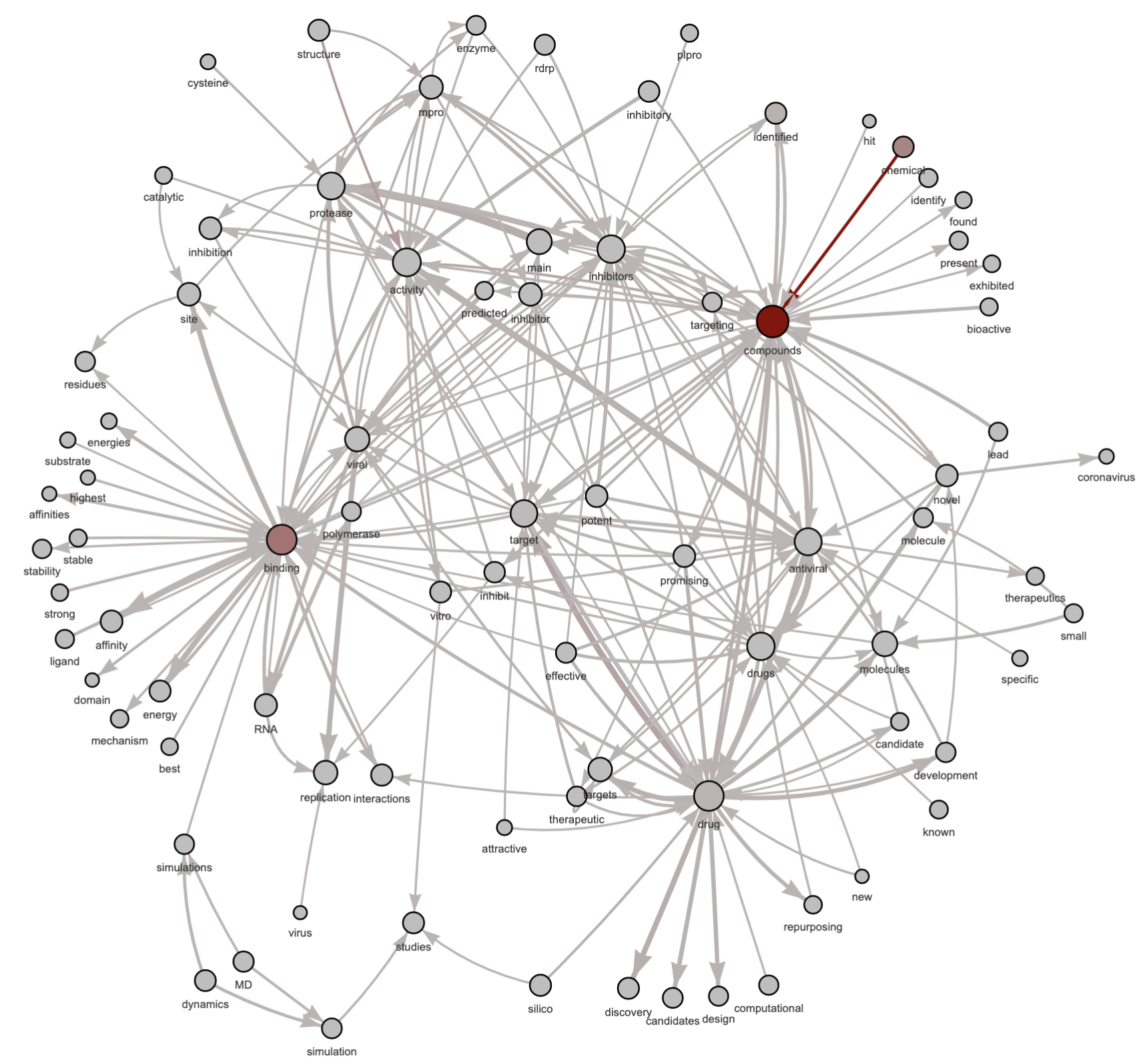}

    \caption{Filtered network of topic $\#88$. Nodes are words (i.e., unigrams) and an (oriented) edge between two words occurs if they form a bigram. Node size is proportional to topic-specific word probability provided by LDA. Edge width is proportional to topic-specific bigram \textit{LDA2Net}-weights. Node and edge color represent their betweenness centrality, ranging from gray ($min$ observed value for nodes/edges) to dark-red ($max$ observed value for nodes/edges).
    }
    \label{fig:wholetopic}
\end{figure}

\begin{figure*}[ht!]
\centering
\begin{subfigure}[t]{0.52\textwidth}
\centering
\includegraphics[width=\textwidth]{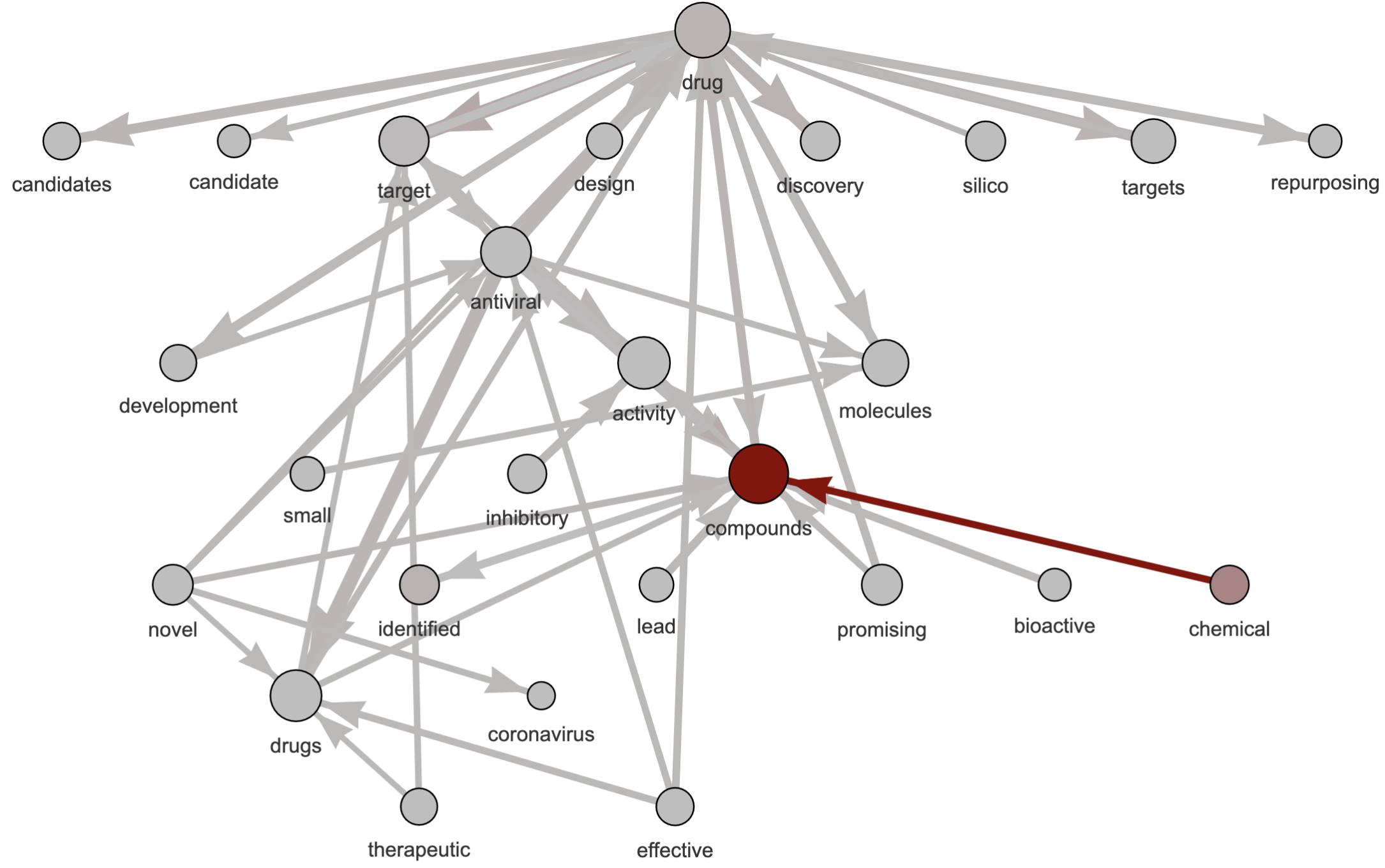}
\caption{Top 50 edges of the $1^{st}$ subtopic of topic $\#88$ (12192 words)}
\end{subfigure}

\begin{subfigure}[t]{0.86\textwidth}
\centering
\includegraphics[width=\textwidth]{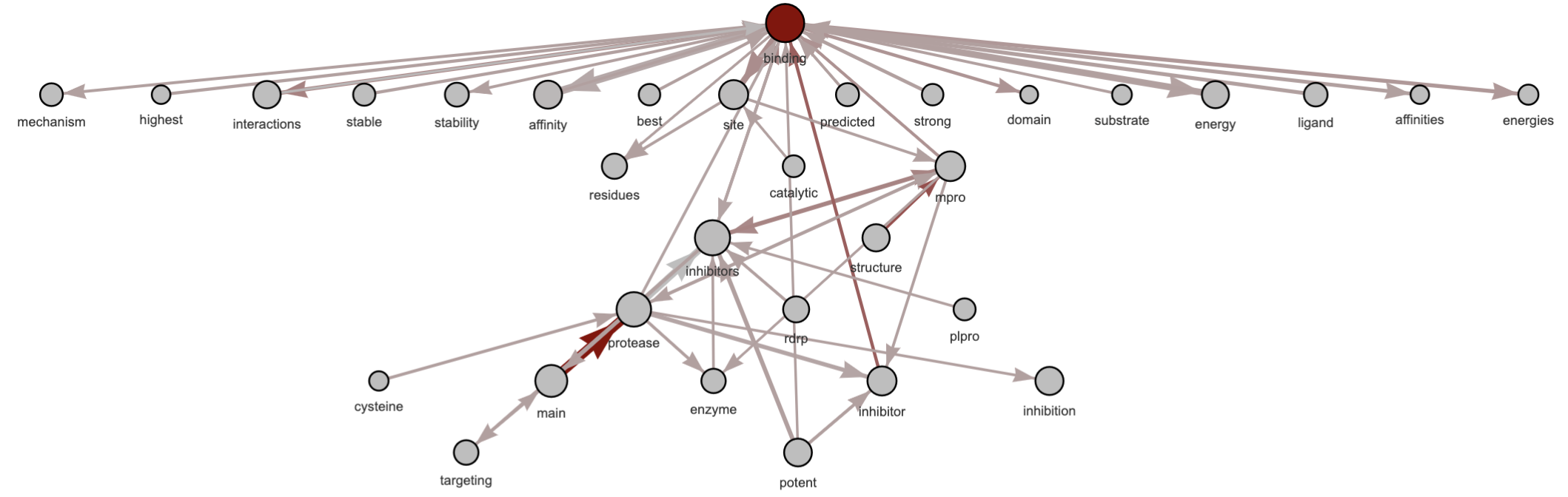}
\caption{Top 50 edges of the $2^{nd}$ subtopic of topic $\#88$ (8443 words)}
\end{subfigure}%

\begin{subfigure}[t]{0.64\textwidth}
\centering
\includegraphics[width=\textwidth]{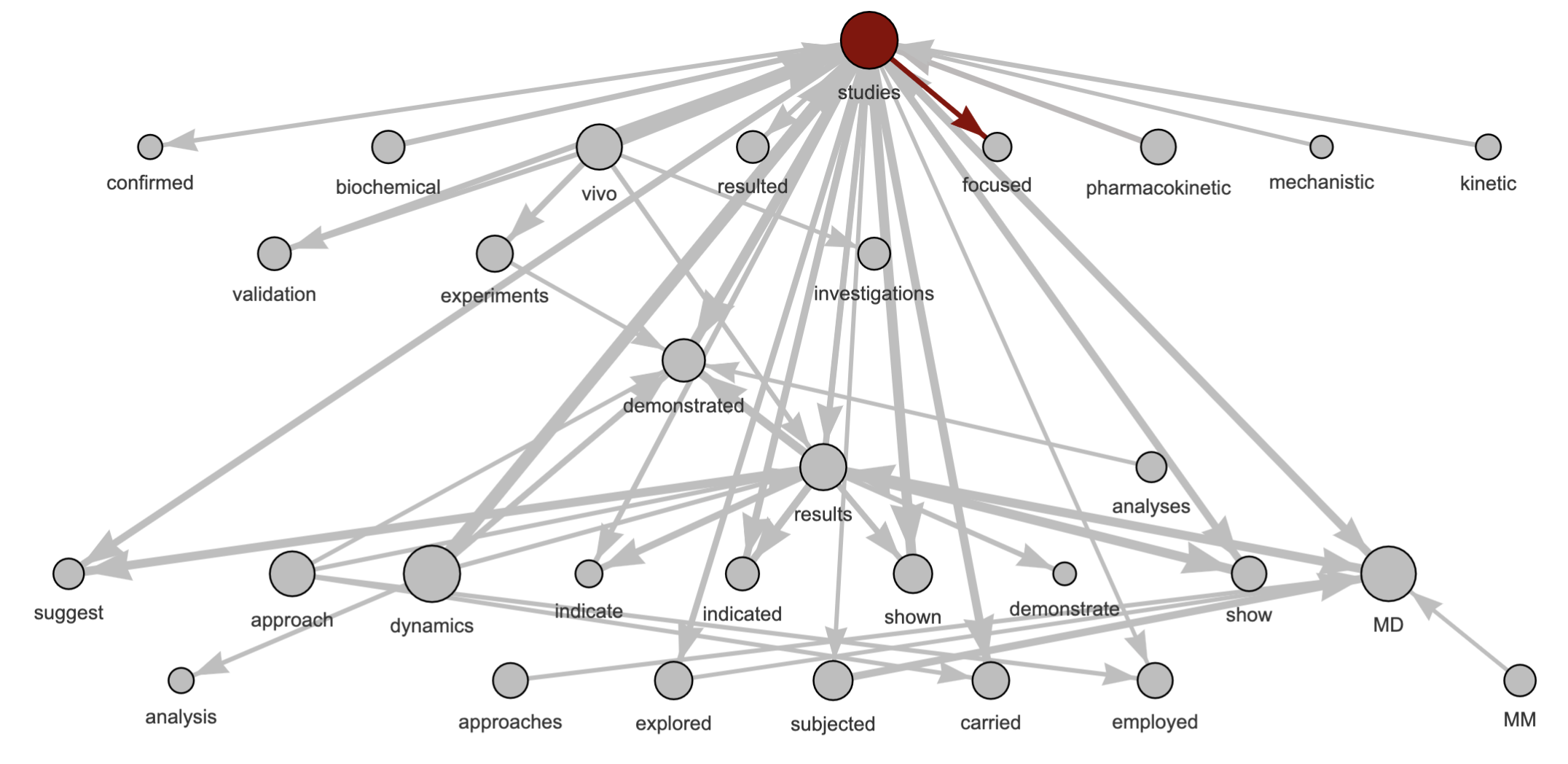}
\caption{Top 50 edges of the $3^{rd}$  subtopic of topic $\#88$ (2965 words)}
\end{subfigure}%

\begin{subfigure}[t]{0.94\textwidth}
\centering
\includegraphics[width=\textwidth]{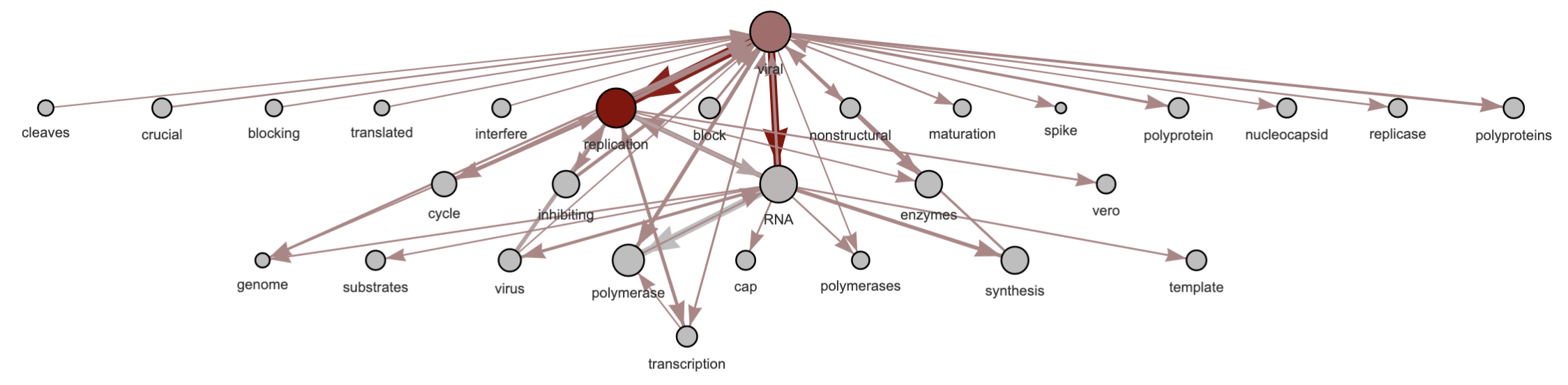}
\caption{Top 50 edges of the $4^{th}$ subtopic of topic $\#88$ (1638 words)}
\end{subfigure}


    \caption{Top-50-edges graphs of the four largest subtopics/communities for Topic 88. Subtopics where constructed using the walktrap comunity algorithm from \cite{pons2005computing}, implemented through the \textsc{igraph} \citep{csardi2006igraph} R library. Node size is proportional to topic-specific word probability provided by LDA. Edge width is proportional to topic-specific \textit{LDA2Net}-weights. Node and edge color represent their betweenness centrality, ranging from gray ($min$ observed value for nodes/edges) to dark-red ($max$ observed value for nodes/edges).
    }
    \label{fig:subtopics_topic_88_graph}
\end{figure*}

\FloatBarrier
 
%

\begin{figure}[!h]
\centering
\includegraphics[width=0.65\textwidth]{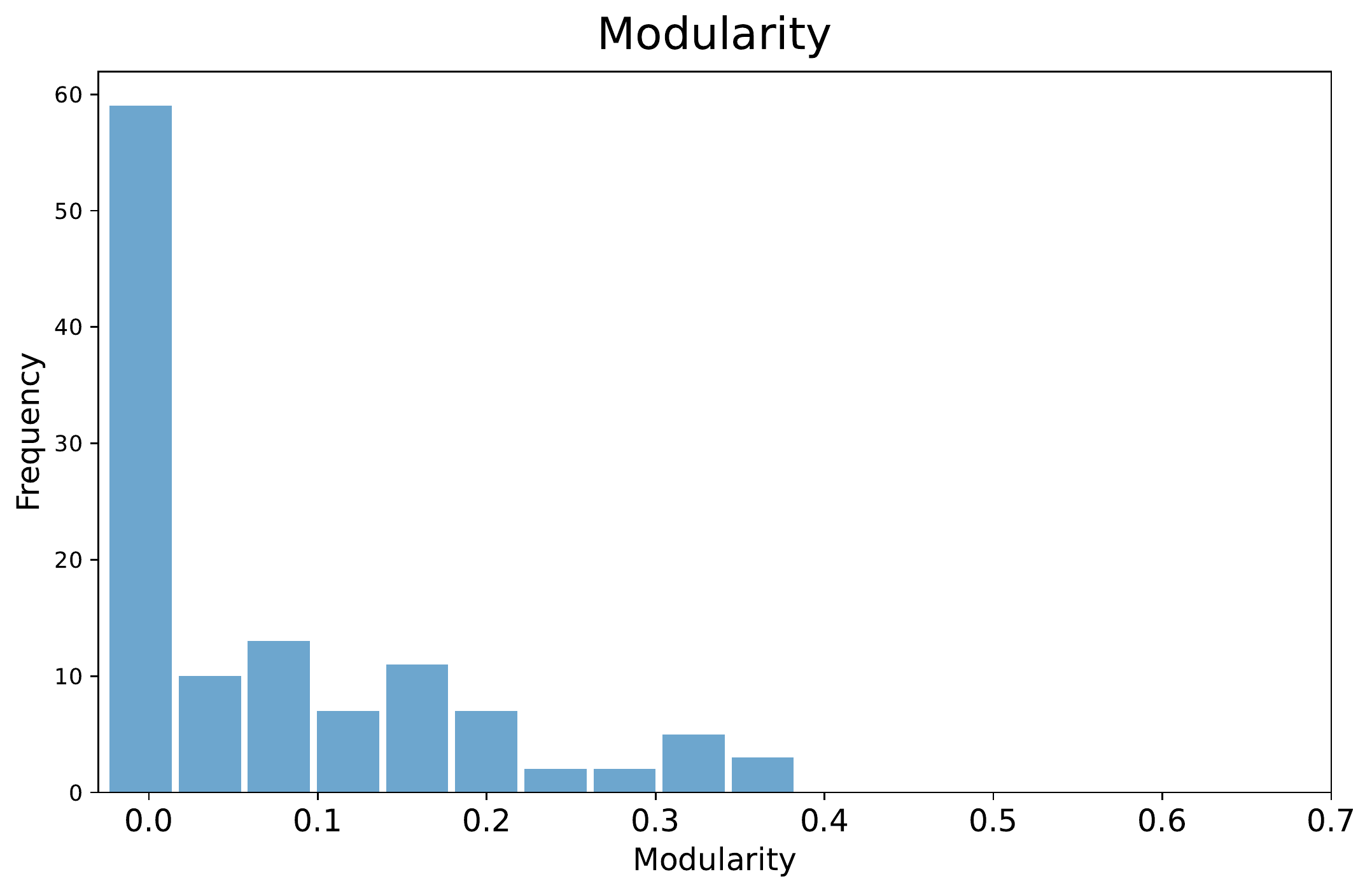}
 
\caption{Histogram of the modularity of topic networks.}\label{fig:modularity}
 
\end{figure}
\begin{figure}[!h]
\centering
\includegraphics[width=0.99\textwidth]{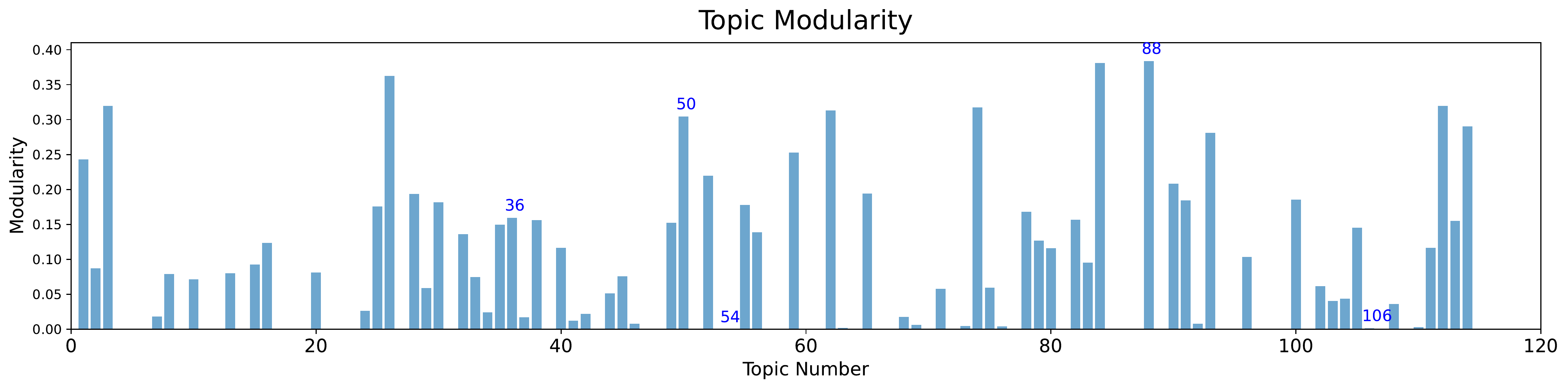}
 
\caption{Bar plot of modularities by topic.}
 \label{fig: }
\end{figure}

 Figure  \ref{fig:modularity} represents the distribution of topics' (weighted) modularity, which is skewed and has a long (and relatively flat) right tail. Notably, about half of the topics have a modularity very close to zero. Recall that, topic networks with high modularity have strong relationships between words within communities and weak relationships between words in different communities. Since topics that have a high network modularity are good candidates for containing closely tied word communities (\ie subtopics), in this subsection, we first partition and then dive-into the network of topic $\#88$, which exhibits the highest modularity among all topics. As shown in Figure \ref{fig:wholetopic}, which represents the filtered network of topic $\#88$ \footnote{Words (represented by nodes) were filtered based on their posterior probabilities, keeping the top 1\% percentile. Bigrams (represented by edges) among top 1\% nodes were filtered, keeping the top 1\%  percentile based on their  \textit{LDA2Net}-weights.}, topic $\#88$ has a modular structure organized around few hubs. Relevant hubs, like the words \texttt{binding}, \texttt{compounds}  and \texttt{drug}, can be easily identified through a visual inspection of the filtered topic network.

To identify and explore the communities of topic $\#88$, we employ the (weighted) walk-trap method proposed by \cite{pons2005computing}\footnote{For this purpose we have employed \href{https://igraph.org/r/doc/cluster_walktrap.html}{\textsc{igraph}'s weighted walktrap community algorithm} implementation, using 4-steps random walks based on \textit{LDA2Net}-weights.}. By so doing, we obtained a set of topic communities each of which can be represented as a network.  In Figure \ref{fig:subtopics_topic_88_graph}, the top 50 bigrams (by \textit{LDA2Net}-weight) of the four largest communities (\ie subtopics) of topic $\#88$ are represented using a hierarchical network layout.
 The first community by size (Figure \ref{fig:subtopics_topic_88_graph} (a) ) focuses on chemical compounds and antiviral drug design. The second community (Figure \ref{fig:subtopics_topic_88_graph} (b) ) is about Mpro inhibitors and binding mechanisms. The third community (Figure \ref{fig:subtopics_topic_88_graph} (c) ) captures some methodological aspects of topic $\#88$ and focuses on health professionals (e.g., \texttt{MD} and \texttt{MA}), study types (e.g., \texttt{vivo$\rightarrow$experiments} and \texttt{pharmacokinetic$\rightarrow$studies}), result-related explanations  (e.g., \texttt{results$\rightarrow$indicate} and \texttt{approach$\rightarrow$demonstrated}).  Finally, the fourth  community (Figure \ref{fig:subtopics_topic_88_graph} (d) ) is about viral RNA replication and transcription. 

Once the network of topic $\#88$ has been partitioned in subtopics, we can automatically generate label candidates for each subtopic through an heuristic that exploits node and edge metrics to sample random walks of different lengths from topic communities. We refer the reader to section \ref{sec:methods} for details about the proposed heuristic.
By observing the distribution of sampled phrase fragments for a specific topic or topic-community one can have a better understanding of the type of contents and phrases that a network can generate.
Many alternative or more sophisticated heuristics may be conceived for specific purposes, for example one could enrich nodes with POS tags and use a grammar to constraint the random walk and produce well formed sentences from topic or subtopic networks. 
Table \ref{tab:labels} enumerates, for the top four subtopics of topic $\#88$, the most frequently sampled n-gram label candidate of length two, three and four.
\begin{table}[ht!]
    \centering
    \footnotesize
    \begin{tabular}{|c|c|c|c|}
    \hline
  \textbf{subtopic}  &  \textbf{2-gram label} & \textbf{3-gram label}  & \textbf{4-gram label } \\\hline
    $1^{st}$    & drug$\rightarrow$target (118) & antiviral$\rightarrow$activity$\rightarrow$compounds (42) &  antiviral$\rightarrow$targets$\rightarrow$identified$\rightarrow$compounds (29) \\\hline
     $2^{nd}$   &  main$\rightarrow$protease (220) & main$\rightarrow$protease$\rightarrow$pro (66) &  binding$\rightarrow$interactions$\rightarrow$evidenced$\rightarrow$AAs (34) \\\hline
       $3^{rd}$  &  MD$\rightarrow$ post (89) & MD$\rightarrow$post$\rightarrow$mortem (23) &  MD$\rightarrow$post$\rightarrow$mortem$\rightarrow$resulted (23) \\\hline
        $4^{th}$  & viral$\rightarrow$RNA (346) &  viral$\rightarrow$RNA$\rightarrow$plus (21) & viral$\rightarrow$RNA$\rightarrow$plus$\rightarrow$nucleocapsid (14) \\
             \hline
    \end{tabular}
    
     \textit{AAs is the acronym for Amino Acids, MD is the acronym for Medical Doctor(s), and RNA is the acronym for Ribonucleic Acid }
     
    \vspace{0.5em}
    \caption{Most frequently sampled label candidate for top four subtopics (in rows) of topic $\#88$, as a function of random walk length (in columns). \\ In parenthesis: frequency of walk out of a sample of 1000 random walks of that lenght.
    }
    \label{tab:labels}
\end{table}
The automatically generated labels in Table \ref{tab:labels} appear to capture relatively well the contents of subtopics represented in Figure \ref{fig:subtopics_topic_88_graph}. Notably, the 3-gram label candidates that have large observed frequencies (out of a sample size of 1000 random walks), like \texttt{antiviral$\rightarrow$activity$\rightarrow$compounds} that stands for the "antiviral activity of compounds"\footnote{During the pre-processing step, stopwords, like \texttt{of}, have been removed.}, represent relatively well the content of topic communities through a short expression that can be used to characterise a subtopic through a sequence of words.

To clean incorrectly formed phrase fragments from the list of label candidates, one can also filter out all n-grams of length 3+ that have never  been observed in the corpus, and hence use the most frequently sampled -and also observed- ones as labels. This post-sampling step may improve the quality of label candidates at the expense of the computational time required to produce them.

The same heuristic can also be applied to generate label candidates at the topic  level. However, since in this work topic networks are extremely large, the proposed method applied at the topic level may produce a very wide range of labels each of which is observed few times, and an almost uniformly distributed set of label candidates can be problematic when the final purpose is to choose only one of them. Therefore, rather than sampling directly topic labels from topic networks, we suggest to use the label candidate of the largest subtopic of a topic also as the topic level label. Accordingly, label candidates have also been generated for all topics by applying the aforementioned heuristic (see Table \ref{tab:topic_labels} in the Appendices).

\FloatBarrier
\section{Discussion}\label{sec:discussion}

This paper applies a novel approach, called \textit{LDA2Net}, to the modelling of topics in the Cord-19 corpus. Our approach builds on top of LDA and allows to reconstruct topic- and subtopic-specific word networks. These networks increase the transparency and readability of identified topics, allowing to appraise and study the structure of COVID-19-related topics in a statistic- and network-informed way.

From the point of view of the COVID-19 literature exploration, \textit{LDA2Net} provides a unique in-depth perspective on key issues discussed in the scientific literature. In particular, \textit{LDA2Net} facilitates the Cord-19 corpus exploration and understanding process for both specialist and non-specialist audiences, acting as a human-facilitation layer between the LDA outputs and the topic model users. In addition, (sub)topic-specific labels, measures and network visualizations can guide the exploration of the Cord-19 corpus and improve COVID literacy through an effective joint usage of LDA and bigrams information. Interpreting topics  hence becomes less discretionary and  more intuitive and accessible. This may allow, for example, to explore and appraise scientific narratives related to drug design, risk factors, or  the economic impact of the virus on tourism, and to disentangle from modular topics the sub-issues therein contained. Even highly experienced and specialized researchers can gain a broader map of the field outside their domain and thematic cross-connections with other domains of expertise. Furthermore, search by relevant sets of words can be greatly facilitated by the inspection of topic word networks.


Through a  broad set of topic metrics, some of which based on the topics' network structure, we were able to disentangle two clusters of topics that are structurally different in terms of their characteristics, making it possible to distinguish cross-cutting topics from specialised ones. Interestingly, it appears that the constraints introduced by \textit{LDA2Net} add relatively more information to cross-cutting topics, for which the information provided by the pairwise product of word probabilities obtained from LDA diverges relatively more (with respect to specialised topics) from that based on observed bigram counts in the corpus.

Moreover, through community detection and labelling algorithms, based on random walks, our method allows to partition topics in subtopics, and use subtopic networks to capture at a finer grain the key issues that are discussed in relation to each topic.

The current framework also offers some technical advantages with respect to (classical) LDA. One of them consists in being more robust to mis-specification of the number of topics, thanks to the possibility of identifying subtopics  ex-post. Notably, through the identification of word-communities in topic networks, modular topics that contain multiple (sub)issues can be identified, labelled and explored at a later stage.

With respect to topic models estimated with bigrams \citep{mackay1995hierarchical, wallach2006topic, wang2007topical, yan2013biterm, nallapati2007sparse}, one of the advantages of \textit{LDA2Net} is that  its results do not depend on any assumption on the distribution family and data generation process of bigrams, being \textit{LDA2Net} based on the observed frequencies of bigrams in documents, which are combined (a-posteriori) with LDA output matrices, as shown in Figure \ref{fig:ldatograph}. 



There are numerous natural developments of the \textit{LDA2Net} approach. One could use Part-Of-Speech tags and/or word Dependency Relations as an alternative to bigrams to construct topic-specific relations among words or lemmas and to typify the nodes and relations in topic networks. 
One could also exploit document-level date-time to create dynamic networks for each topic at the desired granularity, or other document-level metadata to generate covariate-value specific topic networks.
Moreover, we envisage to validate the advantages of \textit{LDA2Net} through a set of experiments with humans, both experts and non-experts subjects, and using different corpora.
Finally, we plan to publish a open-source \texttt{R} library (and possibly also a Python version) to facilitate the deployment, usage and further extension of \textit{LDA2Net}. 
\FloatBarrier
\section{Methods}\label{sec:methods}

In the following sections, we   present our the \textit{LDA2Net} approach from a mathematical and technical viewpoint. The method is also summarized by the scheme in Figure \ref{fig:ldatograph}.

\subsection{Dataset}
The current work is based on the article abstracts of the Cord-19 corpus (Version 93: 21-06-2021), which is made available by Semantic Scholar and the Allen Institute for AI. For details on the Cord-19 corpus plese see \cite{wang2020cord}. The dataset contains coronavirus-related research papers from PubMed, bioRxiv, and medRxiv,  as well as articles from a repository of more than two hundred thousand papers maintained by the World Health Organization. 
As this work focuses on COVID-19,  abstracts uploaded before December 31, 2019 have been filtered out, since mainly concerning SARS and MERS. The resulting dataset contains $398818$ abstracts.
Details about the preprocessing stage can be found in Appendix \ref{sec:datasetprepro}.

\begin{table}[h!]
\caption{Dataset summary information and LDA parameters.}
\centering
\begin{tabular}{|l|c|c|c|c| }
\hline
         &  \#   documents    & \#   topics & \# words (vocabulary) & \# bigrams (vocabulary)   \\
         \hline\hline
$value$   & $398818$            & 120          & 34563                 & 4075820                 \\\hline 
$notation$ & $N_{\mathcal{D}}$ & K            & $N_{\mathcal{W}}$     & $N_{\mathcal{B}}$         \\\hline
\end{tabular}
\end{table}

\subsection{From LDA to Topic Networks}\label{subsec:networkcreation}
\medskip

\paragraph{Latent Dirichlet Allocation} 
This section outlines the  basic concepts about Latent Dirichlet Allocation (LDA) needed for understanding our approach. As LDA analysis is not the crux of this work, we refer the reader to \cite{blei2003latent} and Appendix \ref{sec:app:lda} for a more thorough introduction to LDA, and to Appendix \ref{sec:app:topics} for details about the model estimation and parameters selection.

In layman's terms, LDA is an algorithm that reads through some text documents and automatically outputs the topics discussed. In order to perform this process, LDA takes as input a  collection of documents $\mathcal{D}$ - where $|\mathcal{D}| = N_{\mathcal{D}} $ -    called \textit{corpus}.   Each document  being represented as a set of words belonging to a vocabulary  $\mathcal{W}$,  namely  the list of all unique words in the \textit{corpus}. Formally, $\mathcal{W}= \{ w_{1}, \dots, w_{N_{\mathcal{W}}} \}$, where $N_{\mathcal{W}}= |\mathcal{W}|$. 
In this context, a topic is a set of words that occur frequently together.   The number of topics $K$ to be found is instead an input parameter. With a view to processing the documents, the \textit{corpus} is transformed into a matrix containing the counts of words by document, hereafter called $\mathbf{U}$, of size $N_{\mathcal{D}} \times N_{\mathcal{W}}$. Each entry $\mathbf{U}_{d,i}$ represents the number of times the word $w_i$ appears in the document $d$, where $i = 1, \dots, N_{\mathcal{W}}$ and $d = 1, \dots N_{\mathcal{D}}$.
LDA outputs are distributions namely:
\begin{enumerate*}
    \item[$\sbullet$] a distribution of words for each topic (\ie a weighted list of words);
    \item[$\sbullet$] a distribution of topics for each document.
\end{enumerate*}
In \cite{blei2012probabilistic} a topic is formally defined as a distribution over a fixed vocabulary of words.
 These results are usually delivered in matrix format. Notably, let $\mathbf{M}$ be the topic-word distribution matrix of size $\mathrm{K} \times N_{\mathcal{W}}$ and $\mathbf{Q}$ be the 
document-topic distribution matrix of size  $N_{\mathcal{D}}\times\mathrm{K}$. Then, the entry $\mathbf{M}_{k,i}$, where  $k = 1,\dots K$ and $i = 1, \dots N_{\mathcal{W}}$, is the weight 
of the word $w_i$  within the  topic $k$,   whereas the entry $\mathbf{Q}_{d,k}$, where $d = 1, \dots N_{\mathcal{D}}$, is the proportion of topic $k$ in the document $d$.
For a visual interpretation of the these matrices we refer the reader to Figure \ref{fig:ldatograph}.







\paragraph{Network Construction} In this section, we explain how to convert any topic obtained through LDA into a network. Basically, we aim at converting a weighted list of words into a weighted and directed network, that is a network where each edge has both a direction and a weight and every node has a weight. The ultimate goal is thus providing a way to define the adjacency matrix of the network of a chosen topic (see Appendix  \ref{sec:appendix_nettheory} for notation and definition of adjacency matrix). In other words, we need a way to gauge the direction and weight of the syntagmatic relation between any pair of words,  represented by edges in the network.

As first stage, the \textit{LDA2Net} method requires a data preparatory phase that involves collecting all bigrams present in the \textit{corpus} and arranging them in a fashion akin to the matrix $\mathbf{U}$. That means, building  a  vocabulary for bigrams  $\mathcal{B}= \{ b_{1}, \dots, b_{N_{\mathcal{B}}}\}$, where $N_\mathcal{B} = |\mathcal{B}|$. Unlike $\mathcal{W}$, this vocabulary contains unique ordered pairs of words. Then, bigram frequencies by document are collected into a matrix  $\mathbf{S}$
of size $N_{\mathcal{B}} \times N_{\mathcal{D}}$. Thus,  each entry $\mathbf{S}_{b,d}$ represents the frequency of the bigram $b$ in the document $d$.

The next stage interests the computation of one of the two components of the \textit{LDA2Net}-weights associated to a bigram, conditioned by the topic under analysis. This component is called $counts$-weight.  Given a topic, a bigram will assume a $counts$-weight proportional to both its occurrence (counts) in documents and the proportion of the given topic in documents. This in turn means the same bigram may have a different $counts$-weight depending on the topic under investigation. 
Formally, let  $\mathbf{C}$ be the matrix of size $N_\mathcal{B}\times K$  collecting bigram $counts$-weights by topic; then each entry of  $\mathbf{C}$ is computed as $\mathbf{C}_{b,k} =  \sum_d \mathbf{S}_{b,d}\, \mathbf{Q}_{d,k}$. The same definition expressed for the whole matrix $\mathbf{C}$  is:

\begin{equation}
  \mathbf{C} =   \mathbf{S} \mathbf{Q}
\end{equation}

However, the   quantities   in 
$\mathbf{C}$ do not take into account topic-specific probabilities of words composing the bigram, inferred through LDA. In fact, this information is given by $\mathbf{M}$, the topic-word distribution matrix.
For this reason, in the  computation  of the adjacency matrix of the network we combine the $counts$-weight with a second component, called $probs$-weight, which embeds these values. Specifically, let $A^k$ be the adjacency matrix describing the words network for the topic $k$, of  size equal to $N_\mathcal{W}\times N_\mathcal{W}$. You may notice that,  regardless of the topic under consideration, every network has the same node-set.
Nevertheless, topic networks are not identical as the \textit{LDA2Net}-weights of edges between words, as well as the node weights, distinctively characterize each topic. 
In this regard, let $e_{i,j}^k$  be
the \textit{LDA2Net}-weight of the bigram made by the ordered pair of words $w_i$ and $w_j$ in the network representing topic $k$. Such a quantity exists if and only if the pair of words  $(w_i,w_j)$ is a bigram belonging to the vocabulary of (observed) bigrams $\mathcal{B}$.

Then, we define the adjacency matrix for the topic $k$ as: 

\begin{equation}
    A^k_{i,j}=
    \begin{cases}
      0, & \text{if}\ (w_i, w_j) \notin \mathcal{B} \\
      e_{i,j}^k, & \text{otherwise}
    \end{cases}
  \end{equation}
where $e_{i,j}^k$  is given by the product between the \textit{counts}-weight of the bigram  $(w_i, w_j)$ for the topic $k$, that is $\mathbf{C}_{b,k}$, and its \textit{probs}-weight, which is obtained by multiplying the LDA word probabilities of $w_i $ and $ w_j$, which are respectively given by $\mathbf{M}_{k,i}$ and $\mathbf{M}_{k,j}$. As a result, we have that \textit{LDA2Net}-weight of the bigram  $b=(w_i, w_j)$ is:

\begin{equation}
   e_{i,j}^k = \mathbf{C}_{b,k}  \mathbf{M}_{k,i}   \mathbf{M}_{k,j}
\end{equation}

In this respect, we would like point out that if the bigram composed by $(w_i, w_j)$ does not exist, it does not imply the bigram $(w_j, w_i)$ does not exist as well, \ie  $(w_i, w_j) \neq (w_j, w_i)$. In fact,  a bigram represents an oriented edge between two words.
Finally, we normalise $A^k$ entries, so that, for each $k$, the sum of all entries is equal to 1.


\subsection{Topic Metrics}\label{sec:ovearll}

A set of metrics to characterize each topic has been constructed and employed to classify topics in groups.
These metrics are: 

\begin{itemize}
    \item {\em Topic Mean and Variance}: respectively the average value $\mu_k$ and its variance $\sigma^{2}_k$ of the probability of a topic $k$ to appear in the corpus, formally $\mu_k = \sum_d \mathbf{Q}_{d,k}/N_{\mathcal{D}}$ and $\sigma^2_k = \sum_d (\mathbf{Q}_{d,k}-\mu_k)^2/N_{\mathcal{D}}$
    \item \textit{JS Divergence}: the Jensen–Shannon divergence (please  see Appendix \ref{sec:appendix_JSD}) over the topic-bigram distributions obtained from 
    the two components of the edges' \textit{LDA2Net}-weight for topic $k$, that is, between \textit{counts}-weights ($\mathbf{C}_{b,k}$) and \textit{probs}-weights ($\mathbf{M}_{k,i}\mathbf{M}_{k,i}$)
   \item \textit{Barrat clustering coefficient}: the weighted version of the clustering coefficient for each topic network \citep{barrat2004architecture} (please see Appendix \ref{sec:appendix_nettheory})

\end{itemize}

The Barrat clustering coefficient, which has to do with the statistical level of cohesiveness of the topic's graph, measuring the global density of
interconnected vertex triplets in the network, and the Jensen–Shannon divergence, which informs us of the redundancy among the two components used to compute \textit{LDA2Net}-weights, are in many aspects complementary to the topic mean and variance  metrics. 
As a result, these metrics can be jointly employed for clustering topics (in this work we employ a gaussian finate mixture model for this purpose). This allowed us to distinguish between cross-cutting and specialized topics (see Figure \ref{fig:clustering}). 


\subsection{Automatic Topic Community Labelling}\label{sec:labels}
\begin{figure}[H]
    \centering
    \includegraphics[width=0.99\textwidth]{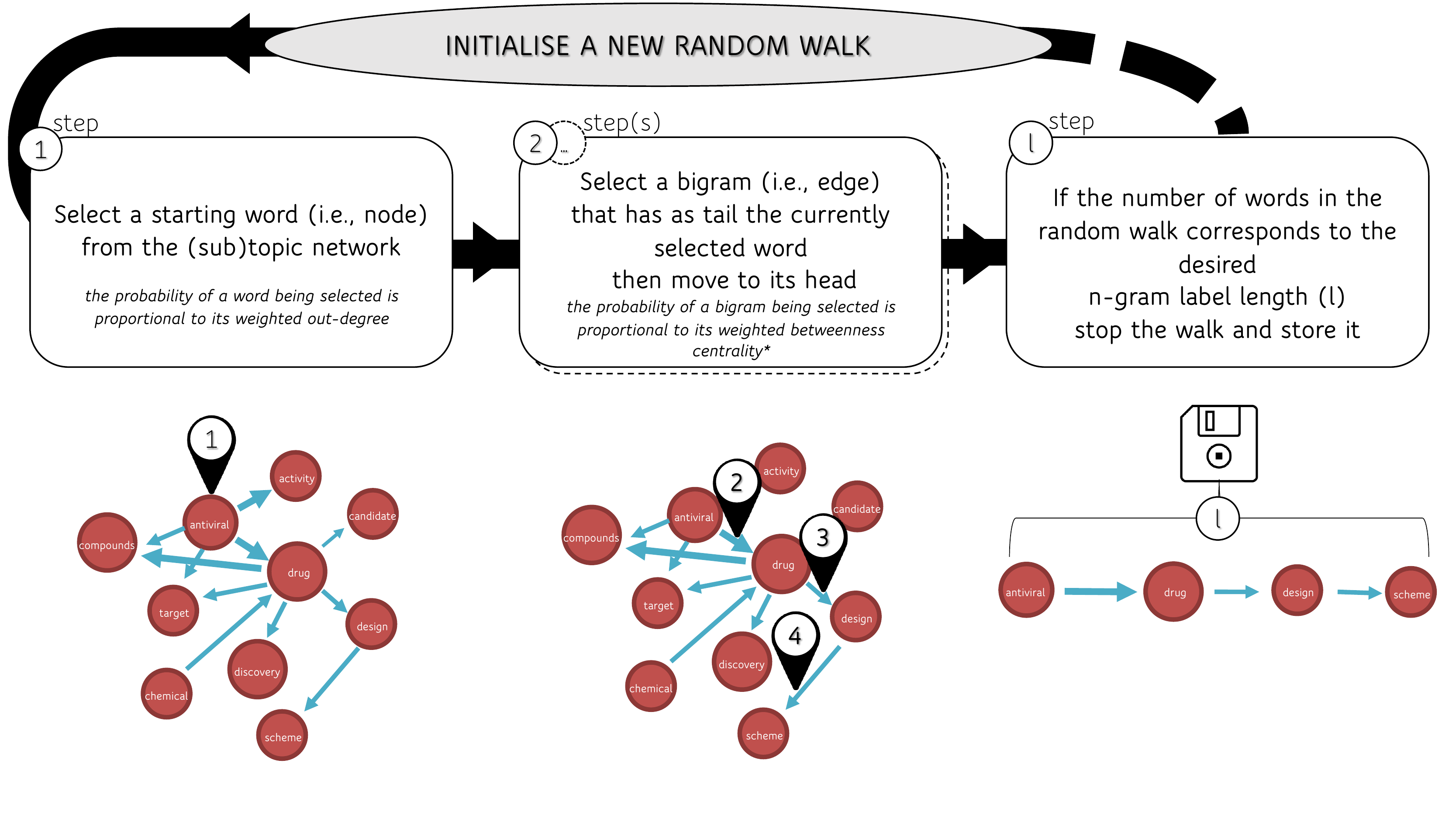}
    {\footnotesize \textit{* if the betweenness of edges having as tail the selected node are null use the edges'  \textit{LDA2Net}-weights instead}}
    \caption{Visual summary of the heuristic employed for generating (sub)topic label proposals.
    }
    \label{fig:heuristic}
\end{figure}

One can leverage the edge and node metrics detailed in the Appendix \ref{sec:centralities}   to create an heuristic capable of generating label propositions with desirable properties from random walks in topic networks or subtopic networks.
The heuristic here proposed for generating labels for (sub)topics using their network structure is summarised in Figure \ref{fig:heuristic}. 
Given the network $\mathbf{G}_{m}^k$ of a community $m$ of a topic $k$, a node is chosen to be the starting vertex of the random-walk. 
The weighted out-degree centrality of nodes is used as the criterion to determine from which word (i.e., unigram) the random-walk is initialised.
This node is sampled from a discrete distribution for which the probability of node $i$ being extracted is proportional to its weighted out-degree, that is $deg^{out}_{i}$. 
Given a topic community network $\mathbf{G}_{m}^k= \mathbf{H}$, we have that   the probability $Pr_{{i}_\mathbf{H}}$  of a node $i$ to be selected is:

  
 \begin{equation}
   Pr_{{i}_\mathbf{H}}=\frac{deg^{out}_i}{\sum_{z}^{|\mathbf{H}|}deg^{out}_{z}}
  \end{equation}
  
One of the advantages of the proposed heuristic initialisation step is that the resulting random-walks are more likely to start from nodes that are hubs from the point of view of the relevance of their outgoing edges, which is a good proxy for being a starting term in phrases related to a given topic or topic community.

Once a starting node has been sampled, for randomly moving from node to node, rather than using the edges' weights contained in
${A}^k$, one can exploit the edge betweenness centrality
of the latter. The reason for this is that the weighted betweenness centrality of edge (i,j), that is called $btw_{(i,j)}$, better captures the overall relevance of a bigram in a network, providing a global rather than local criterion for randomly moving from word to word.
Given the currently selected word $i$, the probability of moving from $i$  to  $j$  is computed as follows:
 \begin{equation}
         Pr_{{(i,j)}_\mathbf{H}}= \frac{btw_{(i,j)}}{\sum_{z}^{|\mathbf{H}|}btw_{(i,z)}}
  \end{equation}

 In case none of the edges having $i$ as tail have a positive betweenness centrality, edge weights $e_{i,j}$, are used instead of the betweenness centrality $btw_{(i,j)}$ as the weighting criterion for randomly moving from node to node. If the selected word $i$ has no outgoing edges the random-walk is halted.

Given a desired \textit{n}-gram label length $l$, through the aforementioned procedure one can generate thousands of random-walks (i.e., \textit{n}-grams) of the same targeted length, then order them by relative frequency and use the most frequently sampled one(s) as label candidate(s) for a topic or a topic community (e.g., subtopic).

\section*{Acknowledgements}

We thank Dr. Martina Gerotto for her expert support in topic labelling.

C.S. and M.W. acknowledge financial support from the European Union Horizon 2020 project ISEED (Grant Agreement No. 960366). C.S. acknowledges financial support from the European Union Horizon 2020 project MUHAI (Grant Agreement No. 951846).

\bibliographystyle{unsrtnat}

\newpage
\appendix
\appendixpage
  
 \section{Pre-processing}\label{sec:datasetprepro}

Only words that have more than two characters are considered in this work. Moreover, all abstracts that don't contain at least ten words that occur at least once every one million tokens have been removed. The resulting filtered corpus contains $398,818$ documents (i.e., article abstracts). The average number of terms per document is close to one hundred ($98.21$) tokens, with a standard deviation of $45.65$.

Before tokenizing the corpus, uppercase strings referring to the abstract sections (e.g., ‘‘KEY RESULTS'') have been removed using a RegEx, as well as arabic and latin numbers. Also, English stopwords from the \textsc{Quanteda} \citep{benoit2018quanteda} \texttt{R} library have been identified through RegEx and removed from the documents' strings.
Excluding acronyms, all tokens have been lower-cased.
Terms that don't appear at least once every million terms have been removed, and therefore do not belong to the words vocabulary $\mathcal{W}$.
Remark that, in \textit{LDA2Net}, tokenization of documents in bigrams takes place once the stopwords have been removed from the documents' strings. Also, during this process, all punctuation characters (except apostrophes and quotation marks) are considered string breaks, which means that two consecutive words that are separated by one or more punctuation characters (different from the aforementioned exceptions) will not be extracted and counted as a bigram. 
Only the bigrams that are composed by two words (unigrams) contained in in the words vocabulary $\mathcal{W}$ are considered. By so doing, bigrams made up by very rare words are also filtered out.

\section{Topic Modeling }

Topic modeling, in natural language processing, is a widely used technique aiming to automatically find themes (or topics) in text. Topic modeling can be performed by vector space models or Probabilistic Graphical Models. PGMs for topic modelling aim at discovering the latent semantic structures of a corpus by relying on a document generation process. The idea behind the document generation process comes from the human written articles. Indeed, when a person writes an article, she or he has some thinking in mind, some topics, and then she or he will extend these themes into some topic related words, which convey the desired meaning. Eventually, these words will be written down to complete an article. Probability topic models simulate the behavior of articles' generating process.


\subsection{Latent Dirichlet
Allocation }\label{sec:app:lda}

Latent Dirichlet
Allocation  (LDA) as defined in \cite{blei2003latent}, is a probabilistic generative model and draws upon this very idea. We define a generative model a machine learning technique generating an output considering the prior distribution of some objects.
LDA assigns a distribution of topics to each document, and a distribution of words to each topic to provide low dimensional, probabilistic descriptions of documents and words. In other words, LDA assumes that documents are mixtures of multiple topics, typically not many, and each document is generated by a process. Dirichlet Distributions encode the intuition that documents are related to a few topics. A topic, in turn, is a distribution over a fixed vocabulary and each topic is assumed to be generated first, before the documents. Only the number of topics is specified in advance.
 
In plain language, the generative process of a document takes two steps. First, a distribution over topics is chosen randomly (that implies a distribution over a distribution for this step). Then, for each word in the document,  a topic from the distribution over topics  is chosen randomly and next a word from the corresponding topic (distribution over the vocabulary) is picked randomly. Note that words are generated independently of other words (bag-of-words model).
Assuming this generative model for a collection of documents, LDA then tries to backtrack from the documents to find a set of topics that are likely to have generated the collection. In  practical terms, given a set of documents, and the number of topics we would like to discover out of this set, the output will be the topic model, that is the documents expressed as a combination of the topics. The algorithm  provides the weight of connections between documents and topics and between topics and words. Topic models are  represented by top-m word lists for facilitating human interpretation.

For details on LDA refer to \cite{blei2003latent}.

\subsection{LDA parameters and model estimation}\label{sec:app:topics}
To select the number of topics $K$ for the LDA model, we employ the method proposed by \cite{nikita2016} and the \texttt{R} library \textsc{ldatuning}. Accordingly, for each $K \in \{5,10,15,20,25,30,40, 50, 60, 70, 80, 90, 100, 110, 120, 130, 140\}$ we estimate 5 LDA models using using the Gibbs Sampling approach with different seeds and $1250$ iterations (with $n.burnin.iter=250$). Then, four criteria \citep{arun2010finding,griffiths2004finding,deveaud2014accurate,cao2009density} are employed to choose the optimal value of $K$.
Based on the average values of the considered criteria, $K=120$ appears to be the a good candidate number of topics. Finally, to ensure convergence we resume the best run of the LDA estimation obtained for $K=120$ running $3000$ additional iterations.

\section{Network Theory Background}\label{sec:appendix_nettheory}

A network is a collection of vertices joined by edges. Vertices and edges are also called nodes and links in computer science. In mathematics, a network is called graph. A graph $G = (V, E)$ is defined as a set of vertices, $V$, which
are connected by a set of edges, $E \in V\times V$, typically represented as a square matrix. Given a graph $G$ with $n$ nodes, the adjacency matrix $A = (a_i,a_j)$ of $G$ is a square $n\times n$ matrix.
The elements $A_{i,j} = a_{i,j}$ of the adjacency matrix $A$  assume values
$a_{i,j} \in \{0,1\}$, such that  

\begin{equation}
    A_{i,j}=
    \begin{cases}
      0, & \text{if}\ (i, j) \notin E \\
      1, & \text{otherwise}
    \end{cases}
  \end{equation}

that is $a_{i,j}=1$ if there exists an edge joining nodes $i$ and $j$, and  $a_{i,j}=0$  otherwise.
A graph is undirected if edges have no direction. If there is an edge from $i$ to $j$ in an undirected graph, then there is also an edge from $j$ to $i$. This means that the adjacency matrix of an undirected graph is symmetric.
A graph is directed if edges have a direction. If there is an edge from $i$ to $j$  in an directed graph, then there is not necessarily an edge from $j$ to $i$, but it might exist. This means that the adjacency matrix of an directed graph is not necessarily symmetric.
For simple graphs without self-loops, the adjacency matrix has $0$s on the diagonal.  
Many of the networks we study are unweighted, that is, they have edges that form simple on/off connections between vertices. Either they are there or they are not. In some situations, however, it is useful to represent edges as having a strength, weight, or value to them. The weights in a weighted network are usually positive real numbers, but there is no reason in theory why they should not be negative. These weighted networks can be represented by giving the elements of the adjacency matrix values equal to the weights of the corresponding connections. 

\subsection{Network Measures}\label{sec:centralities}
In order to characterize the structure of a graph, it is crucial to study and  quantify its properties. These properties can be organized into three levels of abstraction:

\begin{enumerate}
\itemsep-0.1em 
    \item \textit{element-level}: to identify the most important nodes/links of the network
    \item \textit{group-level}: to find cohesive groups of nodes in the network
    \item \textit{network-level}: to study topological properties of networks as a whole
\end{enumerate}

In network analysis, \textit{element-level} properties are used to measure the level of importance of a single component of a graph wit respect to the others.
Element-level descriptors, also called \textit{centrality measures},  are a crucial tool for understanding networks. These topological indicators,  adopted to score both nodes and edges, are  scalar values assigned to each node (edge) in the graph   in order to quantify the node's (edge's) importance based on a certain assumption. In contrast, 
network-level measurements, calculated over the whole network, provide overall indications about the network structure.
For our analysis, we  take into account the following centralities:

\paragraph{\textit{Degree Centrality} (node)} The degree centrality is the degree of a vertex, the number of edges connected to it. In directed networks, vertices have both an in-degree and an out-degree, and both may be useful as measures of centrality in the appropriate circumstances.
 
\paragraph{\textit{Betweenness Centrality} (node)} The betweenness centrality \cite{brandes2008variants} over nodes measures the extent to which a vertex lies on paths between other vertices. This centrality detects the amount of influence a node has over the flow of information in a graph and thus it identifies nodes that serve as a bridge from one part of a graph to another. It is a measure based on shortest paths:  for each vertex it is equal to the number of  shortest (geodesic) paths that pass through the vertex.

\paragraph{\textit{PageRank} (node)} 
The \textit{PageRank} centrality \cite{page1999pagerank}  is the trade name given it by the Google web search corporation, which has adopted it as key part of the web ranking technology. It is particularly suitable for directed network as it accounts for link direction. Each node in a network is assigned a score based on its  indegree. These links are also weighted based on the relative score of its originating vertex. In this way nodes with many incoming links are influential, and nodes to which they are connected share some of that influence. \textit{PageRank} can help uncover influential nodes whose reach extends beyond just their direct connections.
 
\paragraph{\textit{Betweenness Centrality} (edge)} The betweenness centrality  over edges \cite{girvan2002community} is the sum of the fraction of all-pairs shortest paths that pass through that edge. Usually important bridge-like connectors between two parts of a network have high betweenness centrality as they have a large influence on the transfer of information through the network.

\begin{figure}[H]
\centering
\includegraphics[width=0.4\textwidth]{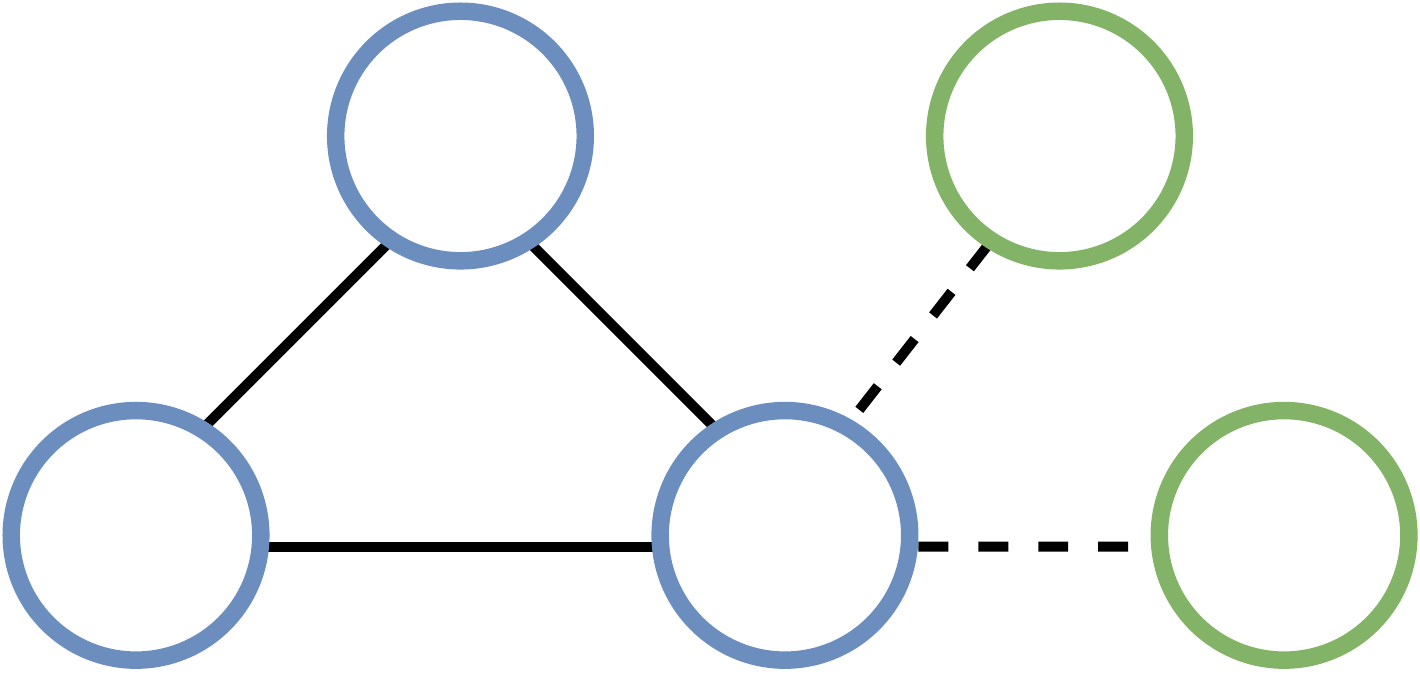}
   
\caption{An example of closed and open triplets.}
\label{fig:triplets}
\end{figure}

\paragraph{\textit{Clustering Coefficient in Weighted Structure} (network)  } The clustering coefficient (CC) is a measure of the degree to which nodes in a graph tend to cluster together. There exist two versions of it:
a global version,  designed to provide an overall indication of the clustering in the network and a local version that gives an idea of the embeddedness of single nodes.
In its global formulation, the clustering coefficient takes into account triplets of nodes. By   triplet we mean  three nodes that are connected by either two (open triplet) or three (closed triplet) undirected ties (see Figure \ref{fig:triplets}); in this context, a triangle is a three closed triplets, one centred on each of the nodes. The global clustering coefficient is the number of closed triplets (or 3 $\times$ triangles) over the total number of triplets (both open and closed). 
A generalisation of the global clustering coefficient to weighted networks (networks with weighted edges) was proposed by \cite{barrat2004architecture}.  Here the authors first propose a novel definition of
the local clustering coefficient
-  measuring the local cohesiveness considering both  
the importance of the clustered structure and the  interaction intensity (e.g., weights) actually found on the local triplets - and then 
the global version,  the weighted clustering coefficient
averaged over all vertices.   
\FloatBarrier

\section{{Jensen-Shannon Divergence} } \label{sec:appendix_JSD}The Jensen-Shannon divergence (JSD) is a measure assessing the difference between probability distributions, that is the ground truth and the simulation, by calculating the mutual information between two probability distributions, quantified by gauging the difference between entropies associated with those probability distributions. The JSD is a symmetrized and smoothed version of the Kullback-Liebler divergence.
In other words, it is a mutual information measure for assessing the similarity between two probability distributions. It is defined as

\begin{equation}
  D_{J S}(P, Q)=\frac{1}{2} D_{K L}(P \| M)+\frac{1}{2} D_{K L}(M \| Q)
\end{equation}

where $D_{K L}$ is the classical Kullback-Leibler divergence  and $ M =\frac{P +Q}{2}$.
 
The generalization of the Jensen-Shannon divergence, in case of more than two probability distributions, is based on the   Shannon entropy as follows 
 
$$
\begin{aligned}
D_{{J S}_{\pi_{1}, \ldots, \pi_{n}}}\left(P_{1}, P_{2}, \ldots, P_{n}\right) &=\sum_{i} \pi_{i} D\left(P_{i} \| M\right) \\
&=H\left(\sum_{i=1}^{n} \pi_{i} P_{i}\right)-\sum_{i=1}^{n} \pi_{i} H\left(P_{i}\right)
\end{aligned}
$$

\noindent where ${\pi_{1}, \ldots, \pi_{n}}$ are the weights of the
probability distributions,  $M=\sum_{i=1}^{n} \pi_{i} P_{i}$ and $H(P)$ is the Shannon entropy for distribution $P$.
 
When using the base $2$ logarithm, the Jensen–Shannon divergence for two probability distributions ranges between $0$ and $1$ \citep{lin1991divergence}.

\section{Community Detection}
 
In the social context, a community is a set of individuals such that those within a group (also referred to as cluster, cohesive subgroup or module) interact with each other more frequently than with those outside the group. In this regard \textit{community detection} is discovering groups where individuals’ group memberships are not explicitly given. Similarly, in network analysis, a community is a group of nodes, which are highly connected to each other than to the rest of the nodes in the network \cite{yang2010discovering}.
In the last years, community structure has increasingly become the
most-studied structural feature of complex networks. 



The approach basically consists in dividing a network into subgroups by grouping nodes that are tightly coupled to each other and  loosely coupled to
the rest of the vertices in the network \cite{fortunato2010community}. 
However, the task of gauging the intuitive concept of community structure is not trivial. 
One of the most effective approaches to  this issue has been   defining a quality function able to estimate
the strength of division of a network into modules \cite{brandes2005network,fortunato2010community}. This strength has been quantified by using the measure known as \textit{modularity} \cite{newman2004finding},  computed as the fraction of the edges that fall within the given groups minus the expected value of the fraction if edges were distributed at random (for more details we refer the reader to \cite{newman2006modularity}).

Although modularity optimization is an NP-hard problem, many community detection algorithms are based on this  principle. The most famous  ones include the greedy method of Newman and its  fast version \cite{newman2004fast}, the  Louvain algorithm \cite{blondel2008fast} and the 
method of Clauset, Newman, and Moore (CNM) \cite{clauset2004finding}.

Another approach to identify communities is by simulating random walks inside the network. 
The general idea is that, given a graph and a starting point, if we select a
neighbour at random and move to the selected neighbour and repeat the same
process till a termination condition, this walk,  namely a random sequence of nodes,   
is more likely to stay within the same community as there are only a few edges that lead outside a given community. In other words, a \textit{random walker} will tend to wander inside densely connected areas of the graph. This concepts inspired  the Walktrap algorithm \cite{pons2005computing}  and   the Infomap algorithm \cite{rosvall2008maps}. Walktrap makes use of short random walks (of 3-4-5 steps) to compute distances between nodes.
Based on (small) intra and (larger) inter-community distances, nodes are being assigned to groups,  via bottom-up hierarchical clustering. In this respect, modularity score can be used to select where to cut the dendrogram. It should be pointed out that this algorithm considers only one community per node, which in some cases can be an incorrect hypothesis.

\newpage
\section{Topic Labels Table}

\centering
{\tiny

\begin{longtable}[ht]{c|l|l|l|l}
 
  \hline
Topic \# & Human label & 2-gram label & 3-gram label & 4-gram label \\ 
  \hline
   \hline
1 & protective devices & protection$\rightarrow$ equipment & protection$\rightarrow$ equipment$\rightarrow$ used & filtration$\rightarrow$ protection$\rightarrow$ equipment$\rightarrow$ used  \\ 
  2 & diagnostics & method$\rightarrow$ developed & method$\rightarrow$ developed$\rightarrow$ used & method$\rightarrow$ developed$\rightarrow$ rapid$\rightarrow$ sensitive  \\ 
  3 & statistical analysis (studies) & analysis$\rightarrow$ used & data$\rightarrow$ analysis$\rightarrow$ used & data$\rightarrow$ analysis$\rightarrow$ used$\rightarrow$ analyzed  \\ 
  4 & mortality rate related to infection & number$\rightarrow$ cases & number$\rightarrow$ cases$\rightarrow$ deaths & number$\rightarrow$ cases$\rightarrow$ deaths$\rightarrow$ per  \\ 
  5 & epidemic models & mathematical$\rightarrow$ model & epidemic$\rightarrow$ used$\rightarrow$ model & reproduction$\rightarrow$ number$\rightarrow$ R0$\rightarrow$ model  \\ 
  6 & retrospective clinical study & hospitalized$\rightarrow$ confirmed & hospitalized$\rightarrow$ confirmed$\rightarrow$ without & hospitalized$\rightarrow$ confirmed$\rightarrow$ without$\rightarrow$ diagnosed  \\ 
  7 & literature review & various$\rightarrow$ aspects & several$\rightarrow$ specific$\rightarrow$ aspects & aspects$\rightarrow$ human$\rightarrow$ papillomavirus$\rightarrow$ related  \\ 
  8 & clinical studies on illness development & mortality$\rightarrow$ rate & mortality$\rightarrow$ rate$\rightarrow$ patients & mortality$\rightarrow$ rate$\rightarrow$ patients$\rightarrow$ risk  \\ 
  9 & guidelines for clinical practice  management & clinical$\rightarrow$ practice & clinical$\rightarrow$ practice$\rightarrow$ management & clinical$\rightarrow$ practice$\rightarrow$ management$\rightarrow$ regarding  \\ 
  10 & statistical analysis  (scale assessment) & scale$\rightarrow$ scores & scale$\rightarrow$ scores$\rightarrow$ used & mean$\rightarrow$ score$\rightarrow$ scale$\rightarrow$ scores  \\ 
  11 & cellular immune response & inflammatory$\rightarrow$ macrophages & inflammatory$\rightarrow$ macrophages$\rightarrow$ including & inflammatory$\rightarrow$ loop$\rightarrow$ inflammation$\rightarrow$ production  \\ 
  12 & online education during pandemic & online$\rightarrow$ learning & online$\rightarrow$ learning$\rightarrow$ students & online$\rightarrow$ de$\rightarrow$ student$\rightarrow$ learning  \\ 
  13 & studies on origins & province$\rightarrow$ since & province$\rightarrow$ city$\rightarrow$ china & province$\rightarrow$ since$\rightarrow$ prevention$\rightarrow$ february  \\ 
  14 & public health response to pandemics & public$\rightarrow$ system & public$\rightarrow$ system$\rightarrow$ crisis & public$\rightarrow$ system$\rightarrow$ crisis$\rightarrow$ health  \\ 
  15 & reviews on recent therapeutical approaches & therapeutic$\rightarrow$ approaches & future$\rightarrow$ strategies$\rightarrow$ various & future$\rightarrow$ strategies$\rightarrow$ various$\rightarrow$ approaches  \\ 
  16 & mechanism of cell infection & ACE2$\rightarrow$ binding & ACE2$\rightarrow$ binding$\rightarrow$ expression & receptor$\rightarrow$ ACE2$\rightarrow$ binding$\rightarrow$ expression  \\ 
  17 &  ND & important$\rightarrow$ key & important$\rightarrow$ key$\rightarrow$ critical & important$\rightarrow$ key$\rightarrow$ critical$\rightarrow$ crucial  \\ 
  18 & study of variants & viral$\rightarrow$ genome & genome$\rightarrow$ sequences$\rightarrow$ identified & viral$\rightarrow$ genome$\rightarrow$ sequencing$\rightarrow$ sequence  \\ 
  19 & management of surgical procedures & surgical$\rightarrow$ procedures & surgical$\rightarrow$ procedures$\rightarrow$ patients & surgical$\rightarrow$ procedures$\rightarrow$ patients$\rightarrow$ procedure  \\ 
  20 & instrumental diagnosis & chest$\rightarrow$ CT & chest$\rightarrow$ CT$\rightarrow$ scan & chest$\rightarrow$ CT$\rightarrow$ scan$\rightarrow$ images  \\ 
  21 & contagion diffusion control & contact$\rightarrow$ tracing & contact$\rightarrow$ tracing$\rightarrow$ individuals & contact$\rightarrow$ tracing$\rightarrow$ persons$\rightarrow$ quarantine  \\ 
  22 & impact of pandemics on people (social relations) & social$\rightarrow$ distancing & social$\rightarrow$ distancing$\rightarrow$ isolation & social$\rightarrow$ distancing$\rightarrow$ isolation$\rightarrow$ support  \\ 
  23 & virus Cov-19 infection & virus$\rightarrow$ human & virus$\rightarrow$ human$\rightarrow$ viruses & SARS$\rightarrow$ MERS$\rightarrow$ virus$\rightarrow$ human  \\ 
  24 & pandemics history & help$\rightarrow$ us & help$\rightarrow$ us$\rightarrow$ make & help$\rightarrow$ us$\rightarrow$ make$\rightarrow$ one  \\ 
  25 & comparatiove  analysis of viremic levels & significantly$\rightarrow$ higher & levels$\rightarrow$ significantly$\rightarrow$ higher & significantly$\rightarrow$ higher$\rightarrow$ lower$\rightarrow$ level  \\ 
  26 & access to health of socio-economic groups & household$\rightarrow$ income & household$\rightarrow$ income$\rightarrow$ inequality & population$\rightarrow$ derived$\rightarrow$ nonprobability$\rightarrow$ sample  \\ 
  27 & Studies, evidences & evidence$\rightarrow$ suggest & evidence$\rightarrow$ suggest$\rightarrow$ studies & evidence$\rightarrow$ suggests$\rightarrow$ studies$\rightarrow$ needed  \\ 
  28 & methodological approaches & approach$\rightarrow$ problem & set$\rightarrow$ theory$\rightarrow$ method & approach$\rightarrow$ problem$\rightarrow$ time$\rightarrow$ algorithm  \\ 
  29 & ND & research$\rightarrow$ limitations & research$\rightarrow$ limitations$\rightarrow$ findings & research$\rightarrow$ limitations$\rightarrow$ development$\rightarrow$ process  \\ 
  30 & development in India & urban$\rightarrow$ population & urban$\rightarrow$ population$\rightarrow$ area & urban$\rightarrow$ population$\rightarrow$ planning$\rightarrow$ space  \\ 
  31 & digital technology & digital$\rightarrow$ version & digital$\rightarrow$ scanning$\rightarrow$ calorimetry & digital$\rightarrow$ scanning$\rightarrow$ calorimetry$\rightarrow$ technology  \\ 
  32 & Covid-19 literature & must$\rightarrow$ considered & one$\rightarrow$ especially$\rightarrow$ important & one$\rightarrow$ especially$\rightarrow$ important$\rightarrow$ reasons  \\ 
  33 & abstract structure & methods$\rightarrow$ results & methods$\rightarrow$ results$\rightarrow$ study & materials$\rightarrow$ methods$\rightarrow$ results$\rightarrow$ study  \\ 
  34 & prevention & prevention$\rightarrow$ measures & control$\rightarrow$ prevention$\rightarrow$ measures & prevention$\rightarrow$ measures$\rightarrow$ spread$\rightarrow$ strategies  \\ 
  35 & review studies & systematic$\rightarrow$ review & systematic$\rightarrow$ reviews$\rightarrow$ protocol & systematic$\rightarrow$ review$\rightarrow$ met$\rightarrow$ literature  \\ 
  36 & economic impact & tourism$\rightarrow$ development & tourism$\rightarrow$ development$\rightarrow$ business & tourism$\rightarrow$ development$\rightarrow$ business$\rightarrow$ market  \\ 
  37 & comparative studies & significantly$\rightarrow$ higher & significantly$\rightarrow$ higher$\rightarrow$ rate & significantly$\rightarrow$ higher$\rightarrow$ rate$\rightarrow$ compared  \\ 
  38 & effects of vitamin somministration & levels$\rightarrow$ supplementation & levels$\rightarrow$ supplementation$\rightarrow$ significantly & levels$\rightarrow$ supplementation$\rightarrow$ significantly$\rightarrow$ reduced  \\ 
  39 & data & data$\rightarrow$ collection & data$\rightarrow$ collection$\rightarrow$ analysis & data$\rightarrow$ collection$\rightarrow$ sources$\rightarrow$ used  \\ 
  40 & publication metadata & without$\rightarrow$ use & copyright$\rightarrow$ international$\rightarrow$ use & copyright$\rightarrow$ international$\rightarrow$ use$\rightarrow$ abstract  \\ 
  41 & impact on life quality \& sexual health & quality$\rightarrow$ life & quality$\rightarrow$ life$\rightarrow$ expectancy & quality$\rightarrow$ life$\rightarrow$ expectancy$\rightarrow$ health  \\ 
  42 & symptoms \& clinical manifestations & clinical$\rightarrow$ symptoms & gastrointestinal$\rightarrow$ infection$\rightarrow$ symptoms & clinical$\rightarrow$ symptoms$\rightarrow$ manifestations$\rightarrow$ reported  \\ 
  43 & effects of alcohool \& addictions & alcohol$\rightarrow$ medications & alcohol$\rightarrow$ medications$\rightarrow$ treatment & alcohol$\rightarrow$ medications$\rightarrow$ treatment$\rightarrow$ cannabis  \\ 
  44 & effects on patients with comorbidity & severe$\rightarrow$ disease & patients$\rightarrow$ severe$\rightarrow$ disease & patients$\rightarrow$ severe$\rightarrow$ disease$\rightarrow$ critical  \\ 
  45 & impact of air pollution & concentration$\rightarrow$ meteorological & concentration$\rightarrow$ meteorological$\rightarrow$ conditions & concentration$\rightarrow$ meteorological$\rightarrow$ conditions$\rightarrow$ observed  \\ 
  46 & correlation with heart disease & heart$\rightarrow$ failure & cardiac$\rightarrow$ arrest$\rightarrow$ patients & heart$\rightarrow$ failure$\rightarrow$ arrhythmias$\rightarrow$ acute  \\ 
  47 & ND & new$\rightarrow$ york & new$\rightarrow$ york$\rightarrow$ city & new$\rightarrow$ york$\rightarrow$ city$\rightarrow$ pandemic  \\ 
  48 & community & community$\rightarrow$ engagement & community$\rightarrow$ engagement$\rightarrow$ development & sustainable$\rightarrow$ strategies$\rightarrow$ partnerships$\rightarrow$ key  \\ 
  49 & stages & early$\rightarrow$ especially & early$\rightarrow$ especially$\rightarrow$ initial & early$\rightarrow$ especially$\rightarrow$ initial$\rightarrow$ results  \\ 
  50 & ND & higher$\rightarrow$ risk & adjusted$\rightarrow$ prevalence$\rightarrow$ ratio & adjusted$\rightarrow$ prevalence$\rightarrow$ ratio$\rightarrow$ among  \\ 
  51 & age and gender & older$\rightarrow$ people & older$\rightarrow$ people$\rightarrow$ individuals & older$\rightarrow$ people$\rightarrow$ individuals$\rightarrow$ higher  \\ 
  52 & interview studies & themes$\rightarrow$ emerged & themes$\rightarrow$ emerged$\rightarrow$ three & themes$\rightarrow$ emerged$\rightarrow$ three$\rightarrow$ key  \\ 
  53 & diagnostic tests efficacy & test$\rightarrow$ positive & test$\rightarrow$ positive$\rightarrow$ results & test$\rightarrow$ positive$\rightarrow$ results$\rightarrow$ tests  \\ 
  54 & french & de$\rightarrow$ patients & de$\rightarrow$ la$\rightarrow$ santé & de$\rightarrow$ la$\rightarrow$ santé$\rightarrow$ plus  \\ 
  55 & pandemic management & pandemic$\rightarrow$ response & pandemic$\rightarrow$ response$\rightarrow$ crisis & pandemic$\rightarrow$ response$\rightarrow$ crisis$\rightarrow$ management  \\ 
  56 & nursing programs residents & programs$\rightarrow$ program & programs$\rightarrow$ program$\rightarrow$ value & programs$\rightarrow$ program$\rightarrow$ value$\rightarrow$ virtual  \\ 
  57 & effective strategies & effective$\rightarrow$ strategies & effective$\rightarrow$ strategies$\rightarrow$ strategy & effective$\rightarrow$ strategies$\rightarrow$ strategy$\rightarrow$ approach  \\ 
  58 & policies for spread containment & control$\rightarrow$ measures & containment$\rightarrow$ interventions$\rightarrow$ implemented & containment$\rightarrow$ interventions$\rightarrow$ implemented$\rightarrow$ reduce  \\ 
  59 & technologies & surface$\rightarrow$ structure & process$\rightarrow$ storage$\rightarrow$ efficiency & process$\rightarrow$ storage$\rightarrow$ parameters$\rightarrow$ ph  \\ 
  60 & susceptibility to viral infection in epathitys & viral$\rightarrow$ infection & viral$\rightarrow$ infection$\rightarrow$ infections & viral$\rightarrow$ infection$\rightarrow$ infections$\rightarrow$ caused  \\ 
  61 & data for the study & analysis$\rightarrow$ used & data$\rightarrow$ analysis$\rightarrow$ used & data$\rightarrow$ analysis$\rightarrow$ used$\rightarrow$ study  \\ 
  62 & sociopolitical aspects & political$\rightarrow$ cultural & political$\rightarrow$ questions$\rightarrow$ social & ethical$\rightarrow$ issues$\rightarrow$ legal$\rightarrow$ political  \\ 
  63 & cancer therapy & cancer$\rightarrow$ treatment & cancer$\rightarrow$ treatment$\rightarrow$ patients & cancer$\rightarrow$ patients$\rightarrow$ diagnosis$\rightarrow$ therapy  \\ 
  64 & pediatry & children$\rightarrow$ adolescents & children$\rightarrow$ adolescents$\rightarrow$ young & children$\rightarrow$ adolescents$\rightarrow$ young$\rightarrow$ families  \\ 
  65 & oftalmic & ocular$\rightarrow$ surface & eye$\rightarrow$ ocular$\rightarrow$ surface & ocular$\rightarrow$ surface$\rightarrow$ conditions$\rightarrow$ presence  \\ 
  66 & comorbidity diabetes & chronic$\rightarrow$ diseases & chronic$\rightarrow$ diseases$\rightarrow$ disease & chronic$\rightarrow$ diseases$\rightarrow$ disease$\rightarrow$ diabetes  \\ 
  67 & rehabilitation after intervention & intervention$\rightarrow$ program & intervention$\rightarrow$ program$\rightarrow$ improve & intervention$\rightarrow$ program$\rightarrow$ improve$\rightarrow$ organizational  \\ 
  68 & pandemics in the us & state$\rightarrow$ local & state$\rightarrow$ local$\rightarrow$ federal & state$\rightarrow$ local$\rightarrow$ federal$\rightarrow$ level  \\ 
  69 & immunity \& specificity of response & igg$\rightarrow$ antibodies & igg$\rightarrow$ antibodies$\rightarrow$ antibody & igg$\rightarrow$ antibodies$\rightarrow$ antibody$\rightarrow$ response  \\ 
  70 & online (mis)information & information$\rightarrow$ sources & information$\rightarrow$ sources$\rightarrow$ related & information$\rightarrow$ sources$\rightarrow$ regarding$\rightarrow$ SDM  \\ 
  71 & variability & three$\rightarrow$ different & different$\rightarrow$ patterns$\rightarrow$ populations & three$\rightarrow$ different$\rightarrow$ patterns$\rightarrow$ populations  \\ 
  72 & healthcare & health$\rightarrow$ facilities & health$\rightarrow$ facilities$\rightarrow$ system & health$\rightarrow$ facilities$\rightarrow$ system$\rightarrow$ planning  \\ 
  73 & paper info & web$\rightarrow$ response & web$\rightarrow$ response$\rightarrow$ one & web$\rightarrow$ response$\rightarrow$ disclosure$\rightarrow$ including  \\ 
  74 & clinical cases reports & patient$\rightarrow$ presented & patient$\rightarrow$ presented$\rightarrow$ developed & patient$\rightarrow$ presented$\rightarrow$ developed$\rightarrow$ acute  \\ 
  75 & numbers & one$\rightarrow$ three & one$\rightarrow$ three$\rightarrow$ study & one$\rightarrow$ three$\rightarrow$ study$\rightarrow$ five  \\ 
  76 & UK pandemic timeline & first$\rightarrow$ time & first$\rightarrow$ time$\rightarrow$ UK & first$\rightarrow$ time$\rightarrow$ UK$\rightarrow$ weeks  \\ 
  77 & time \& duration & median$\rightarrow$ time & median$\rightarrow$ time$\rightarrow$ period & median$\rightarrow$ time$\rightarrow$ period$\rightarrow$ day  \\ 
  78 & geographic (countries) & world$\rightarrow$ health & around$\rightarrow$ world$\rightarrow$ health & around$\rightarrow$ world$\rightarrow$ health$\rightarrow$ response  \\ 
  79 & impact (on energy market) & paper$\rightarrow$ shows & paper$\rightarrow$ shows$\rightarrow$ impact & paper$\rightarrow$ shows$\rightarrow$ impact$\rightarrow$ examine  \\ 
  80 & antiviral treatments & antiviral$\rightarrow$ drugs & antiviral$\rightarrow$ drugs$\rightarrow$ used & antiviral$\rightarrow$ drugs$\rightarrow$ used$\rightarrow$ treatment  \\ 
  81 & medical organization and management & emergency$\rightarrow$ medicine & medical$\rightarrow$ personnel$\rightarrow$ emergency & emergency$\rightarrow$ medicine$\rightarrow$ doctors$\rightarrow$ management  \\ 
  82 & mental health & psychological$\rightarrow$ stress & stress$\rightarrow$ disorder$\rightarrow$ symptoms & psychological$\rightarrow$ stress$\rightarrow$ disorder$\rightarrow$ symptoms  \\ 
  83 & clinical studies on illness development & clinical$\rightarrow$ trial & clinical$\rightarrow$ trial$\rightarrow$ outcomes & clinical$\rightarrow$ trial$\rightarrow$ outcomes$\rightarrow$ registered  \\ 
  84 & questionnaire studies & study$\rightarrow$ participants & study$\rightarrow$ participants$\rightarrow$ conducted & study$\rightarrow$ participants$\rightarrow$ conducted$\rightarrow$ aimed  \\ 
  85 & damage mechansims & nervous$\rightarrow$ system & nervous$\rightarrow$ system$\rightarrow$ involvement & nervous$\rightarrow$ system$\rightarrow$ damage$\rightarrow$ including  \\ 
  86 & anti-inflammatory treatment & patients$\rightarrow$ received & patients$\rightarrow$ received$\rightarrow$ treatment & treatment$\rightarrow$ receiving$\rightarrow$ tocilizumab$\rightarrow$ systemic  \\ 
  87 & effects if pandemic & pandemic$\rightarrow$ impact & pandemic$\rightarrow$ impact$\rightarrow$ affected & pandemic$\rightarrow$ impact$\rightarrow$ affected$\rightarrow$ changes  \\ 
  88 & antiviral drug molecules & drug$\rightarrow$ target & antiviral$\rightarrow$ activity$\rightarrow$ compounds & antiviral$\rightarrow$ targets$\rightarrow$ identified$\rightarrow$ compounds  \\ 
  89 & risk factors & risk$\rightarrow$ factors & risk$\rightarrow$ factor$\rightarrow$ exposure & risk$\rightarrow$ factor$\rightarrow$ exposure$\rightarrow$ factors  \\ 
  90 & neurological and cognitive impairment & brain$\rightarrow$ including & neurologic$\rightarrow$ manifestations$\rightarrow$ including & neurologic$\rightarrow$ manifestations$\rightarrow$ including$\rightarrow$ brain  \\ 
  91 & respiratory failure & patients$\rightarrow$ severe & patients$\rightarrow$ severe$\rightarrow$ forms & patients$\rightarrow$ local$\rightarrow$ contestation$\rightarrow$ different  \\ 
  92 & predictive models & predictive$\rightarrow$ value & predictive$\rightarrow$ value$\rightarrow$ model & predictive$\rightarrow$ value$\rightarrow$ model$\rightarrow$ used  \\ 
  93 & behavioural studies & behavior$\rightarrow$ intention & positive$\rightarrow$ relationship$\rightarrow$ perceived & positive$\rightarrow$ relationship$\rightarrow$ perceived$\rightarrow$ behavioral  \\ 
  94 & sampling viral rna & viral$\rightarrow$ RNA & viral$\rightarrow$ RNA$\rightarrow$ load & viral$\rightarrow$ RNA$\rightarrow$ loads$\rightarrow$ samples  \\ 
  95 & pandemic impact & current$\rightarrow$ pandemic & current$\rightarrow$ pandemic$\rightarrow$ caused & current$\rightarrow$ pandemic$\rightarrow$ caused$\rightarrow$ coronavirus  \\ 
  96 & orthopedic traumas & wound$\rightarrow$ healing & wound$\rightarrow$ healing$\rightarrow$ repair & wound$\rightarrow$ healing$\rightarrow$ repair$\rightarrow$ attempt  \\ 
  97 & vaccine and immunity & vaccine$\rightarrow$ development & vaccine$\rightarrow$ development$\rightarrow$ efficacy & vaccine$\rightarrow$ development$\rightarrow$ candidate$\rightarrow$ vaccines  \\ 
  98 & icu and patinet care & ICU$\rightarrow$ patients & care$\rightarrow$ unit$\rightarrow$ patients & patients$\rightarrow$ admitted$\rightarrow$ hospitalized$\rightarrow$ hospital  \\ 
  99 & cellular mechanisms & gene$\rightarrow$ expression & gene$\rightarrow$ expression$\rightarrow$ genes & gene$\rightarrow$ expression$\rightarrow$ genes$\rightarrow$ involved  \\ 
  100 & pandemic & coronavirus$\rightarrow$ disease & novel$\rightarrow$ coronavirus$\rightarrow$ disease & coronavirus$\rightarrow$ disease$\rightarrow$ virus$\rightarrow$ spread  \\ 
  101 & health care personnel and pandemic & healthcare$\rightarrow$ workers & healthcare$\rightarrow$ workers$\rightarrow$ work & healthcare$\rightarrow$ workers$\rightarrow$ work$\rightarrow$ providers  \\ 
  102 & transplant & dental$\rightarrow$ treatment & clinical$\rightarrow$ course$\rightarrow$ dental & clinical$\rightarrow$ course$\rightarrow$ dental$\rightarrow$ treatment  \\ 
  103 & structural dynamics & complex$\rightarrow$ interactions & complex$\rightarrow$ interactions$\rightarrow$ structure & complex$\rightarrow$ interactions$\rightarrow$ structure$\rightarrow$ dynamics  \\ 
  104 & treatment of cardiovascular complications & stroke$\rightarrow$ acute & stroke$\rightarrow$ acute$\rightarrow$ severe & ischemic$\rightarrow$ thrombotic$\rightarrow$ outcomes$\rightarrow$ complications  \\ 
  105 & pandemic in selected countires & italy$\rightarrow$ spain & italy$\rightarrow$ spain$\rightarrow$ one & italy$\rightarrow$ spain$\rightarrow$ one$\rightarrow$ france  \\ 
  106 & spanish & de$\rightarrow$ la & de$\rightarrow$ la$\rightarrow$ pandemia & de$\rightarrow$ la$\rightarrow$ pandemia$\rightarrow$ esta  \\ 
  107 & AI models & deep$\rightarrow$ learning & deep$\rightarrow$ learning$\rightarrow$ de & deep$\rightarrow$ learning$\rightarrow$ de$\rightarrow$ performance  \\ 
  108 & physical activity and exercise & home$\rightarrow$ time & home$\rightarrow$ time$\rightarrow$ confinement & home$\rightarrow$ time$\rightarrow$ confinement$\rightarrow$ le  \\ 
  109 & statistics & control$\rightarrow$ groups & control$\rightarrow$ groups$\rightarrow$ significant & control$\rightarrow$ groups$\rightarrow$ significant$\rightarrow$ difference  \\ 
  110 & animal models \& in vitro & stem$\rightarrow$ reduced & stem$\rightarrow$ reduced$\rightarrow$ lung & stem$\rightarrow$ reduced$\rightarrow$ lung$\rightarrow$ human  \\ 
  111 & bacterial resistance to antibacterials & bacterial$\rightarrow$ infections & bacterial$\rightarrow$ infections$\rightarrow$ isolated & bacterial$\rightarrow$ pathogens$\rightarrow$ identified$\rightarrow$ found  \\ 
  112 & time studies & total$\rightarrow$ number & significant$\rightarrow$ increase$\rightarrow$ number & ED$\rightarrow$ emergency$\rightarrow$ visits$\rightarrow$ total  \\ 
  113 & body weight and metabolism & healthy$\rightarrow$ eating & BMI$\rightarrow$ obesity$\rightarrow$ metabolic & overweight$\rightarrow$ BMI$\rightarrow$ obesity$\rightarrow$ metabolic  \\ 
  114 & neo/perinatal infections & included$\rightarrow$ among & intrauterine$\rightarrow$ transplacental$\rightarrow$ flux & intrauterine$\rightarrow$ insemination$\rightarrow$ doses$\rightarrow$ low  \\ 
  115 & telemedicine & patient$\rightarrow$ care & telemedicine$\rightarrow$ visits$\rightarrow$ telehealth & telemedicine$\rightarrow$ visits$\rightarrow$ telehealth$\rightarrow$ access  \\ 
  116 & coronavirus disease progression & coronavirus$\rightarrow$ disease & coronavirus$\rightarrow$ disease$\rightarrow$ progression & coronavirus$\rightarrow$ de$\rightarrow$ un$\rightarrow$ autre  \\ 
  117 & report & reported$\rightarrow$ among & reported$\rightarrow$ among$\rightarrow$ less & reported$\rightarrow$ less$\rightarrow$ likely$\rightarrow$ report  \\ 
  118 & severe respiratory syndrome & acute$\rightarrow$ respiratory & acute$\rightarrow$ respiratory$\rightarrow$ syndrome & acute$\rightarrow$ respiratory$\rightarrow$ syndrome$\rightarrow$ coronavirus  \\ 
  119 & related works & related$\rightarrow$ pandemic & related$\rightarrow$ pandemic$\rightarrow$ including & status$\rightarrow$ related$\rightarrow$ pandemic$\rightarrow$ including  \\ 
  120 & research (publications) & scientific$\rightarrow$ research & scientific$\rightarrow$ studies$\rightarrow$ research & research$\rightarrow$ field$\rightarrow$ community$\rightarrow$ development  \\ 
   \hline
    \caption{Human labels and automated n-gram label proposals based on the proposed heuristic (see Section \ref{sec:methods}). Human labels have been defined by a medical expert by looking (only) to the list of the top 25 words of the topic.
    ND stands for not determined.
    }
    \label{tab:topic_labels}
\end{longtable}
}

\end{document}